\documentclass[aps, prb, superscriptaddress, amsmath, amssymb, reprint, floatfix]{revtex4-2}
\usepackage{siunitx, physics, mathtools, mathrsfs, bm, cancel, xfrac, nth} 
\usepackage{amsthm}
	\newtheorem{lemma}{Lemma}
	
\usepackage[dvipsnames]{xcolor}
\usepackage{graphicx, circuitikz, tikz,  tkz-euclide}
	\usetikzlibrary{calc, angles, positioning, intersections, quotes, decorations.markings}
	\usetikzlibrary{shapes, arrows, fadings}
\usepackage{pgfplots}
	\pgfplotsset{compat=newest}
\usepackage{hyperref, bookmark, cleveref} 
\hypersetup{
	colorlinks=true,
	linkcolor=blue,
	breaklinks=true,
	filecolor=blue,      
	urlcolor=blue,
	citecolor=blue
}
\usepackage{ulem}

\newcommand{\ycl}[1]{\textcolor{black}{#1}}

\newcommand{\jmax}{\mathsf{d}}
\newcommand{\Zjmax}{\mathbb{Z}_{\jmax}}

\renewcommand{\emph}[1]{\textit{#1}}

\begin{document}
\title{Circuit Quantisation from First Principles}
\author{Yun-Chih Liao}
\email{uqyliao1@uq.edu.au}
\affiliation{ARC Centre of Excellence for Engineered Quantum Systems}
\affiliation{School of Mathematics and Physics, The University of Queensland,  Brisbane, Queensland 4072, Australia}

\author{Ben J. Powell}
\affiliation{School of Mathematics and Physics, The University of Queensland,  Brisbane, Queensland 4072, Australia}

\author{Thomas M. Stace}
\email{stace@physics.uq.edu.au}
\affiliation{ARC Centre of Excellence for Engineered Quantum Systems}
\affiliation{School of Mathematics and Physics, The University of Queensland,  Brisbane, Queensland 4072, Australia}

\begin{abstract}
Superconducting circuit quantisation conventionally starts from classical Euler-Lagrange circuit equations-of-motion.  Invoking the correspondence principle yields a canonically quantised  description of circuit dynamics over a bosonic Hilbert space. This approach has been very successful for describing experiments, but implicitly starts from the classical Ginsberg-Landau mean field theory for the circuit.  Here we employ a different approach which starts from a microscopic fermionic Hamiltonian for interacting electrons, whose ground space is described by the Bardeen-Cooper-Schrieffer (BCS) many-body wavefuction that underpins conventional superconductivity. We introduce the BCS ground-space as a subspace of the full fermionic Hilbert space, and show that projecting the electronic Hamiltonian onto this subspace yields the standard Hamiltonian terms for Josephson junctions, capacitors and inductors, from which standard quantised circuit models follow. 
This approach does not impose a spontaneously broken symmetry so that it consistently describes quantised circuits that support superpositions of phases, and the canonical commutation relations between phase and charge are derived from the underlying fermionic commutation properties, rather than imposed.
By expanding the projective subspace, this approach can be  extended to  describe phenomena outside the BCS ground space, including  quasiparticle excitations.
\end{abstract}

\maketitle

\section{Introduction}

Superconducting devices are one of the leading candidates for building quantum computers, showing great flexibility for implementing qubits and gates \cite{ref:SCcircuit_QI_outlook, ref:SCqubits_wilhelm}.
The low working temperature reduces the noise, providing for long coherence time \cite{ref:SC_coherent}.
The strong coupling between superconducting qubits and electromagnetic fields allows us to  control and operate the system \cite{ref:1Dtransmon_control}.

The conventional approach to quantising superconducting circuits \cite{ref:decoherence_SCqubits, ref:circuit_quantization} starts with the classical circuit equations of motion given by Kirchoff's laws, which are derived from  conservation laws and constitutive relations for the circuit.  For conservative circuits, these equations are the Euler-Lagrange equations, which can be generated from a Lagrangian.  Canonical quantisation then proceeds, via the correspondence principle, by postulating commutation relations between classically conjugate coordinates describing the circuit phase space.  

These classical equations describe the dynamics of the Ginsburg-Landau (GL) superconducting order parameters, $\phi_j$, defined at the nodes $j$ of the circuit.  The GL order parameter is itself a classical mean field, approximating an underlying microscopic electronic Hamiltonian, and the (re-)quantisation procedure of the classical Euler-Lagrange equations for $\phi_j$, described above, yields the conventional bosonic theory for $\hat{\phi}_j$ with the conjugate coordinates being the circuit charges, $\hat{n}_j$.

\begin{figure}[!]
    \centering
    \includegraphics[width=0.7\columnwidth]{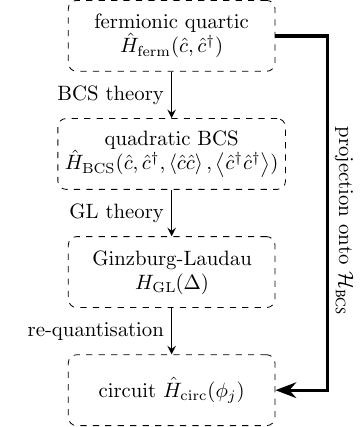}
    \caption{
        The conventional approach to circuit quantisation, along the left pathway, conceptually begins with a microscopic electronic theory of interacting fermions in a metal, which proceeds via a BCS ground-state ansatz to a classical Ginsburg-Landau mean-field theory (GL-MFT) of superconductivity in a metal, which is finally re-quantised to produce a quantised theory of a circuit.  The approach described here (right pathway) starts at the same point, but uses the GL-MFT as inspiration to define a low-energy BCS subspace, $\mathcal{H}_{\rm BCS}$ (see \cref{eq:HBCS}) and then directly projects the dynamics onto this subspace, resulting in the effective low-energy circuit Hamiltonian. 
    }
    \label{fig:flowchart}
\end{figure}

This standard approach, which is illustrated in \cref{fig:flowchart}, is practically satisfactory for modelling the vast majority of experiments, however it is conceptually unfulfilling for several reasons. Firstly, we already have a very successful microscopic quantum theory of superconductivity, namely, Bardeen-Cooper-Schrieffer (BCS) theory, which describes the dynamics of the microscopic electronic Hilbert space of interacting electrons in a metal.  As such, there should be a ``direct'' pathway from the microscopic electronic Hamiltonian to the circuit Hamiltonian, without leaving the fermionic Hilbert space.  In this goal, there is a close analogy in modelling low-dimensional semiconducting systems (such as quantum dots \cite{ref:quantumDot_Milburn} or quantum Hall fluids \cite{ref:mesoscopic_QO, ref:mesoscopicChannel_QHE}), in which a microscopic Fermi-liquid theory of a semiconductor is projected onto a low-dimensional subspace spanned by the low-energy bound states of a spatially inhomogeonous, weakly-confining potential landscape, yielding effective descriptions in terms of mass-renormalised, few-electron (or hole) systems \cite{ref:mesoscopic_QO}. 

Secondly,  GL mean-field theory (MFT) does not typically include phenomena such as quasiparticles or quantum fluctuations in the gap parameter, $\Delta$, which are otherwise added `by-hand' after re-quantisation \cite{ref:Higgs_dynamics_Floquent, ref:quasiparticles_SCqubits}.  

Thirdly, when re-quantising GL mean-field theory (MFT), the bosonic canonical commutation relation, $[\hat\phi,\hat n]=i$, is postulated \cite{widom1979,Widom1981,https://doi.org/10.1002/cta.2359}, or see \cite{Pandey_2024} for variations in this approach. However, it is not clear that this is consistent with the Stone-von Neumann theorem when the spectrum of $\hat\phi$ has compact support on the interval $[-\pi,\pi)$, and the spectrum of $\hat n$ is $\mathbb{Z}$. In fact, we expect to be able to derive the correct commutation relations directly from the underlying fermionic degrees of freedom.  

Fourthly, GL mean-field theory assumes spontaneous symmetry breaking of the superconducting order parameter.  However in the quantised circuit, we expect coherent quantum superpositions of (relative) phases \cite{Mizel2024}, which is conceptually at odds with the spontaneous realisation of only one amongst the large manifold of allowed phases.  

Lastly, the classical GL MFT for the order parameter $\phi$ is indifferent to the support of $\phi$: descriptions in which $\phi$ is compact, $\phi\in [-\pi,\pi)$, and non-compact, $\phi\in\mathbb{R}$, are entirely equivalent at a classical level.  In contrast, the quantum descriptions of these two alternatives, or anything in between, are not equivalent to one another \cite{ref:compact-noncompact}.  Starting from the classical GL MFT therefore cannot resolve which alternative provides the more accurate description. 

The objective of this work is to provide a somewhat formal derivation of the Hamiltonian for superconducting circuits, starting from the microscopic theory of interacting electrons.  To do this, we reformulate BCS theory slightly, by introducing a low energy Hilbert space of the of the set of BCS ground states, $\mathcal{H}_{\rm BCS}=\text{span}\{\ket{\Psi(\phi_j)}\}_{\phi_j\in \mathcal{Z}\subset[-\pi,\pi)}$, where $\ket{\Psi(\phi_j)}$ is a BCS ground state (defined later), parameterised by a BCS phase angle $\phi_j$.  We define a projector, $P$ onto $\mathcal{H}_{\rm BCS}$, and use this projector to directly construct the low energy Hamiltonian from the microscopic electronic Hamiltonian, as illustrated in \cref{fig:flowchart}. 

We note that our projector-based approach to circuit quantisation has some similarity to early work by \citet{Leggett1966} and \citet{Ambegaokar1982}, which evaluate expectations of microscopic interaction terms directly.  The underlying analytical approach we take is similar to the early work by  Widom \emph{et al.} \cite{widom1979,Widom1981} and  recently  by \citet{Mizel2024}, starting with a low-energy vector space defined by the BCS ground states.

We apply this projector-based approach to re-analyse the conventional superconducting toolbox: capacitors, inductors, and Josephson junctions, and we show that we recover the expected results in the limit of large systems.  

In what follows, we formalise this approach, and derive a number features of the model, including
the emergent, discrete basis for the compact Hilbert subspace $\mathcal{H}_{\text{BCS}}$, the corresponding dimensionality of the low-energy subspace, the canonical commutation relations, and we identify the relevant low-energy Hilbert space of standard lumped-element superconducting components.  We  specifically address the difference between the effective Hilbert space for  inductors and Josephson junctions.
With this approach, we describe the connection between the microscopic  BCS theory of superconductors \cite{ref:BCS_superconductivity_theory} and the mesoscopic electronic description of a Cooper-pair-box (CPB) or transmon  \cite{ref:CPB}, as well as devices that include an inductive element.

This paper is organised as follows:
We firstly compute the overlap of BCS states and discuss the requirements for a Hilbert space that is approximately spanned by them in Section~\ref{sec:LowEspace}.
In Section~\ref{sec:LCJprojection}, we project a microscopic description of an inductor-capacitor-junction (LCJ) circuit onto the low energy subspace, showing consistency with the conventional result.
Finally we discuss some implications   that follow from this analysis, and conclude with potential directions for further work.

\begin{figure}[!]
	\centering
    \includegraphics[width=0.49\columnwidth]{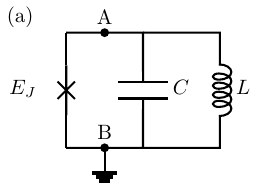}
    \includegraphics[width=0.49\columnwidth]{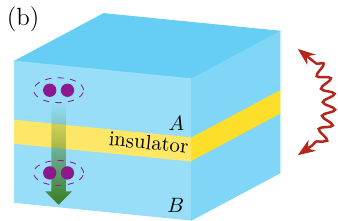}
	\caption{
	(a) 
	A standard LCJ circuit consists of a inductor $L$, a capacitor $C$, and a Josephson junction $E_J$ made by superconductors.
	(b) 
	A cartoon of the microscopic elements of a Josephson junction, consisting of two superconducting islands (blue) separated by an insulating tunnel barrier (yellow). 
	At low temperature, the electrons bind into Cooper pairs which tunnel coherently across the (green arrow).
	In addition, the charges inside the two superconducting islands have a capacitive Coulomb interaction (red arrow). 
	}
	\label{fig:system}
\end{figure}

\subsection{ Microscopic Electronic Hamiltonian}
To introduce notation, we  consider two pieces of superconducting metal separated by an insulator, which is corresponds to the capacitor and Josephson junction parts of the LCJ circuit shown in \cref{fig:system}. 
For the Josephson junction illustrated in \cref{fig:system}b, the superconducting islands $A$ and $B$  are described by their respecitve Hamiltonians  \cite{ref:annett}, 
\begin{equation}    \label{eq:Hamiltonian_SCwoMFA}
    H_{A,B} = H_{K_{A,B}}+H_{I_{A,B}},
\end{equation} where the microscopic electron kinetic and interaction terms are given respectively in the momentum basis as
\begin{align} 
\label{eq:HKA}
H_{K_A}&=
	\sum_{\bm{k}_A \sigma_A}
	\left(
		\epsilon_{\bm{k}_A}	-	\mu_A
	\right)
	c_{\bm{k}_A \sigma_A}^\dagger	c_{\bm{k}_A \sigma_A},
\\
H_{I_A}&=
-
	\abs{g}^2
	\sum_{\bm{k}_A\bm{k}_A'}
	c_{\bm{k}_A\uparrow}^\dagger	c_{-\bm{k}_A\downarrow}^\dagger
	c_{-\bm{k}_A'\downarrow}		c_{\bm{k}_A'\uparrow},
\end{align}
and similarly for $H_B$, where 
$\epsilon_{\bm{k}}$ is the single-particle kinetic energy of an electron with momentum $\bm{k}$, 
$\mu$ is the chemical potential,  $\sigma$ labels the spin index, $g$ is the effective coupling strength between electrons, and
$c_{x}^\dagger$ and $c_{x}$ are  creation and annihilation operators for electron mode $x$.

Assuming point-like metallic islands, the Coulomb interaction between electrons on island $A$ and $B$ is given by
\begin{align}
	H_\text{Coul}
&=
	\lambda_C \sum_{\bm{k}_A  \sigma_A}
	c_{\bm{k}_A \sigma_A}^\dagger  c_{\bm{k}_A \sigma_A}
	\sum_{ \bm{k}_B \sigma_B}\nonumber
	c_{\bm{k}_B \sigma_B}^\dagger  c_{\bm{k}_B \sigma_B},\\
	&=4\lambda_C  \hat{N}^{\rm CP}_A\hat{N}^{\rm CP}_B\label{eq:Hamiltonian_coulomb}
\end{align}
where \mbox{$
    \hat{N}^{\rm CP}_A
=
    \sum_{\bm{k}_A\sigma_A}
			c_{\bm{k}_A\sigma_A}^{\dagger}   c_{\bm{k}_A\sigma_A}
 /2
$}
and $\lambda_C$ is the effective Coulomb interaction strength between the two point-like superconductor.

The electrons also tunnel  through the insulating barrier, with tunneling Hamiltonian
\begin{equation} \label{eq:Hamiltonian_tunneling}
	H_T
=
	\displaystyle\sum_{\bm{k}_A \bm{k}_B  \sigma_A \sigma_B}
	t_{\bm{k}_{A}\bm{k}_{B}}
	c_{\bm{k}_A \sigma_A}^{\dagger}  c_{\bm{k}_B \sigma_B}
	+ {\rm h.c.},
\end{equation} 
where 
$t_{\bm{k}_A\bm{k}_B}$ is the tunneling matrix element between modes in the different islands.  
The microscopic  Hamiltonian for an electrical circuit including tunnel junctions is then 
\begin{equation}
	H_{\text{LCJ}}
=
	H_A + H_B + H_\text{Coul} + H_T. 
\end{equation}

Conventionally, the  quartic Hamiltonian \cref{eq:Hamiltonian_SCwoMFA} describing each metallic island is treated approximately by factoring into a quadratic Hamiltonian with coefficients computed in mean-field theory, yielding
\begin{align}    
    H_{\text{BCS}}
=&
    \sum_{\bm{k} s}
	(
		\epsilon_{\bm{k}}\!	-	\!\mu
	)
	c_{\bm{k} s}^\dagger	c_{\bm{k} s}\nonumber
\\
&-
    \abs{g}^2
    \sum_{\bm{k}\bm{k}'}
    \big(\langle{
		c_{\bm{k}\uparrow}^\dagger	c_{-\bm{k}\downarrow}^\dagger
	}\rangle
	c_{-\bm{k}'\downarrow}	c_{\bm{k}'\uparrow}
+{\rm h.c.}\big),
\label{eq:Hamiltonian_BCSquadratic}\\
=&
    \sum_{\bm{k} s}
	\left(
		\epsilon_{\bm{k}}	\!-\!	\mu
	\right)
	c_{\bm{k} s}^\dagger	c_{\bm{k} s}
 \!-\!
    \sum_{\bm{k}}
    (
        \Delta^*    c_{-\bm{k}\downarrow}   c_{\bm{k}\uparrow}
        +{\rm h.c.} 
    ).
\end{align}
where $
    \Delta
=
    \abs{g}^2   \sum_{\bm{k}}
    \ev{c_{-\bm{k}\downarrow}   c_{\bm{k}\uparrow}}
=|\Delta| e^{i \phi}$ is the well-known complex-valued superconducting order parameter.
The quadratic Hamiltonian is diagonalised using the Bogoliubov-Valatin transformation to construct the BCS ground state
\begin{equation} 
	\ket{\Psi(\Delta)}
=
	\prod_{\bm{k}}
	\left(
		u_{\bm{k}}(|\Delta|) + v_{\bm{k}}(|\Delta|) e^{i \phi}
		c_{\bm{k}\uparrow}^{\dagger} c_{-\bm{k}\downarrow}^{\dagger}
	\right)
	\ket{0},\label{eq:BCSGS}
\end{equation}
where $u_{\bm{k}}$, $v_{\bm{k}}$ are variational parameters used to minimise the total energy,  $\ket{0}$ is the vacuum of the electronic Fock space, and \mbox{$\phi\equiv\arg(\Delta)\in[-\pi,\pi)$} is an undetermined phase angle \cite{ref:annett, ref:Coleman}. The self-consistent mean field  theory is found by requiring that the expectation in the definition of $\Delta$ is taken with respect to \cref{eq:BCSGS}. 
The Bogoliubov coefficients that minimise the self-consistent energy of \cref{eq:Hamiltonian_BCSquadratic} are  given by 
\begin{align} \label{eq:BogoCoeff}
u_{\bm{k}}^2
=&\big(1 + ({\epsilon_{\bm{k}} - \mu})/{E_{\bm{k}}}\big)/2\text{ and}\nonumber\\
v_{\bm{k}}^2=&\big(1 - ({\epsilon_{\bm{k}} - \mu})/{E_{\bm{k}}}\big)/2,
\end{align}
where 
\mbox{$
E_{\bm{k}} 
\equiv 
\sqrt{(\epsilon_{\bm{k}} - \mu)^2 + \Delta^2}
$} \cite{ref:annett}.

 We use this standard approach to motivate a choice of low energy basis states, but depart from it in that we do not impose the symmetry breaking implied by taking the partial expectations that transforms the microscopic quartic Hamiltonian in \cref{eq:Hamiltonian_SCwoMFA} into the quadratic form in \cref{eq:Hamiltonian_BCSquadratic}.

\section{Low-Energy BCS Subspace} \label{sec:LowEspace}
In this section, we use the \mbox{phase-symmetry-broken} BCS ground state $\ket{\Psi(\Delta)}$ to motivate the introduction of a low-energy Hilbert space spanned by a set of linearly independent BCS ground states $\{\ket{\Psi(\phi)}\}_{\phi\in[-\pi,\pi)}$.

To understand the structure of this Hilbert space, we analyse the overlap between BCS states from this set, and determine conditions under which we can construct an approximately orthogonal and complete basis. 
We then construct ground-space phase and number operators, and connect the resulting phase-number commutator to the well-known Pegg-Barnett theory \cite{ref:Pegg&Barnett}.

\subsection{BCS Ground Space} \label{sec:BCSstate}
Motivated by \cref{eq:BCSGS}, we  define a family of BCS-type ground state wavefunctions for a superconducting island as \cite{ref:BCS_superconductivity_theory,Mizel2024}
\begin{equation} \label{eq:BCS_state}
	\ket{\Psi(\phi_j)}
=
	{\prod}_{\bm{k}}
	\left(
		u_{\bm{k}} + v_{\bm{k}} e^{i\phi_j}
		c_{\bm{k}\uparrow}^{\dagger} c_{-\bm{k}\downarrow}^{\dagger}
	\right)
	\ket{0},
\end{equation}
where $\phi_j\in [-\pi,\pi)$ parameterises the allowed set of BCS states.    
We adopt the subscript $j$ to preempt our intent to restrict the allowed phases to a discrete subset \mbox{$\mathcal{Z}=\{\phi_j\}_{j\in \Zjmax}$} where $\jmax$ is the effective dimension of the Hilbert space, which we will compute below. 


The core of our approach here is to define a basis of BCS ground states $\{\ket{\Psi(\phi_j)}\}_{\phi_j\in \mathcal{Z}}$ and the corresponding low-energy Hilbert space 
\begin{equation}
    \mathcal{H}_{\rm BCS}=\text{span}\{\ket{\Psi(\phi_j)}\}_{\phi_j\in \mathcal{Z}}.\label{eq:HBCS}
\end{equation}  
Projecting the microscopic electronic Hamiltonian into this subspace allows us to construct the low-energy circuit theory without passing through the conventional GL MFT  followed by a re-quantisation procedure.

For a system consisting of multiple superconducting islands, the  total Hilbert space is formed from a tensor product of the island subsystems.  For example the state of  a system  consisting of metallic islands $A$ and $B$ is spanned by the tensor product of single-island states as
\begin{align} \label{eq:BCS_state2}
	\ket{\Psi_{AB}(\phi_{A},\phi_{B})} 
\nonumber
&=
	\ket{\Psi_A(\phi_A)} \otimes \ket{\Psi_B(\phi_B)} 
\\\nonumber
&=
	\prod_{\bm{k}_A, \bm{k}_B}
	( 
		u_{\bm{k}_A} + v_{\bm{k}_A} e^{i\phi_{A}} 
		c_{\bm{k}_A\uparrow}^{\dagger}  c_{-\bm{k}_A\downarrow}^{\dagger}  
	)
\\
&\quad\quad\times
	( 
		u_{\bm{k}_B} + v_{\bm{k}_B}  e^{i\phi_{B}} 
		c_{\bm{k}_B\uparrow}^{\dagger}  c_{-\bm{k}_B\downarrow}^{\dagger}  
	)
	\ket{00},\nonumber\\
 &=
	\prod_{\bm{k}_A, \bm{k}_B}
	( 
		u_{\bm{k}_A} + v_{\bm{k}_A} e^{i\phi_{AB}} 
		c_{\bm{k}_A\uparrow}^{\dagger}  c_{-\bm{k}_A\downarrow}^{\dagger}  
	)\nonumber
\\
&\quad\quad\times
	( 
		u_{\bm{k}_B} + v_{\bm{k}_B}  
		c_{\bm{k}_B\uparrow}^{\dagger}  c_{-\bm{k}_B\downarrow}^{\dagger}  
	)
	\ket{00},\nonumber\\
 &=\ket{\Psi_{AB}(\phi_{AB})},
\end{align}
where $\ket{00}$ denotes the electron vacuum of the two islands, and
a gauge transformation on the electronic operators allows us to express the joint state as a function of  the relative phase between the two superconducting islands  $\phi_{AB} \equiv \phi_A - \phi_B$.

\subsection{BCS Ground-Space Projection Operator}
This work is focused on the dynamics of an electronic system restricted to the BCS ground space, so we define an (approximate) BCS ground space projection operator $P$  onto $\mathcal{H}_{\text{BCS}}$ \cite{ref:POM} as
\begin{equation}    \label{eq:projector}
	P
	=
	{\sum}_j
	\dyad{\Psi(\phi_j)},
\end{equation}
with the complementary projector defined as \mbox{$\bar{P}=\mathbb{I} - P$}. 
The degree to which this is a well-defined, idempotent projector satisfying $P^2=P$ depends on the orthogonality of the low-energy phase basis, which we address below. 

We will use the projection operator to construct effective operators acting on the low energy space.  For example, in a system of two superconducting islands with a microscopic electronic Hamiltonian consisting of intra-island, and inter-island Coulomb and tunnelling terms,
\begin{equation}
    H_{\rm el}=H_A + H_B + H_\text{Coul}+H_T,
\end{equation}  
the low-energy effective Hamiltonian is given by
\begin{align} \label{eq:Heff}
	PH_{\text{eff}}P
\nonumber
&\approx
	PH_0P
	+
	PH_1P
\\
&\quad+
	PH_1\bar{P}
	\left(
	\bar{P}E_0
	-
	\bar{P}H_0\bar{P}
	\right)^{-1}
	\bar{P}H_1P,
\end{align}
up to the second-order in the tunneling, where \mbox{$H_0 = H_A + H_B + H_\text{Coul}$} and  $H_1 = H_T$ is treated to second order in perturbation theory \cite{ref:POM,Ambegaokar1982}.

\subsection{BCS Subspace Orthogonality and Completeness} \label{sec:H0_space}

The overlap of two  BCS states $\ket{\Psi(\phi)}$ and $\ket{\Psi(\phi')}$  is given by 
\begin{align} \label{eq:BCS_overlap}
	\mathcal{W}(\varphi) 
\equiv
	\ip{\Psi(\phi)}{\Psi(\phi')} 
&=
	{\prod}_{\bm{k}}
	\left(
		u_{\bm{k}}^2 + v_{\bm{k}}^2 e^{i\varphi}
	\right),
\end{align}
where $\varphi \equiv \phi' - \phi$.  Such overlaps have been considered by \citet{ONISHI1966367}, and are discussed in \cite{Mizel2024}.

To characterise the functional behaviour of $\mathcal{W}$, we assume a symmetric electronic band consisting of \mbox{$n\gg1$} single-particle modes with bandwidth \mbox{$\pm \mathcal{B}$} at half-filling, $N_{\rm el}=n/2$, so that \mbox{$-\mathcal{B}\leq \epsilon_{\bm k} \leq \mathcal{B}$} and $\mu=0$.  We non-dimensionalise the bandwidth  relative to the superconducting gap as   $b=\mathcal{B}/\Delta$.  

From BCS theory \cite{ref:annett}, $\Delta=2\,\mathcal{B}\,e^{-1/\lambda}$ where the dimensionless quantity \mbox{$\lambda={\abs{g}^2 n}/({2\mathcal{B}})\ll1$}  
is the  electron-phonon coupling strength in units of the single-particle density states  (which we re-derive  in \cref{sec:BCSenergy}). It follows that 
\begin{equation}
b=\mathcal{B}/\Delta=e^{1/\lambda}/2\gg1.    \label{eqn:bBCS}
\end{equation}
 In typical metallic systems, $\mathcal{B}\sim 1{\rm eV}$ \cite{ref:SC_bandwidth} and  $\Delta\sim {\rm 1 meV}$ \cite{ref:SC_gap_Cs3C60},
so $b\sim1000\gg1$. 

\begin{figure}[!]
	\centering
	\includegraphics[width=.49\columnwidth]{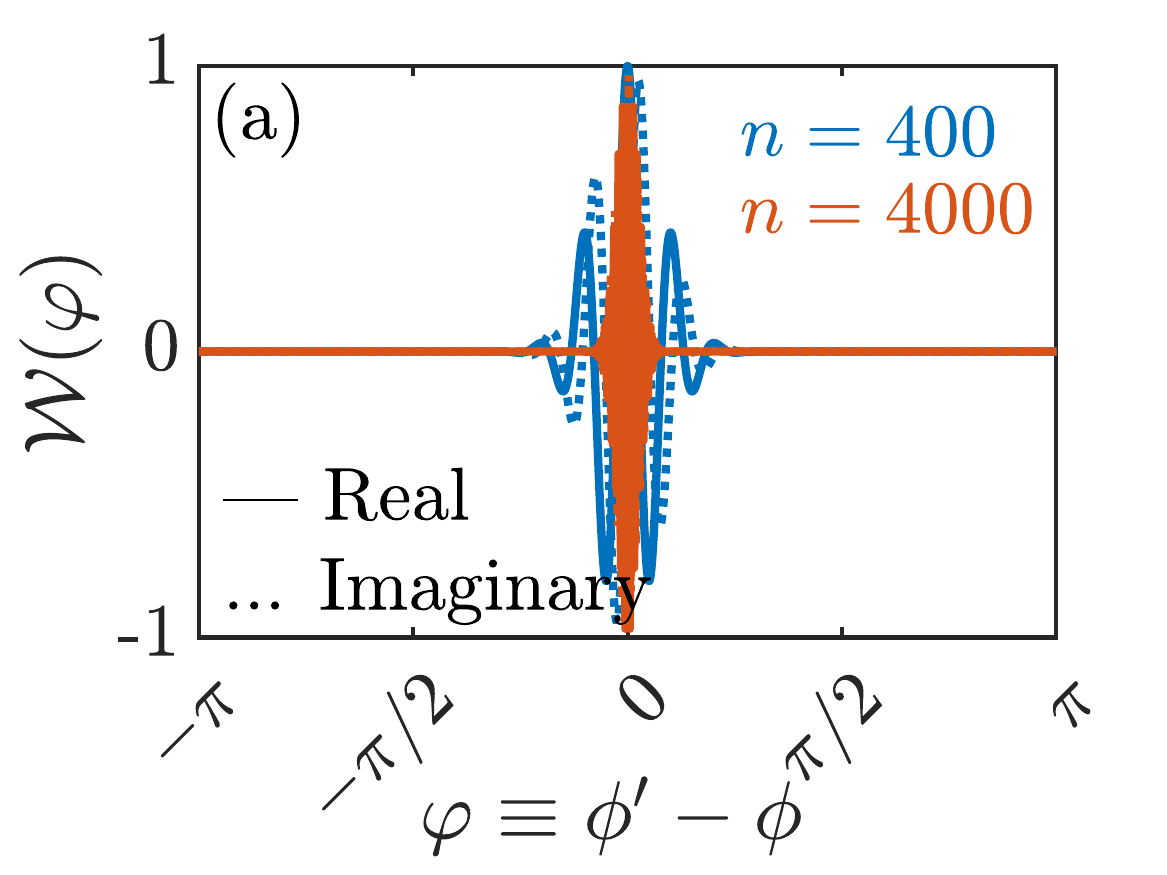}
	\includegraphics[width=.49\columnwidth]{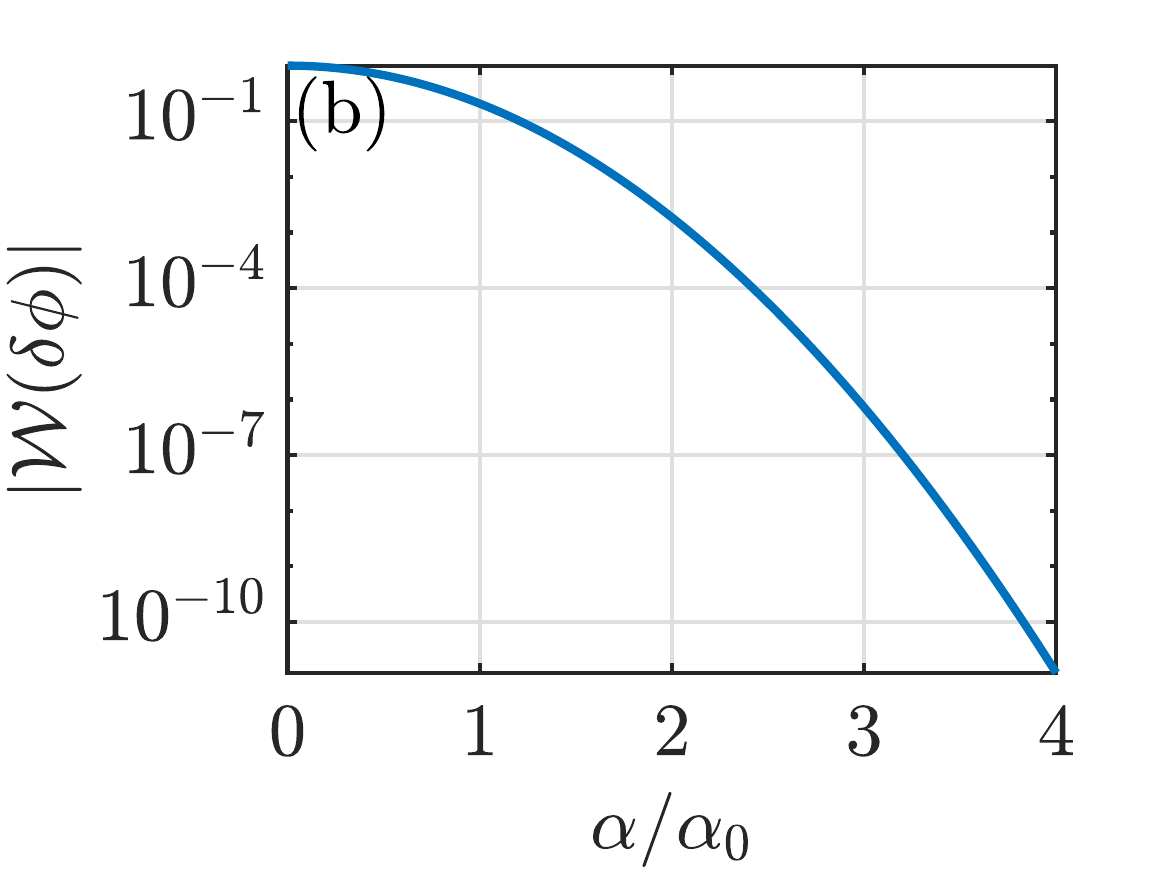}
	\caption{
		(a)
		Overlap of the BCS states $\mathcal{W}(\varphi) = \ip{\Psi(\phi)}{\Psi(\phi')}$ as a function of the phase difference $\varphi \equiv \phi' - \phi$.
        Increasing the number of electron modes, $n$,  leads to  faster oscillations over a narrower range, and the overlap $\mathcal{W}(\varphi)$ tends to a pseudo-Kronecker delta function, 
        $
            \mathcal{W}(\phi_j-\phi_{j'})
            \xrightarrow{n\rightarrow\infty}
            \delta_{j j'}
        $. 
		For this plot we have set  $b=10$.
		\ycl{(b)}
		The overlap amplitude of  adjacent basis states given in \Cref{eq:orthogonality}.
		As $\alpha$ grows, neighbouring phase states become increasingly orthogonal.
	}
	\label{fig:BCS_overlap&orthogonality}
\end{figure}

We take the logarithm of  \cref{eq:BCS_overlap} and replace the resulting momentum sum with an integral over the  Bogoliubov energies $E=\pm\sqrt{\epsilon^2+\Delta^2}$, i.e.\ 
\mbox{$
\sum_{\bm{k}}\!\bullet
\rightarrow
\int_{-\mathcal{B}}^{\mathcal{B}}dE \,
\rho(E)\bullet
$},
where   the density of Bogoliubov energies is given by \cite{ref:mahan}
\begin{equation} \label{eq:DoS}
	\rho(E)
=\dfrac{n}{2\Delta\sqrt{b^2 - 1}}
	\dfrac{|E|}{\sqrt{E^2 - \Delta^2}}\Theta(|E|-\Delta),
\end{equation} 
and $\Theta$ is the unit-step function. 
We then find
\begin{align}
	\ln\mathcal{W}
&=\nonumber
	\sum_{\bm{k}}
	\ln\!
	\big(
		\tfrac{1}{2}
		(1+\tfrac{\epsilon_{\bm{k}}}{E_{\bm{k}}})
+\tfrac{1}{2}e^{i\varphi}
		(1-\tfrac{\epsilon_{\bm{k}}}{E_{\bm{k}}})
	\big),
\\\nonumber
&=
	\int_{-\mathcal{B}}^{\mathcal{B}}	\!\!d E\,
	\rho(E)
		\ln\!\big(
		\tfrac{1}{2}
	(1+\tfrac{\epsilon_E}{E})
		+\tfrac{1}{2}e^{i\varphi}
		(1-\tfrac{\epsilon_E}{E})
	\big),\nonumber
\\
&=
	\int_\Delta^{b\Delta}	\!\!d E\,
	\rho(E)\,
		{\ln} \big({e^{i \varphi } (1-\tfrac{\Delta^2}{2 E^2} (1-\cos\varphi))}\big),\nonumber
\\
&=n\big(i \,\varphi/2 -\pi\sin(\varphi/4)^2/b\big) +O(b^{-2})
\nonumber
\\
  &=i\,n\,\varphi/2 -\pi n\,\varphi^2/(16b)+O(\varphi^3,b^{-2}),
\end{align}
where $\epsilon_E =  
\sqrt{E ^2 - \Delta^2}$, and we have retained the leading terms in  $b^{-1}$.  It follows that 
\begin{align} 
	\mathcal{W}(\varphi)
&\approx e^{
			i\,n\,\varphi/{2}
			 - \pi n\,  \varphi^2/(16b)}\label{eq:BCSoverlap_approx},
\end{align}
so that the overlap is approximately a Gaussian envelope of width $\sim\sqrt{b/n}$, modulated by an oscillating function.   
We plot $\mathcal{W}(\varphi)$ in \cref{fig:BCS_overlap&orthogonality}(a), showing that $\mathcal{W}(0)= 1$, and that $|\mathcal{W}|$ decreases quickly for $\varphi>\sqrt{b/n}$.

We now use the functional form of $\mathcal{W}$  to establish conditions under which we can construct an (approximately) orthogonal  basis of phase states which is also (approximately) complete in the low-energy subspace. 

\subsubsection{Basis Orthogonality}   \label{sec:orthogonality}
An orthogonal basis of phase states would satisfy $\mathcal{W}(\phi_{j} - \phi_{j'}) = \delta_{jj'}$ \cite{Leggett1966}.  However, \mbox{$|\mathcal{W}(\phi' - \phi)|>0$} everywhere, and so any set  $\{\ket{\Psi(\phi}\}$ is not precisely orthogonalised. Nonetheless  
\cref{fig:BCS_overlap&orthogonality}a shows that $|\mathcal{W}(\varphi)|$ is a rapidly decreasing function of $|\varphi|$, and so we construct an approximately orthogonal basis by discretising  \mbox{$\phi\in[-\pi,\pi)$}
with discretisation interval $\delta\phi$, to construct the discrete set of phases \mbox{$\phi_j=-\pi+\delta\phi\, j$} with \mbox{$ j\in\Zjmax\equiv\{0,1,...,\jmax=\lfloor 2\pi/\delta\phi\rfloor\}$}, so that 
\begin{equation}
   \phi_j\in \mathcal{Z}\equiv 2\pi(\Zjmax/\jmax-1/2)\subset[-\pi,\pi).\label{eq:Zset}
\end{equation} 

\Cref{eq:BCSoverlap_approx} provides a foundation on which to choose the discretisation interval. The overlap amplitude of successive basis states is given by
\begin{equation}\label{eq:overlap_adjacentBCSstate}
	\abs{\mathcal{W}(\phi_{j+1} - \phi_j)}=\abs{\mathcal{W}(\delta\phi)}
= e^{-\delta\phi^2 n \,\pi/(\ycl{16} \,b)}.
\end{equation} 
We define $\alpha=\delta\phi {\sqrt{n}}/(2\pi)$ and \ycl{$\alpha_0 = \sqrt{2b}/\pi$ }
so that
\begin{align}\label{eq:orthogonality}
	\abs{\mathcal{W}(\delta\phi)}&= e^{-\pi^3 \alpha^2/(\ycl{4} b)}=e^{-\pi \alpha^2/\ycl{2\alpha_0^2}}
\end{align}
is nominally independent of $n$.  With these definitions, the effective dimension of the BCS subspace is \mbox{$\jmax=2\pi/\delta\phi=\sqrt{n}/\alpha$}, and the set $\mathcal{Z}$ becomes dense in the interval $[-\pi,\pi)$ as $n$ grows.  In addition, $\jmax\ll n$, so that the effective dimension of the BCS subspace is much smaller than the number of single-particle modes in the metallic island.

The  ratio $\alpha/\alpha_0$ controls the approximate orthogonality of the BCS phase basis discretisation: basis orthogonality improves with larger $\alpha$.
From \cref{fig:BCS_overlap&orthogonality}b we see that choosing \mbox{$\alpha \gtrsim 2\alpha_0$} produces a nearly orthogonal basis with $\abs{\mathcal{W}(\delta\phi)}<e^{-2 \pi}\sim10^{-3}$.  With this choice, we have the discretisation interval and effective Hilbert space dimension given respectively by 
\begin{align}
    \delta\phi &= \ycl{4}\sqrt{\ycl{2}b/n},\text{ and}\label{eqn:deltaphi}\\
     \jmax_{\rm eff}&=\frac{\pi}{2}\sqrt{\frac{n}{2b}}\label{eqn:deff1}.
\end{align}
This formalises the assertion \cite{Leggett1966} that distinct BCS phase states can be approximated as being orthogonal in a suitable limit, and provides an estimate for the effective, coarse-grained dimensionality of the phase basis, which was also addressed in detail by \citet{Mizel2024}.

In \cref{sec:experimental}, we will see that for some typical junction parameters, $n/(2b)\sim 10^3$, so that $\jmax_{\rm eff}\sim50$ is the (surprisingly small) effective Hilbert space dimension of a point-like superconducting element.  This scale is similar to the number of microscopic constituents that has been estimated to participate in superposition states in small, experimentally accessible flux qubits  \cite{Korsbakken_2009}.

\subsubsection{Basis Completeness}
Given a discrete set of phases $\mathcal{Z}$, we cannot perfectly represent an arbitrary state $\ket{\Psi(\phi)}$ where \mbox{$\phi\notin\mathcal{Z}$}.  Here we analyse how well the Hilbert space \mbox{$\mathcal{H}_{\rm BCS}=\text{span}\{\ket{\Psi(\phi_j)}\}_{{\phi_j}\in \mathcal{Z}}$} covers all possible phase states $\ket{\Psi(\phi)}$ by bounding the distance $d$ between $\ket{\Psi(\phi)}$ and $\mathcal{H}_{\rm BCS}$.  We find 
\begin{align}	\label{eq:completeness_distance}
	d(\ket{\Psi(\phi)}, \mathcal{H}_{\rm BCS}), \nonumber
	&=
	\underset{\{a_j\}}{\min}\;
	d\big(
	\ket{\Psi(\phi)}, {\sum}_j a_j \ket{\Psi(\phi_j)}
	\big),
	\\\nonumber
	&=
	\underset{\{a_j\}}{\min}\;
	\Big|{
		\ket{\Psi(\phi)} 
		- 
		{\sum}_j
		a_j  \ket{\Psi(\phi_j)}
	}\Big|,
	\\\nonumber
	&\leq
	\underset{\{a_j\}}{\min}\;
	\sqrt{2\big(
	1 - {\sum}_j a_j \,|\mathcal{W} (\phi_j - \phi)| \big)},
	\\\nonumber
	&\leq
	\sqrt{2
	\min_j(1- |\mathcal{W} (\phi_j - \phi)| )},
	\\
	&\leq
	\sqrt{2(1-|\mathcal{W} (	\delta \phi)|)},\nonumber\\
 &\approx
	\sqrt{2(1-e^{-\pi \alpha^2/(\ycl{2\alpha_0^2})})}.
\end{align}
For  $\alpha\ll\alpha_0$, we see that $d(\ket{\Psi(\phi)}, \mathcal{H}_{\rm BCS})\approx \sqrt{\pi}\alpha/\alpha_0$. That is, smaller values of $\alpha$ generate a basis that is capable of representing a more complete set of phases.  This is contrary to the requirement of basis orthogonality, which favours larger $\alpha$.  There is therefore a trade-off between the orthogonality of the phase basis and the completeness of the representation.

\subsubsection{Projection Error}
\begin{figure}[!]
		\centering
		\includegraphics[width=0.49\columnwidth]{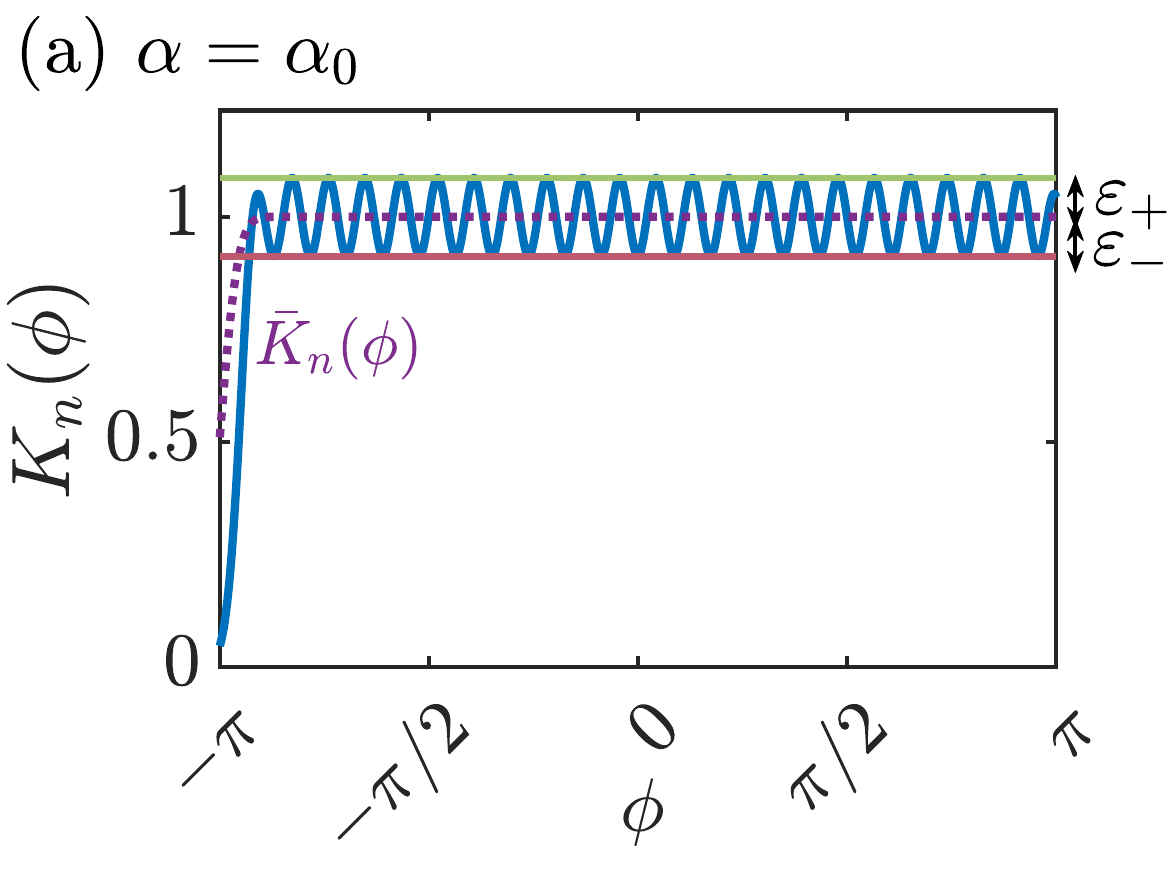}
		\centering
		\includegraphics[width=0.49\columnwidth]{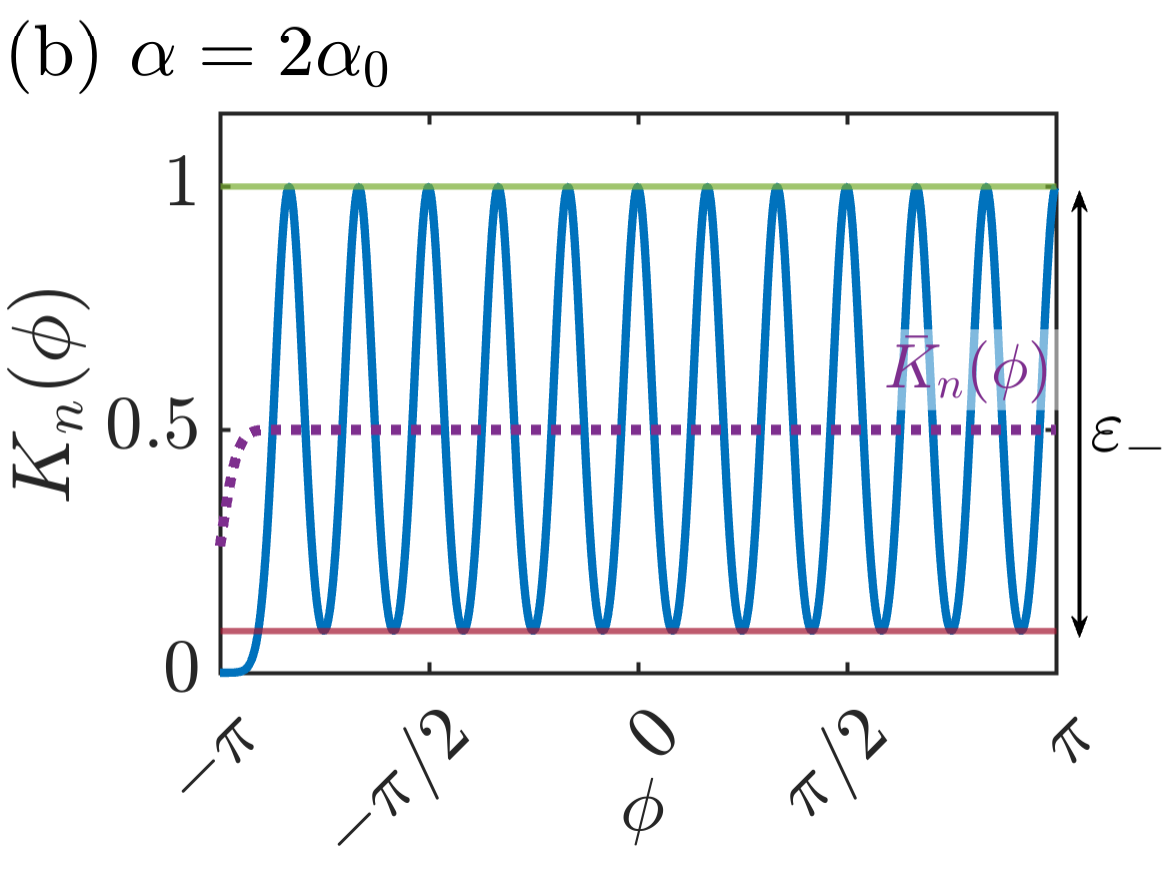}
	\caption{
		Projections $K_n$(oscillatory lines) and their integral approximations $\bar K_n$ (dotted lines) for (a) $\alpha=\alpha_0$, in which the approximate BCS projection is close to unity everywhere but is a strongly nonorthogonal basis   with $K_n>1$,  and (b) $\alpha=2\alpha_0$ where the basis is very nearly orthogonal, but with poorly represented states at $\phi=(\phi_{j+1}-\phi_j)/2$.
		The parameters are chosen to be $n=10^3$, $b=10$.
        }
	\label{fig:projErr}
\end{figure}

Given the tradeoff between basis orthogonality and completeness when  choosing the discretisation parameter $\alpha$, we synthesise both considerations by computing the error in projecting an arbitrary phase state onto the low energy BCS subspace.  
We quantify the degree to which the discretised subspace $\mathcal{H}_{\rm BCS}$ can represent an arbitrary phase state $\ket{\Psi(\phi)}$,  by computing its projection  onto $\mathcal{H}_{\rm BCS}$
\begin{equation}	\label{eq:projection_discrete}
K_n(\phi)\equiv	\ev{P}{\Psi(\phi)}
=
	{\sum}_{\phi_j \in\mathcal{Z}}
	\abs{\mathcal{W} (\phi - \phi_j)
	}^2.
\end{equation}
This quantity is oscillatory, with maxima when $\phi \in \mathcal{Z}$, and minima when $\phi=(\phi_{j}+\phi_{j+1})/2$, and is plotted in \cref{fig:projErr} for  $\alpha = \alpha_0$ and for $\alpha = 2\alpha_0$. 
The mean of the oscillatory projection is approximated by turning the sum into an integral,
\begin{align}   \label{eq:projection_continuum}
	\bar K_n(\phi)
&=
	\int_0^\jmax	\dd{j}
	e^{
		-n\left( \phi + \pi - \delta\phi\, j \right)^2
		/
		(4\alpha_0^2)
	},
\\
&\approx
	\tfrac{\alpha_0}{2\alpha}
    \Big(
    \ycl{
        \erf\Big[
            \tfrac{\sqrt{n}(\pi+\phi)}{2\alpha_0\sqrt{\pi}}
        \Big]
        -
        \erf\Big[
            \tfrac{\sqrt{n}(\pi+\phi) - 2\pi\sqrt{n-1}}{2\alpha_0\sqrt{\pi}}
        \Big]
    }
	\Big),\nonumber
\end{align}
which averages the maxima and minima of $K_n$. 
The integral approximation $\bar K_n$ is also shown in in \cref{fig:projErr}.  

For small $\alpha$, $K_n\sim1$ everywhere indicating that it can well represent arbitrary phase states, however, $K_n>1$ for some values of $\phi$ indicating that it is not well orthogonalised.  Conversely, for larger $\alpha$, the minima in $K_n$ are much lower than unity, indicating that the subspace cannot well represent arbitrary phase states. The maxima and minima of $K$ reflect the tradeoff between completeness and orthogonality, described above.  We conclude that $\alpha\approx 2\alpha_0$ is a reasonable tradeoff between basis completeness and orthogonality.  

For very small metallic islands where $n\sim b$, there may be observable physical consequences arising from small $n$ effects, but we do not pursue this analysis any further here.

\subsubsection{Orthonormalisation and Effective Dimensionality}

The choice $\alpha=2\alpha_0$ was somewhat heuristic, so we provide some additional analysis that gives an independent estimate for the effective dimensionality of the phase basis $\jmax_{\rm eff}$.  The approach  starts with a very large phase-basis set $\mathcal{Z}$, then orthogonalises the phase-basis overlap matrix, and then computes the effective, reduced dimensionality of the Hilbert  space.

As discussed above,  the phase basis is normalised but not orthogonalised.  Given an initial choice of the phase basis size, $|\mathcal{Z}|=\jmax$, we define the $\jmax\times\jmax$ overlap matrix $\uuline{W}$ with matrix elements 
\mbox{$W_{lm}=\bra{\Psi(\phi_{l})}\Psi(\phi_{m})\rangle=\mathcal{W}(\phi_{l} - \phi_m)$} so that
\begin{equation}
    \uuline{W}=\begin{bmatrix}
\mathcal{W}(0) & \mathcal{W}(-\delta\phi) &\\
\mathcal{W}(\delta\phi) & \mathcal{W}(0) & \\
\mathcal{W}(2\delta\phi) & \mathcal{W}(\delta\phi)& \dots\\
\vdots & \vdots
\end{bmatrix}.
\end{equation}
Since $\mathcal{W}(-x)=\mathcal{W}(x)^*$, 
$\uuline{W}$ is Hermitian, so it is unitarily diagonalisable with real eigenvalues.  Further, it is a circulant matrix, in which each column vector is a cyclic permutation of its predecessor, and so $\uuline{W}$ is generated by its first column vector, $\uline{w}$, i.e.\ $\uuline{W}=\big[\uline{w},\, \pi_1\uline{w},\, \pi_2\uline{w},...\big]$ where $\pi_j$ is the cyclic permutation by $j$ slots. 
Diagonalising $\uuline{W}$ provides a linear transformation that orthogonalises the original phase basis.

It is important to note that $\uuline{W}$ is invertible for all $\jmax$, so its matrix rank is $\jmax$; that is, formally the Hilbert space dimension of the phase basis is always given by $\jmax$.  However, for large $\jmax\gtrsim\sqrt{n/b}$ the phase basis states become highly overlapping, reflected in the fact that   $\uuline{W}$ becomes increasingly ill-conditioned. So there is a sense in which the effective dimensionality of the Hilbert space defined by $\mathcal{Z}$ can be much smaller than $\jmax$.  We address this here.

Circulant matrices are diagonalised by the unitary discrete Fourier transform matrix $\uuline{F}$ \cite{Mizel2024} whose matrix elements are $F_{lm}=e^{2\pi i(l-1)(m-1)/\jmax}/\sqrt{\jmax}$, so 
\begin{equation}
\uuline{W}=\uuline{F}^\dagger.\uuline{D}.\uuline{F},
\end{equation}
where $\uuline{D}={\rm diag}(\uline{\Lambda})$ is a diagonal matrix of real and positive eigenvalues.  In addition, for a $\jmax\times\jmax$ circulant matrix,  the eigenvalues are proportional to the discrete Fourier transform of the first column vector, 
\begin{equation}
\uline{\Lambda}=\sqrt{\jmax}\,\uuline{F}.\uline{w},\label{eqn:Devals}
\end{equation}
which we compute below.

Representing a column vector of  basis states as \mbox{$\uline{\ket{\phi}}=[\cdots,\ket{\Psi(\phi_j)},\cdots]^\mathsf{T}$}, 
we seek a (non-unitary) basis transformation $\uuline{E}$ that orthonormalises the phase basis  $\{\ket{\Psi(\phi_j)}\}$ into  an orthonormal basis $\{\ket{e_j}\}$, i.e.\
\begin{equation}
    \uline{\ket{e}}=\uuline{E}.\uline{\ket{\phi}}\label{eqn:ephi}
\end{equation}
with $\bra{e_i}e_j\rangle={\sum}_{lm}E_{il}^* W_{lm}E_{jm}=\delta_{ij}$.  That is, $\uuline{E}^*.\uuline{W}.\uuline{E}^{\mathsf{T}}=\mathbb{I}$.  Noting that $ \uuline{F}^{\dagger}.\uuline{F}=\mathbb{I}$, $\uuline{F}=\uuline{F}^{\mathsf{T}}$ and $\uuline{F}^*=\uuline{F}^{\dagger}$ it is straightforward to verify that 
\begin{equation}
    \uuline{E}=\uuline{F}.\uuline{D}^{-1/2}.\uuline{F}^\dagger=(\uuline{W}^\mathsf{T})^{-1/2}.
\end{equation}

Defining the Fourier-transformed bases \mbox{$\uline{\ket{f}}=\uuline{F}^\dagger.\uline{\ket{e}}$}, which is  orthonormal, and  $\uline{\ket{n}}=\uuline{F}^\dagger.\uline{\ket{\phi}}$, which is merely orthogonal, \cref{eqn:ephi}  becomes
\begin{equation}
    \uline{\ket{n}}=\uuline{D}^{1/2}.\uline{\ket{f}}\label{eqn:fn}.
\end{equation}
That is, $\ket{n_k}=\sqrt{\Lambda_k} \ket{f_k}$, so that the `number' basis satisfies \mbox{$\bra{n_k}n_q\rangle=\Lambda_k \delta_{kq}$}. 
The Fourier transform matrix thus maps the nonorthgonal-but-normalised phase basis $\uline{\ket{\phi}}$ into the orthogonal-but-unnormalised number basis $\uline{\ket{n}}$, where $\sqrt{\Lambda_k}$ determines the unnormalised length of the corresponding number state.

The normalisation $\sqrt{\Lambda_k}$ provides a measure of the `significance' of each number state relative to the original non-orthogonal phase basis: those with large normalisation have large support on the orthonormal basis  $\uline{\ket{f}}$, while those with small normalisation are relatively insignificant and can be truncated without much error.  We therefore define the `relative significance' of $\ket{n_k}$ as \mbox{$s_k=\sqrt{\Lambda_k/\Lambda_{\rm max}}$}, where $\Lambda_{\rm max}={\max}_k \Lambda_k$.  

The relative significance allows us to compute the effective dimensionality of the Hilbert space defined by the phase basis. By analogy, suppose that $\uuline{W}$ was idempotent with $s_k=0$ or 1, defining an ideal projector. The dimension of the projected space would be 
\begin{equation}
    \jmax_{\rm eff}={\sum}_k {s_k}.
\end{equation}
We generalise this to be the definition of the `effective dimensionality' for any basis overlap matrix $\uuline{W}$.  We now compute estimates for the dependence of $\jmax_{\rm eff}$ on the choice of the basis size $\jmax$.

In the limit $\jmax\ll\sqrt{n/b}$, the phase discretisation \mbox{$\delta\phi=2\pi/\jmax$} will be `large', and $|\mathcal{W}(\delta\phi)|\ll1$,  so that the phase basis will itself be very close to an orthonormal basis.  This is the scenario where the phase basis is relatively sparse on the interval $[-\pi,\pi)$, illustrated in \cref{fig:projErr}b, and  poorly represents  phase values where $K_n(\phi)$ is small.  In this limit,   $\uuline{W}\approx\mathbb{I}$, so $\Lambda_k\approx 1$ for all $k$, and the effective dimensionality of the phase basis is $\jmax_{\rm eff}={\sum}_k {s_k}\approx\jmax$, as expected for a nearly orthonormal basis of size $\jmax$.  Since  $\jmax=|\mathcal{Z}|$ is essentially an arbitrary choice made for calculational purposes, it is not itself a physically significant property of the system.

Conversely, in the limit $\jmax\gg\sqrt{n/b}$, the phase discretisation  \mbox{$\delta\phi$} is relatively small, so the discrete phase basis can well-represent the continuum of phase values, but neighbouring phase-basis states will strongly overlap one-another, and $\uuline{W}$ will be far from the identity. This is the scenario  illustrated in \cref{fig:projErr}a.  In this limit, we use \cref{eqn:Devals}  and take the Fourier transform of  \cref{eq:BCSoverlap_approx} to estimate the  eigenvalues
\begin{align}
    \Lambda_k
   &=
   \sum_{m=0}^{\jmax-1} e^{2\pi i \,k \,m} \,\mathcal{W}(m\,\delta\phi),\nonumber\\
   &\approx \frac{\jmax}{2\pi} 
   \int_{-\infty}^\infty d\phi\, e^{i\,\phi \,k}\,\mathcal{W}(\phi),  \nonumber\\
   &=  \frac{2\jmax}{\pi}\sqrt{\frac{b}{n}} \, e^{
   -4b{( k+n/2)^2}/({\pi n})},\label{eqn:lambdaGauss}
\end{align}
where {$k\in\{ -n/2-\jmax/2,...,-n/2+\jmax/2-1\}\cong\mathbb{Z}_\jmax$}, and in the second line we approximate $\mathcal{W}$  as negligibly small for  $|\phi|>\pi$, which is a good approximation resulting from \cref{eq:BCSoverlap_approx}.  

Thus if $\jmax$ is chosen to be sufficiently large, then $\Lambda_k$ is approximately Gaussian, with  $\Lambda_{\rm max}=2\jmax\sqrt{b/n}/{\pi } $ at $k=-n/2$.  The effective dimensionality is then
\begin{align}
    \jmax_{\rm eff}&={\sum}_k {s_k}
    \approx \int_{-\infty}^\infty dk\, {s_k}
     =\pi\sqrt{\frac{n}{2b}}.\label{eqn:deff}
\end{align}
This  matches \cref{eqn:deff1}, up to a factor of 2.  

Importantly, \cref{eqn:deff} is independent of the initial choice of the size of the phase basis set, $\jmax$.  In addition, through \cref{eqn:bBCS}, $\jmax_{\rm eff}$ depends only on physical parameters from the underlying microscopic electronic theory.  That is, in the large-$\jmax$ limit, $\jmax_{\rm eff}$ is an intrinsic, physically-relevant  quantity that emerges naturally from the structure of the low-energy BCS subspace.  

Finally, we conclude this section by noting that \cref{eqn:deff} arises from the kinematics of the phase basis structure.  It is not determined by energetic considerations, which require the introduction of low-energy dynamics that we discuss later.

\subsection{Phase and Charge Operators} \label{sec:canonical}

Having constructed a  low energy Hilbert space from a discrete basis of phase states, in this section we define a basis of ``number'' states through the Fourier transform.  These two bases implicitly define the low-energy phase and charge operators, $\hat\phi$ and $\hat N$, and from these we compute commutation relations.  For simplicity, we will assume that the phase basis is an orthonormal basis, notwithstanding the preceding discussion. 

The phase operator is straightforwardly defined as 
\begin{equation}
	\hat{\phi}=
	\sum_{\phi_j\in\mathcal{Z}}
	\phi_j \dyad{\Psi(\phi_j)},\label{eqn:phase_operator}
\end{equation}
recalling that $\phi_j=\phi_0+\delta\phi\, j$, and we choose the convention $\phi_0=-\pi$.

The BCS states $\ket{\Psi(\phi_j)}$ are not  eigenstates of the electron number. Here  we define low-energy ``number'' eigenstates $\ket{\Psi(N)}$ via a discrete Fourier transform
\begin{align}\label{eq:|N>_def}
			\ket{\Psi(N)}
            &=
			\frac{1}{\sqrt{\jmax}}
			\sum_{\phi_j\in\mathcal{Z}}			e^{-iN\phi_j}
			\ket{\Psi(\phi_j)}
	\end{align}

We show in \cref{append:N_eigenstate} that $\ket{\Psi(N)}$ are eigenstates of the microscopic Cooper pair number operator 
\mbox{$
    \hat{N}^{\rm CP}
=
    \sum_{\bm{k}\sigma}
			c_{\bm{k}\sigma}^{\dagger}   c_{\bm{k}\sigma}/2
$}, i.e.
\begin{equation} \label{eq:|N>_eigenEq}
	\hat{N}^{\rm CP}	\ket{\Psi(N)}
=
	N \ket{\Psi(N)},
\end{equation}
establishing that the Fourier conjugate states are indeed charge-number eigenstates, and the set $\{\ket{\Psi(N)}\}$ 
defines a conjugate charge-number basis with  \mbox{$N\in\{\frac{n}{2} - \frac{\jmax}{2},..., \frac{n}{2} + \frac{\jmax}{2}\}\subset\mathbb{Z}$}.  The centre of this interval corresponds to half filling of the band.
The number basis implicitly defines the number operator in the low-energy subspace,
\begin{equation}
    \hat{N}
=P \hat{N}^{\rm CP} P=
    \sum_{N=\frac{n}{2} - \frac{\jmax}{2}}^{\frac{n}{2} + \frac{\jmax}{2}}
    N   \dyad{\Psi(N)}.
\end{equation}

Pegg and Barnett have shown that the phase-number commutator follows the commutator-Poisson-bracket correspondence for physical states \cite{ref:Pegg&Barnett}.
Here we compute the canonical commutation relation between the matter phase operator $\hat{\phi}$ and number operator $\hat{N}$ 
in a single superconducting island; we then generalise to the relative phase of a system consisting of two superconductors.

We represent the phase operator $\hat{\phi}$ in the number basis
\begin{widetext}
\begin{align} \label{eq:phaseOP_Nbasis}
	\hat{\phi}
&\equiv
	\sum_{j=0}^{\jmax}
	\phi_j \dyad{\Psi(\phi_j)},
\nonumber
\\\nonumber
&=
	\phi_0 \mathbb{I}
	+
	\dfrac{2\pi\alpha}{\sqrt{n}}
		\sum_{j=0}^{\jmax}
		\dfrac{j}{\sqrt{n-1}+\alpha} 
		\sum_{NN'}
		e^{i(N-N')(\phi_0 + \frac{2\pi\alpha}{\sqrt{n}}j)}
		\dyad{\Psi(N)}{\Psi(N')},
\\
&=
	\left(
	\phi_0 + \dfrac{\pi\alpha}{\sqrt{n}}
	\right) \mathbb{I}
	+
	\dfrac{2\pi\alpha}{\sqrt{n}}
	\dfrac{\sqrt{n-1}}{\sqrt{n-1}+\alpha}
	\sum_{N\neq N'} 	e^{i\phi_0(N-N')} 
	\dfrac{e^{2\pi i\alpha(N-N')/\sqrt{n-1}}}
	{e^{2\pi i\alpha(N-N')/\sqrt{n-1}} - 1}
	\dyad{\Psi(N)}{\Psi(N')}.
\end{align}
In the number basis, we compute the phase-number commutator matrix elements, and for  $n \gg |N'-N|$ we find
\begin{equation}	\label{eq:PhaseNumberComm_mtxElem}
	\mel{\Psi(N)}{[{\hat{\phi}},{\hat{N}}]}{\Psi(N')}
=
	\begin{cases}
		0,	&\quad	N=N'
		\\
		\dfrac{2\pi\alpha\sqrt{n-1}}{\sqrt{n}(\sqrt{n-1}+\alpha)}
		(N' - N) e^{i\phi_0(N-N')} 
		\dfrac{e^{2\pi i\alpha(N-N')/\sqrt{n-1}}}
		{e^{2\pi i\alpha(N-N')/\sqrt{n-1}} - 1},
		&\quad N\neq N'
	\end{cases}.
\end{equation}
\end{widetext}
For finite dimensional operators, the commutator is traceless, and in the limit $n\rightarrow\infty$, 
we find
\begin{equation} \label{eq:comm_discrete}
	\mel{\Psi(N)}{[{\hat{\phi}},{\hat{N}}]}{\Psi(N')}
\approx
	i \left(1 - \delta_{NN'}\right)
	e^{i\phi_0(N-N')}.
\end{equation}

The $\phi_0$ dependence is consistent with the results of Pegg-Barnett  theory \cite{ref:Pegg&Barnett} for phase-number commutators. 
Transforming back to the $\phi$-basis, the commutator becomes
\begin{equation}
	[{\hat{\phi}},{\hat{N}}]
\approx
	-i
	(
		\mathbb{I} 
		- 
		\dfrac{\sqrt{n-1} +\alpha}{\alpha} 
		\dyad{\Psi(\phi_0)}
	),
\end{equation}
which, given that
$(\sqrt{n-1}+\alpha)/\alpha 
= 
	2\pi/\delta\phi=\jmax$ the commutator is traceless, as required by the Stone-Von Neumann theorem for finite dimensional operators.  This commutator structure is notably different from $[{\hat{\phi}},{\hat{N}}]=-i$ postulated in other circuit quantisation procedures \cite{widom1979,https://doi.org/10.1002/cta.2359}.
    
    The expectation value of the commutator with respect to an arbitrary phase state $\ket{\Psi(\phi)}$ is given by
\begin{align}\label{eq:comm_evPhase}
	\ev{[{\hat{\phi}},{\hat{N}}]}{\Psi(\phi)}
\nonumber
&\approx
	-i
	\big(
		1
		-
        \jmax
		\abs{\mathcal{W}(\phi_0 - \phi)}^2
	\big),
\\
&\approx
	-i
	\big(
		1
		-
		\jmax
		\delta_{\phi_0, \phi}
	\big).
\end{align}

In Josephson junctions consisting of two superconducting islands $A$ and $B$, we are interested in the commutator of the phase and number differences between the two islands. 
Assuming the junction is symmetric such that $n_A\approx n_B \equiv n$ and $\alpha_A \approx \alpha_B \equiv \alpha$, we have
\begin{align}
	[{\hat{\phi}_{AB}},{\hat{N}_{AB}}]
&\equiv
	[{\hat{\phi}_A - \hat{\phi}_B},{\hat{N}_A - \hat{N}_B}],\nonumber
\\
&\approx
	-2i 
		\mathbb{I}_{AB} 
	+	i\,
    \jmax
	\big(
		\dyad{\Psi_A(\phi_0^{A})}	\otimes	\mathbb{I}_B
		\nonumber\\
        &\hspace{2cm}+
		\mathbb{I}_A	\otimes	\dyad{\Psi_B(\phi_0^{B})}
	\big).\nonumber
\end{align}

The Hilbert space of the system is now a tensor product space spanned by states of the form
\mbox{$
    \ket{\Psi(\phi_{AB})}
=
    \ket{\Psi_A (\phi_A)}   \otimes \ket{\Psi_B (\phi_B)}
$}. 
With the reference phase difference defined as $\phi_0^{AB} \equiv \phi_0^{A} - \phi_0^{B}$, we obtain the expectation value
\begin{align}	
	\bra{\Psi(\phi_{AB})}&
    \big[{\hat{\phi}_{AB}},{\hat{N}_{AB}}\big]
    \ket{\Psi(\phi_{AB})}
\nonumber
\\
&\approx
	-2i
	(
		1-
\jmax\,\delta_{\phi_0^{AB},\phi_{AB}}	
).\label{eq:commEV_junction}
\end{align}

\section{Projection of an L-C-J Circuit} \label{sec:LCJprojection}

Having derived an approximate projector onto the low-energy Hilbert space $\mathcal{H}_{\rm BCS}$ we now derive the effective low-energy Hamiltonian for the circuit starting from the microscopic electronic Hamiltonian, as sketched in \cref{fig:flowchart}.  We show that this approach gives results that are consistent with the conventional re-quantisation of GL theory for systems with a large number of electrons. 

We evaluate the projectors in \cref{eq:Heff} for an LCJ circuit  up to the second-order in electron tunnelling through the junction barrier separating metallic islands $A$ and $B$. 
The lowest-order terms in the projection are
\begin{equation}
	PH_0P=PH_AP + PH_BP+PH_\text{Coul}P.\label{eqn:firstorder}
\end{equation}
Below we evaluate the superconducting island terms, $H_A$ and $H_B$, and the Coulomb interaction, $H_\text{Coul}$, separately.

\subsection{On-Site Energy: Intra-Island Terms}
\label{sec:BCSenergy}

We compute the low-energy projections of the Hamiltonian term for island $A$ from \cref{eq:Hamiltonian_SCwoMFA}, with a similar expression for  $    PH_{B}P $. We find
\begin{align}
    PH_{A}P 
&= 
    \sum_{jj'} 
    \mel{\Psi(\phi_j)}{H_{A}}{\Psi(\phi_{j'})} 
    \dyad{\Psi(\phi_j)}{\Psi(\phi_{j'})},\nonumber\\
    &= \sum_{jj'} \big(\mathcal{K}(\varphi_{j'j})    -   \abs{g}^2 \mathcal{V} (\varphi_{j'j})\big)  \dyad{\Psi(\phi_j)}{\Psi(\phi_{j'})},\label{eqn:PHAP}
\end{align}
where   $\varphi_{j'j}=\phi_{j'}-\phi_j$ and
\begin{align}
\mathcal{K}(\varphi)&=2\sum_{\bm{k}_A}
    	\frac{\left(	\epsilon_{\bm{k}_A}	-	\mu_A	\right)	v_{\bm{k}_A}^2}
    	{u_{\bm{k}_A}^2	+	v_{\bm{k}_A}^2	e^{i\varphi}}e^{i \varphi}\mathcal{W}(\varphi),\label{eqn:K}\\
 \mathcal{V} (\varphi)	&= \Big(
        \sum_{\bm{k}_A}
        \dfrac{u_{\bm{k}_A}	v_{\bm{k}_A}}
        {u_{\bm{k}_A}^2	+	v_{\bm{k}_A}^2	e^{i \varphi}}
    \Big)^2e^{i \varphi}\mathcal{W}(\varphi)\label{eqn:V}
\end{align}

\begin{figure}[!t]
		\centering
		\includegraphics[width=0.49\columnwidth]{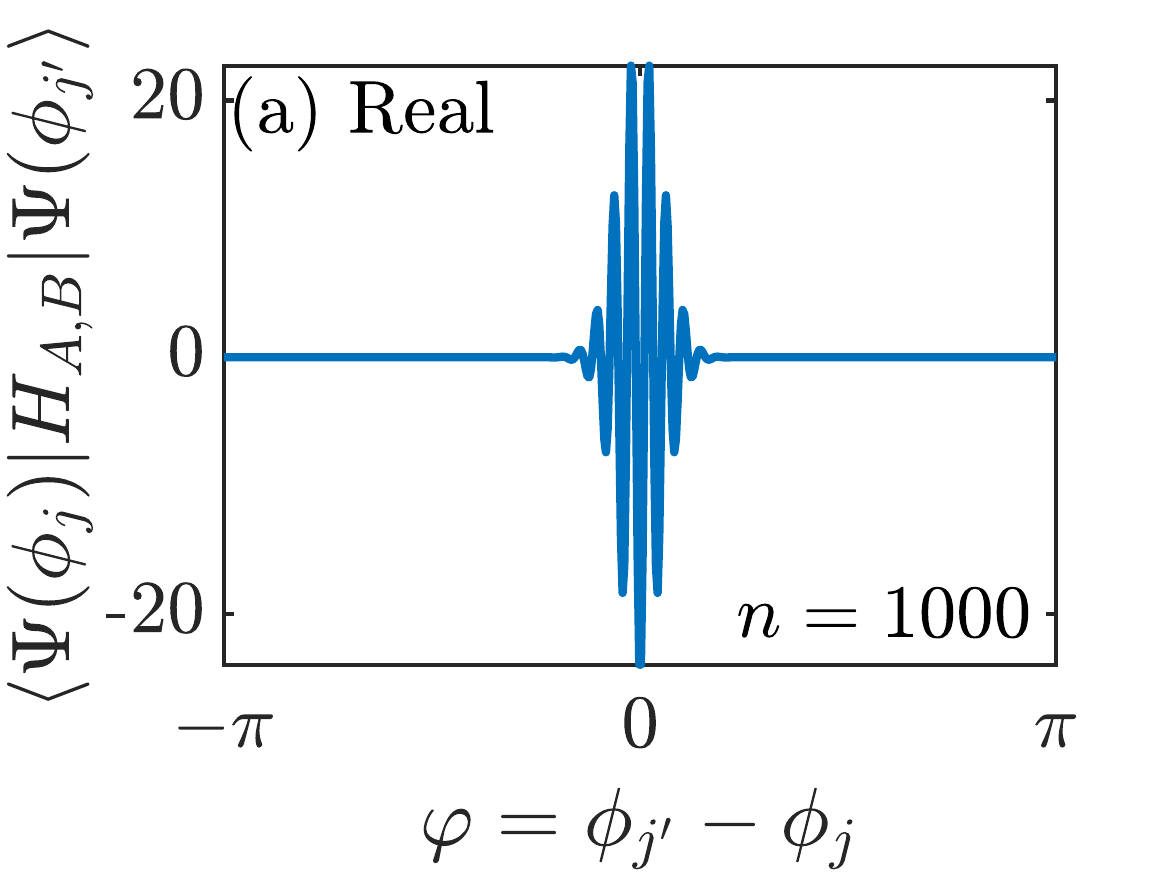}
		\centering
		\includegraphics[width=0.49\columnwidth]{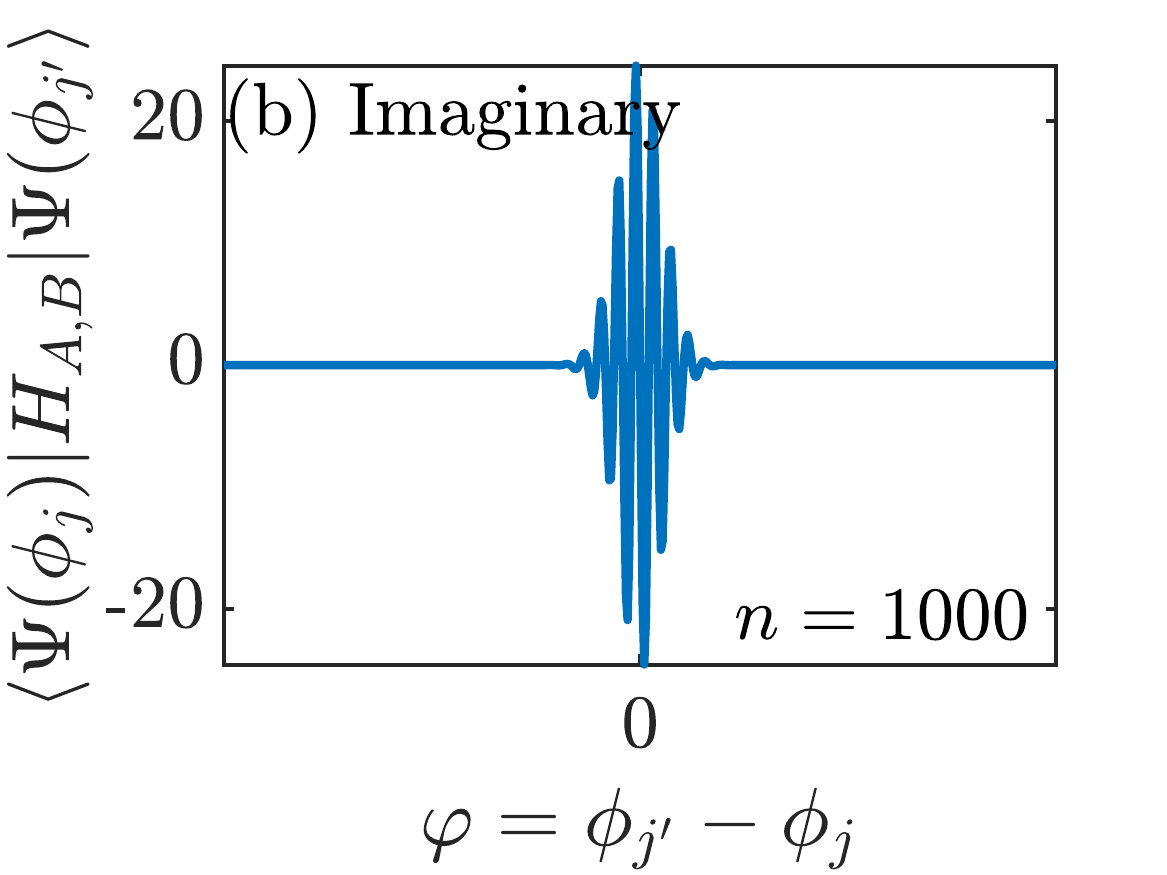}
	\caption{
		Matrix elements $\mel{\Psi(\phi_j)}{H_{A,B}}{\Psi(\phi_{j'})}$ for $n=10^3$, $\Delta = 1$, $b=10$.
		The real and imaginary parts are illustrated in panels (a) and (b), respectively.
	}
	\label{fig:HAelem_woMFapprox}
\end{figure}

We give a detailed calculation of the complex-valued functions $\mathcal{K}(\varphi)$ and $\mathcal{V}(\varphi)$  in \cref{append:Melem_islandHamiltonain}.
\Cref{fig:HAelem_woMFapprox} shows the real and imaginary parts of the matrix element
$\mel{\Psi(\phi_j)}{H_{A}}{\Psi(\phi_{j'})} $, which, similar to
$\mathcal{W}(\varphi)$,  has a maximum at $\varphi=0$ and rapidly tends to zero for $\varphi\neq0$ and as $n\rightarrow\infty$.
The diagonal terms in  $PH_AP$ give the mean ground-state energy:
\begin{align} 
			\bra{\Psi(\phi)}&H_A\ket{\Psi(\phi)}
			\nonumber\\
&			=
			\mathcal{K}(0) - \abs{g_A}^2 \mathcal{V}(0),\nonumber
			\\
			&\approx-
			\frac{n_A\Delta_A^2}{2\mathcal{B}_A}
			\Big(
			\ln\!\tfrac{\Delta_A}{2\mathcal{B}_A}+
			\lambda_A \ln\!{\big(\tfrac{\Delta_A}{2\mathcal{B}_A}\big)^2}
			\Big)-{\mathcal{B}_An_A}/2,\nonumber\\
			&\equiv 
			E_A(\mathcal{B}_A,\Delta_A, \lambda_A)-E_F,\label{eq:Hl_diagElem}
	\end{align}
where we have separated out the Fermi energy \mbox{$E_F={\mathcal{B}_An_A}/2$}, and we have defined the electron-phonon coupling constant \cite{ref:annett}
\begin{equation}
	\lambda_A 
	=
	\abs{g_A}^2\varrho_F
	=
{\abs{g_A}^2n_A}/({2\mathcal{B}_A}),
\end{equation}
where $\varrho_F=n_A/({2\mathcal{B}_A})$ is the non-interacting electronic density of states at the Fermi energy. 
The diagonal intra-island energy given in \cref{eq:Hl_diagElem} is independent of  $\phi_j$, and corresponds to the ``Mexican hat'' potential. 
Interestingly, the on-site energy of an island, $E_A$, is not analytic at $\Delta_A = 0$, unlike the conventional quartic form used in  GL theory.

\subsubsection*{Minimum Gap Energy}

The intra-island energy $E_A(\mathcal{B}_A,\Delta_A,\lambda_A)$ has a minimum value $E_A^\text{min}$  at  $ \Delta_A^\text{min}$, given  at leading order in $1/\lambda_A$  by
\begin{align} 
		\Delta_A^{\text{min}}
		&=
		2\mathcal{B}_A e^{-1/\lambda_A},
		\label{eq:quarticH_minGap}		\\
		E_A^{\text{min}}
		&=
		-n_A \mathcal{B}_A e^{-2/\lambda_A},
		\label{eq:quarticH_minE}
\end{align} 
and similarly for island $B$. 
These values correspond to the standard GL MFT results \cite{ref:annett}. 

For the majority of the following discussion, we will assume  that the system relaxes to this energy minimum, and that there are no `Higgs'-like excitations away from this condition.  We return briefly to this in \cref{sec:Higgs_modes}.

In the limit of large $n$, we have $\mathcal{W}(\phi_{j}-\phi_{j'}) \approx \delta_{jj'}$, so that \cref{eqn:PHAP} reduces to
\begin{equation} \label{eq:P.HA.P&P.HB.P}
	PH_AP  =    E_A^\text{min} P,
\end{equation}
and similarly $	PH_BP  =    E_B^\text{min} P$.
Thus, the first two terms of \cref{eqn:firstorder} are diagonal,  constant energy offsets in the BCS subspace.

\subsection{Capacitance: Inter-island Coulomb Interaction}	\label{sec:proj_Coulomb}

The last term in \cref{eqn:firstorder}  is the inter-island Coulomb interaction.
Using the definition of $H_\text{Coul}$ in \cref{eq:Hamiltonian_coulomb} and the result of \cref{eq:|N>_eigenEq}, we evaluate the low-energy projection of
$H_\text{Coul}$ in the number-basis.  For a fixed total charge $\hat{N}_A +\hat{N}_B =N_{\rm tot}$, we have
\begin{align}
	PH_\text{Coul}P
	&=
	4\lambda_C (P_A\hat{N}^{\rm CP}_AP_A )(P_B \hat{N}^{\rm CP}_BP_B),\nonumber\\
	&=4\lambda_C
	\hat{N}_A \hat{N}_B,\nonumber\\
		&=\lambda_C
	(N_{\rm tot}^2- \hat{N}_{AB}^2),\label{eq:P.Hcoul.P_op}
\end{align}
where $\hat{N}_{AB}=\hat{N}_{A}-\hat{N}_{B}$ is the inter-island charge difference. 
The is the usual quadratic capacitive coupling of two metallic islands.

\begin{figure}[t]
	\centering
    \includegraphics[width=0.7\columnwidth]{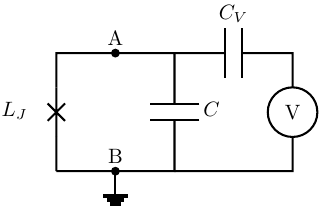}
	\caption{
	A circuit consisting of a  Josephson junction with a bias voltage $V$ connected capacitively.
    The bias voltage creates an offset charge $n_g$ to the capacitance Hamiltonian.
	}
	\label{fig:LCV_circuit}
\end{figure}

\paragraph*{Bias voltage:} We now add a bias voltage,  $n_g \neq 0$, as shown in \cref{fig:LCV_circuit}.
The total charge  on  each island is shifted:
\(
	N_A
\rightarrow
	N_A' = N_A + n_e,
\)
\(
	N_B
\rightarrow
	N_B' = N_B - n_e,
\)
where $n_e = CV/(2e)$.
The effective charge on the capacitor $N_{AB}' = N_A' - N_B'$ becomes 
\mbox{\(
	N_{\text{tot}}^2
-
	4(
		N_A N_B - n_e^2 - n_e(N_A - N_B)
	).
\)}
From the conservation of the total charge $N_{\text{tot}}' = N_{\text{tot}}$, we obtain
\begin{equation}
	(N_{AB}' - (2n_e))^2
=
	-4N_AN_B + N_{\text{tot}}^2.
\end{equation}
The offset charge $n_g = 2n_e$, results from the electron pairing in superconductors.
In summary, the capacitance term in \cref{eq:P.Hcoul.P_op} shifts for a non-zero charge bias.

\subsubsection*{Lowest order Terms} 
Combining the results in Eqs.~\eqref{eq:P.HA.P&P.HB.P},~\eqref{eq:P.Hcoul.P_op}, the lowest-order projected Hamiltonian terms gives
\begin{equation} \label{eq:0order}
	PH_0P
	\approx	\left(E_A^\text{min} + E_B^\text{min} \right) P+
	4\lambda_C \hat{N}_A \hat{N}_B,
\end{equation}
which are the island and capacitive energies respectively.

\subsection{Josephson Effect: Tunnelling}
The lowest-order tunnelling term is given by $H_T$ \cite{ref:POM}, which changes the parity of the electron numbner on the two sides of the junction.
Electronic  wavefunctions of different charge parity are orthogonal, so the matrix element vanishes:
$
\mel{\Psi(\phi_j)}{H_T}{\Psi(\phi_{j'})}
=
0$. It follows that the first-order projection of the linear tunnelling term vanishes,
\begin{equation} \label{eq:1order}
	PH_TP
	=
	0.
\end{equation}
This can be verified by detailed calculation.

The second-order tunnelling contribution in the low-energy BCS subspace is given by \cite{ref:POM} 
\begin{align}
    P &H_T(E_0\bar{P} - \bar{P}H_0\bar{P})^{-1}H_T P\nonumber\\
&=
    {\sum}_{jj'}
    \mel{\Psi(\phi_j)}{H_T(E_0\bar{P} - \bar{P}H_0\bar{P})^{-1}H_T}{\Psi(\phi_{j'})}\nonumber\\
  &\hspace{3.5cm}\times  \dyad{\Psi(\phi_j)}{\Psi(\phi_{j'})},\label{eqn:2ndorder}
\end{align}
where $E_0 = E_A + E_B + \lambda_C N_A^e N_B^e$ is the unperturbed energy of the metallic island. 
The diagonal matrix elements inside the sum (for $j=j'$) can be evaluated from the Fourier transform of the Greens' function \cite{ref:bruus}:
\begin{align}
\bra{\Psi(\phi_j)}{H_T&(E_0\bar{P} - \bar{P}H_0\bar{P})^{-1}H_T}\ket{\Psi(\phi_j)}
\nonumber
\\
&=
	\mathscr{F}_{\tau\rightarrow E_0}
	\langle{
		\mathcal{T}_\tau
		\left[
			H_T(\tau)	H_T
		\right]	
	}\rangle\nonumber\\
&=
	2\sum_{\bm{k}_A\bm{k}_B}
	t^2
	e^{i\phi_j}
	\mathcal{F}_{\uparrow\downarrow} (\bm{k}_A, \tau)
	\mathcal{F}_{\downarrow\uparrow} (\bm{k}_B, \tau)
	+
	\text{c.c.},
\end{align}
where $\mathscr{F}_{\tau\rightarrow E_0}$ is the Fourier transform from time to energy, and we assume a uniform tunnelling matrix element, $t_{\bm{k}_{A}\bm{k}_{B}}=t$, and
$\mathcal{F}_{\uparrow\downarrow}$ and $\mathcal{F}_{\downarrow\uparrow}$ are anomalous Greens functions in reciprocal space, defined by
\mbox{$
\mathcal{F}_{\sigma\sigma'} (\bm{k}, \tau)
\equiv
-
\langle{\mathcal{T}_{\tau}
	c_{-\bm{k}\sigma}^{\dagger} (\tau)
	c_{\bm{k}\sigma'}^{\dagger}\rangle
}.
$}

To compute off-diagonal matrix elements, we also include a phase displacement operator $D_\varphi$, defined for a single island as 
\begin{align} \label{eq:Phase_displacement_OP_def}
		D_{\varphi}=
        \exp\big(
        i{\varphi}
        \hat{N}^{\rm CP}/2
        \big),
\end{align}
which displaces phase states in the usual way 
$
		D_{\varphi} \ket{\Psi(\phi)}
		=
		\ket{\Psi(\phi + \varphi)}.
$

With this definition, the off-diagonal matrix elements in \cref{eqn:2ndorder} are given by
\begin{align}	 
	\bra{\Psi(\phi_j)}&{H_T (E_0\bar{P} - \bar{P}H_0\bar{P})^{-1} H_T}\ket{\Psi(\phi_{j'})}
\nonumber
\\
&=
	\bra{\Psi(\phi_j)}{H_T (E_0\bar{P} - \bar{P}H_0\bar{P})^{-1} H_TD_{\varphi_{j'j}}}\ket{\Psi(\phi_j)},\nonumber\\
&=
	\langle{H_T(\tau) H_T D_{\varphi_{j'j}}}\rangle,\nonumber
\\
&=
	2\sum_{\bm{k}_A\bm{k}_B}
	t^2
	e^{i\phi_j}
	\mathcal{F}_{\uparrow\downarrow} (\bm{k}_A, \tau)
	\mathcal{F}_{\downarrow\uparrow} (\bm{k}_B, -\tau)\nonumber
\\
&\hspace{3cm}\times
	\ip{\Psi(\phi_j)}{\Psi(\phi_{j'})}
    +
	\text{c.c.},\label{eq:time_ordering_2FF}\nonumber\\
	&=
    2t^2 e^{i\phi_j}
    \dfrac{2 N_A^e N_B^e}{b^2 - 1}
    \mathcal{I} (E_0,\Delta)
    +
    \text{c.c.},
\end{align}
where
\begin{equation} \label{eq:double_integral_2order}
	\mathcal{I} (E_0, \Delta) =
\begin{cases}
	\dfrac{\pi^2}{4\Delta}
	\left(
		1
		-
		\dfrac{E_0^2}{16\Delta^2}
	\right),
&
	\text{for }E_0\ll\Delta
\\
	\dfrac{\pi\ln(2E_0/\Delta)}{2E_0},
&
	\text{for }E_0\gg\Delta
\end{cases}.
\end{equation}
Calculational details are given in \cref{append:GreensFun}.

The unperturbed energy $E_0$ is roughly the Coulomb interaction between the two superconductors, \mbox{$E_0 \sim \lambda_C = E_C$}, where
$
E_C = (2e)^2/(2C)
$.
For a typical  capacitance $C\sim10\;\si{\femto\farad}$  of a Cooper-pair-box system we find that
\(
E_C \sim 10^{-5}\;\si{\electronvolt}
\)
\cite{ref:2CPB_circuit}.
In general, the superconductor gap is $\sim10^{-3}\;\si{\electronvolt}$, such that $E_0/\Delta\sim10^{-2}$, so we obtain $\mathcal{I} \approx \pi^2/2\Delta$ and then the second-order tunnelling contribution  is 
\begin{align}	
	&PH_T
	\left(
		E_{0}\bar{P}-\bar{P}H_{0}\bar{P}
	\right)^{-1}
	H_TP
\nonumber
\\\nonumber
&\approx
	\frac{\pi^2N_A^eN_B^e}{b^2}
	\frac{t^2}{\Delta}\sum_{j,j'}
	e^{i\phi_j}
	\underbrace{
		\ip{\Psi(\phi_j)}{\Psi(\phi_{j'})}
	}_{\approx\delta_{j j'}}
	\dyad{\Psi(\phi_j)}{\Psi(\phi_{j'})}
\\\nonumber
&\quad+
	\text{c.c.},
\\
&\equiv E_J \cos\hat{\phi},\label{eq:2order}
\end{align}
where we have defined
\mbox{$
\cos\hat{\phi}
=
\sum_j
\cos\phi_j
\dyad{\Psi(\phi_j)}
$} and 
\begin{equation}
E_J={t^2}{2\pi^2N_A^eN_B^e}\Delta/\mathcal{B}^2.\label{eqn:EJ}
\end{equation} 
\Cref{eq:2order} is the expected functional form describing the effective Hamiltonian for the well-known Josephson junction.  We compare the estimate for $E_J$  to physical values of the Josephson energy in  \cref{sec:experimental}.

Importantly, since the junction is point-like, the phase coordinate is compact, $\phi\in\mathcal{Z}\subset[-\pi,\pi)$.

\subsection{Inductance: Spatially-Varying Phase Torsion}
 \label{sec:inductance}

Here we analyse the  inductance of a superconducting wire.  The inductive energy arises from two contributions: the kinetic energy of the electrons in motion, and the potential energy in the  magnetic field induced by the current. Both terms contribute to the effective nodal flux, $\Phi$, at each point in a circuit, which is itself determined by the superconducting phase, $\phi$.

\begin{figure}[!]
	\centering
    \includegraphics[width=\columnwidth]{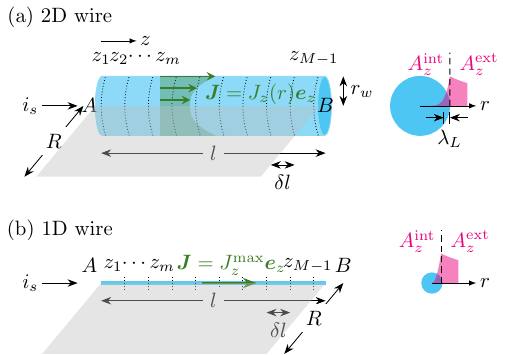}
	\caption{
    (a)
    A cylindrically symmetric superconducting wire.
    The wire carries a supercurrent $i_s = \int_{\bm{\sigma}} \bm{J} \vdot \dd{\bm{\sigma}}$, which induces an internal field inside the superconductor (\cref{append:Afield_SCwire}).
    The external magnetic field penetrates a small depth $\lambda_L = \sqrt{m / (\mu_0 D e^2)}$ into the superconductor, where $D$ is the electron density in the material and $m$ is electron mass.
    (b)
    The 1D approximation model to (a). We consider the effective cross-section area 
    \(
        \sigma_\text{eff} = \pi r_w^2 - \pi (r_w - \lambda_L)^2
        \approx 2\pi r_w\lambda_L,
    \)
    i.e., the area that the applied vector potential can penetrate into.
    The supercurrent in the 1D model is the maximal value in the 2D model: $\bm{J} = J_z^\text{max} \bm{e}_z$.
	}
	\label{fig:SC_straightWire}
\end{figure}

We consider a superconducting wire with finite cross-section area shown in \cref{fig:SC_straightWire}(a). For notational simplicity, we will take the wire to be cylindrical, so that the problem is a 2D system (while realistic microelectronic wires are 3D), and the superconducting phase and the vector potential depend on both the axial and radial coordinates, $z$ and $r$.  However we seek to describe the problem as a quasi-1D wire, shown in \cref{fig:SC_straightWire}(b), characterised by an axially varying phase, $\phi(z)$. 

Our goal is to derive the relevant Hilbert space and the constitutive relation for a lumped inductive element from the microscopic wavefunction of a distributed, spatially-varying phase in a superconducting wire of length $l$. We start with a brief review of the classical analysis of a superconducting inductive wire, with more detail given in \cref{append:Afield_SCwire}.

Classically, the current density, $\bm{J}(\bm{r})$, in a superconductor is related to the vector potential, $\bm{A}^\text{int} (\bm{r})$,  \cite{ref:Jackson_electrodynamics}
\begin{equation}    \label{eq:JA}
    \bm{J} (\bm{r})
=
    \dfrac{1}{\mu_0}
    \curl{\curl{\bm{A}^\text{int} (\bm{r})}},
\end{equation}
where $\bm{A}^\text{int} (\bm{r})$ is the vector potential on the interior of the wire, and $\bm{J} (\bm{r}) = J_z (r) \bm{e}_z$ is axially symmetric. For convenience we choose a gauge in which the vector potential is also axially symmetric, so that 
\begin{equation}    \label{eq:Agauge}
    \bm{A} (\bm{r})
=
    A_z(r) \bm{e}_z, 
\end{equation}
and we give more details about the form of $A_z(r)$ in \cref{eq:Afield_int}. 
In an axially symmetric gauge, we solve \cref{eq:JA} to find 
\begin{equation}
    A^\text{int}_z(r)
=
    -\mu_0\int_0^r dr' \,r'\ln(r/r')J_z(r').
\end{equation}
That is, $A^\text{int}_z$ is a linear functional of the current density.

More generally, the gauge field inside the superconductor is a linear functional of the current density, 
\begin{equation}    \label{eq:AJfunctional}
    \bm{A}^\text{int}(\bm{r})
=
    \bm{A}^\text{int}[\bm{J}],
\end{equation} which we use in subsequent analysis, below.    
The total current in the wire is the integral of the current density over a cross-section $\bm{\sigma} = \pi r_w^2 \bm{e}_z$, i.e. 
\(
    i_s
\equiv
    \int_{\bm{\sigma}} \bm{J} \vdot \dd{\bm{\sigma}}
\).  
Since the system is linear, the current density will scale linearly with total current in the wire $\bm{J} (\bm{r}) =i_s \, \bm{j}(\bm{r})$, where $\bm{j}$ is the normalised current density profile satisfying $\int_{\bm{\sigma}} \bm{j} \vdot \dd{\bm{\sigma}}=1$, which carries information about the specific geometry of the wire. Putting these functional forms together, we have the general result
\begin{equation}
    \bm{A}^\text{int}[\bm{J}]
=
    i_s \bm{A}^\text{int} [\bm{j}(r,z)],\label{eqn:Aintis}
\end{equation}
indicating that the vector potential inside the superconductor scales linearly in the total current carried by the wire.

\subsubsection{BCS Subspace and Projector for a 1D Wire} 

We assume that a wire of finite diameter can be approximated by a quasi-1D wire shown in \cref{fig:SC_straightWire}(b), which is characterised by an axially-varying  phase, $\phi(z)$.
 Formally, we  discretise the wire into $M \gg 1$ segments, so that the segmented Hilbert space of the wire is \mbox{$\mathcal{H}_\text{wire} = \mathcal{H}_A \otimes \mathcal{H}_{z_1} \otimes \cdots  \otimes \mathcal{H}_B$}, in a  basis spanned by states of the form
\begin{align}
    \ket{\Psi (\phi(z))}
&=
    \ket{\Psi_A (\phi_A)}
    \otimes
    \ket{\Psi_{z_1} (\phi_{z_1})}
    \otimes
    \cdots
    \otimes
    \ket{\Psi_B (\phi_B)},\nonumber\\
    &\equiv\bigotimes_z
    \ket{\Psi_{z} (\phi(z))},\label{eq:fieldprojector}
\end{align}
where $z$  labels both the tensor-product element along the discretised wire, and the phase coordinate at each position.   In this formulation, there is a local compact Hilbert space associated to each segment

Compared to the point-like Josephson junction and capacitor model, the (distributed) inductive element has a significantly larger Hilbert space, which will become important below.  We use the basis spanned by states of the form in \cref{eq:fieldprojector} to define the low energy projector over the length of the wire,
\begin{align}   \label{eq:projector_inductor}
    \Pi&=\sum_{\phi(z)}\bigotimes_z\dyad{\Psi_z (\phi(z))},\nonumber\\
    &=\bigotimes_z\sum_{\phi(z)}\dyad{\Psi_z (\phi(z))},\nonumber\\
    &=\bigotimes_z P_z,
\end{align}
where $P_z=\sum_{\phi(z)}\dyad{\Psi_z (\phi(z))}$ defines a projector at each of the segments, $z$, along the wire,  generalising \cref{eq:projector}.  Similarly, we define the phase-field, and phase-gradient operators 
\begin{align}
    \hat \phi(z)&=
    \bigotimes_z\sum_{\phi(z)} \phi(z) \dyad{\Psi_z (\phi(z))},\\
    \hat \phi'(z)&=
    \bigotimes_z\sum_{\phi(z)} \phi'(z) \dyad{\Psi_z (\phi(z))},\label{eqn:phi'}
\end{align}
where the derivative in \cref{eqn:phi'} is understood on the segmented wire as as a finite difference. 
We note that the summations in these field operators represent path-integral-like objects \cite{Ambegaokar1982}.

\subsubsection{Inductive Constitutive Law} 
Using the second quantised form of the field operators $\psi_{\bm{r}} = e^{i\phi(z)/2} c_{\bm{r}}$ (where we suppress the spin label $s$), the (quantised) supercurrent density operator is given by \cite{ref:Tinkham_superconductivity, ref:Kertterson}
\begin{equation}   \label{eq:current_densityOP}
    \hat{\bm{J}} (\bm{r})
=
    \dfrac{e}{2m}
    \left(
        \psi^\dagger_{\bm{r}} (\hat{\bm{p}} - 2e \hat{\bm{A}}^\text{int})  \psi_{\bm{r}}
        +
        \text{h.c.}
    \right),
\end{equation}
where the quantised verision of \cref{eq:AJfunctional}
\begin{equation}
\hat{\bm{A}}^\text{int}(r) = \bm{A}^\text{int}[\hat{\bm{J}}]\label{eqn:Afunctional}
\end{equation} 
is itself a linear functional of the current density operator, which can be computed from \cref{eq:JA} for any given geometry. \Cref{eq:current_densityOP}  therefore  implicitly relates $\hat{\bm{J}}$ to the microscopic electronic degrees of freedom.

In the effective low-energy BCS subspace, the supercurrent density operator is given by the projection onto the low energy subspace
\begin{align}
    \hat{\bm{J}}^\text{eff}(r,z)
\nonumber
&=
    \Pi \hat{\bm{J}}  \Pi
\\\nonumber
&=
\bigotimes_{z}\sum_{\phi(z),\bar\phi(z)}
    \mel{\Psi_z(\phi(z))}{\hat{\bm{J}} (\bm{r})}{\Psi_z(\bar\phi(z))}
\\
&\hspace{2.5cm}
   \times
    \dyad{\Psi_z(\phi(z))}{\Psi(\bar\phi(z))},\nonumber
\\\nonumber
&\approx
  \bigotimes_{z}\sum_{\phi(z)}  
    \ev{\hat{\bm{J}} (r,z)}{\Psi_z(\phi(z))}
\\\nonumber
&\hspace{2.5cm}
\times\dyad{\Psi_z(\phi(z))},
\\
&=
    \bigotimes_{z}\sum_{\phi(z)}
    {\bm{J}} (r,z)
    \dyad{\Psi_z(\phi(z))},\label{eq:Jeff}
\end{align}
where we have used the (approximate) orthonormality of the set $\{\ket{\Psi(\phi(z))}\}$ for each $z$. 
The  current density matrix element is computed in \Cref{append:spatialBCS} to be
\begin{align}   
    \bm{J}(r,z)
&=
    \ev{\hat{\bm{J}} (r,z)}{\Psi(\phi)},\nonumber
\\
    &=
    \dfrac{De^2}{m}
    \Big(
        \dfrac{\Phi_0}{2\pi}
        \phi'(z) \, \bm{e}_z
        +
        \bm{A}^\text{int} (r)
    \Big),\label{eq:Jexpectation} \\
    &=
    \dfrac{De^2}{m}
    \Big(
        \dfrac{\Phi_0}{2\pi}
        \phi'(z) \, \bm{e}_z
        +
        \bm{A}^\text{int} [{\bm J}]
    \Big),\label{eq:Jexpectation2} 
\end{align}
where $m$ is the electron mass, $e$ is the electron charge, and $D$  is the electron density. 
 The current-operator matrix element \cref{eq:Jexpectation}  is identical to the mean-field theory current density in a wire induced by gauge-invariant phase torsion \cite{ref:Tinkham_superconductivity}. \Cref{eq:Jexpectation2} makes clear that this expression represents an implicit definition of the projected current density matrix elements in terms of the BCS phase gradient, $\phi'$.

Putting together equations (\ref{eq:current_densityOP}), (\ref{eqn:Afunctional}), (\ref{eq:Jeff}) and (\ref{eq:Jexpectation2}), we find the local current density operator in the low-energy BCS subspace is given by
\begin{align}\label{eq:Jeffective}
    \hat{\bm{J}}^\text{eff}(r,z)
&=\dfrac{De^2}{m}
    \Big(
        \dfrac{\Phi_0}{2\pi}
        {\hat \phi}'(z) \, \bm{e}_z
        +
        \bm{A}^\text{int}[\hat{\bm{J}}^\text{eff}]
    \Big).
\end{align}

We are mainly interested in the mesoscopic properties of the wire, in particular, the net current  integrated over the wire cross section,  $\bm{\sigma}$. Integrating both sides of \cref{eq:Jeffective} gives
\begin{align}    \label{eq:totalcurrent}
    \hat i_s(z)
&=
    \int_{\bm{\sigma}}
    \hat{\bm{J}}^\text{eff}(r,z)  \cdot \dd{\bm{\sigma}},\nonumber\\
    &= \frac{De^2}{m}
    \Big(\sigma_{\cross}
        \dfrac{\Phi_0}{2\pi}
        {\hat \phi}'(z) 
        +
        \hat i_s(z) \int_{\bm{\sigma}} \bm{A}^\text{int}[{\bm{j}(r,z)}]
    \Big),
\end{align}
where $\sigma_{\cross}$ is the cross sectional area, and we have used \cref{eqn:Aintis}  in the second line.  Solving for $\hat i_s(z)$ yields
\begin{align}   \label{eq:supercurrentOP0}
    \hat{i}_s(z)
=
    \tfrac{\Phi_0}{2\pi}
    {\hat{\phi}}'(z)
    /{\bar{L}},
\end{align}
where the total linear inductive density (i.e.\ inductance per unit length) is given by $\bar{L} = \bar L_K + \bar L_G$, which consists of the kinetic and geometric inductive densities 
\begin{align}
\bar L_K = \frac{m}{D e^2\sigma_{\cross}} \textrm{ and } 
\bar L_G =-\frac{1}{\sigma_{\cross}}\int_{\bm{\sigma}} \bm{A}^\text{int}[{\bm{j}(r,z)}].\label{eqn:barL}
\end{align}

For a concrete example, we consider a  uniform cylindrical wire, in which we have an explicit form for the current density, given in \cref{eq:supercurrent_density}.  In that case, the total integrated current operator is
\begin{align}
    \hat{i}_s(z)
\nonumber
&=
  2\pi\!
    \int_{0}^{r_w}\! dr \, r 
    \hat{J}_z(r),
\nonumber
\\
&=
    \dfrac{De^2}{2m}
    \Big(
      {\Phi_0}  {\hat{\phi}'(z)}
        r_w^2
        -
       \hat{i}_s(z) {\mu_0  \lambda_L}
        \big(            2\lambda_L\!-\!
            \tfrac{r_w}{I_1(r_w/\lambda_L)}   
        \big)
    \Big),\label{eq:is}
\end{align}
where $\lambda_L = \sqrt{m/(\mu_0 De^2)}$ is the London penetration depth. In the limit $\lambda_L \ll r_w$, $I_1(r_w/\lambda_L)^{-1}$ is exponentially suppressed. Solving for $\hat{i}_s(z)$ gives
\begin{align}   \label{eq:supercurrentOP}
    \hat{i}_s(z)
\approx
    \tfrac{\Phi_0}{2\pi}
    {\hat{\phi}}'(z)
    \frac{1}
    {
        \tfrac{m}{De^2}
        \tfrac{1}{\pi r_w^2}
        +
        \tfrac{\mu_0}{8\pi}
    }
=
    \tfrac{\Phi_0}{2\pi}
    {\hat{\phi}}'(z)
    /{\bar{L}},
\end{align}
where the total inductive density is given by \mbox{$\bar{L} = \bar L_K + \bar L_G$}, where  $\bar L_G = \mu_0  / (8\pi)$ (which agrees with the usual calculation of $\bar L_G$ up to a logarithmic factor in \cref{eq:inductance_geometric}), and $\bar L_K$ is given in \cref{eqn:barL}.

Finally, we make a lumped element approximation, in which  ${\hat{\phi}}'(z)$ is  assumed to be uniform over the length of the wire, and is determined by the boundary values $\hat\phi_{A,B}\equiv\hat \phi(z_{A,B})$.  That is, the phase-field operator simply interpolates linearly between the phases at each end of the wire ${\hat{\phi}}'(z)=({\hat \phi_B-\hat\phi_A})/{l}$, so that there is only a single effective current operator associated to the wire, given by
\begin{equation}   \label{eq:supercurrentOP_QM}
    \hat{i}_s
=
    ({\hat \Phi_B-\hat\Phi_A})/L,
\end{equation}
where $L={\bar{L}l}$ is the total inductance, and 
\begin{equation}
    \hat\Phi_{A,B}\equiv\tfrac{\Phi_0}{2\pi}\hat\phi_{A,B}\label{eqn:FluxPhase}
\end{equation}
defines the  lumped-element nodal fluxes associated to the end points of the wire. 
\Cref{eq:supercurrentOP_QM} agrees with the usual constitutive inductive relation between current and flux. Importantly in \cref{eqn:FluxPhase}, $\hat\Phi$ is derived from the underlying phase operator, $\hat\phi$.

\subsubsection{Effective Hamiltonian of an Inductor}
Here we project the spatially varying fermionic Hamiltonian \cref{eq:Hamiltonian_SCwoMFA} onto the low energy subspace $\mathcal{H}_\text{BCS}$, using a generalised projector $\Pi$ that extends over the length of the wire. 
In the position basis in the presence of a vector potential, the electronic kinetic term in the island Hamiltonian $H_A$ from \cref{eq:HKA} is given by
\begin{align} \label{eq:continuumBCS}
	H_{K_A}
&=
	\int_\text{SC} \dd{V}
		c_{\bm{r}s}^\dagger
		\left[
			\dfrac{1}{2m}
			\left(
				\bm{p} + q \bm{A}^\text{int}
			\right)^2
			-
			\mu
		\right]
		c_{\bm{r}s},
\end{align}
where the volumetric integral is over the interior of the superconducting metal. 
The effective  Hamiltonian for an inductive wire is given by the projection 
\begin{equation}
H_{\rm wire}=\Pi (H_{K_A} + H_{I_A}) \Pi,
\end{equation}
 in which the electronic interaction term $H_{I_A}$ contributes a constant $\mathcal{V} (\varphi)$, given in \cref{eqn:V}, which sets an energy offset  in the inductive Hamiltonian.

To proceed, we remove the $z$-dependence  by introducing the gauge transformation $\mathcal{G}_{\phi(z)} = e^{i\phi(z)\hat{N}/2}$ and compute the projection
\begin{align*}
 H_{\rm wire}&=   \Pi H_{K_A} \Pi\nonumber\\
&=
    \bigotimes_{z}\sum_{\phi(z),\bar\phi(z)}
    \mel{\Psi_z(\phi(z))}{H_{K_A}}{\Psi_z(\bar\phi(z))}
\\
&\hspace{2.5cm}\times
    \dyad{\Psi_z(\phi(z))}{\Psi(\bar\phi(z))},\nonumber
\\
&\approx
    \bigotimes_{z}\sum_{\phi(z)}  
    \ev{H_{K_A}}{\Psi_z(\phi(z))}
\\
&\hspace{2.5cm}
    \times\dyad{\Psi_z(\phi(z))},
\\
&=
    \dfrac{1}{2}
    \dfrac{De^2}{m}
    \int_\text{SC}  \dd{V}
    \Big(
        \dfrac{\Phi_0}{2\pi}    \hat{\phi}'(z)
        -
        A^\text{int}_z  (\hat{i}_s)
    \Big)^2,
\end{align*}
where $A^\text{int}_z$ is the vector potential inside the superconducting wire that couples to the current operator \cref{eq:supercurrentOP}, ``SC" denotes the interior volume of the superconductor, and the last line is a volumetric integral.  Details of the derivation are given in Equations (\ref{eq:projection_inductance_mel}) to (\ref{eq:HL_elem}).
For a superconducting wire with cylindrical symmetry the vector-potential operator is given by \mbox{$A^\text{int}_z (\hat{i}_s) = \hat{i}_s \frac{\mu_0\lambda_L}{2\pi r_w} \frac{1 - I_0(r/\lambda_L)}{I_1(r_w/\lambda_L)}$}, computed in \cref{eq:Afield_int}.  Expressing $\hat{i}_s$ in terms of $\hat{\phi}'$ using \cref{eq:supercurrentOP}, we obtain
\begin{align}\label{eq:Pi.HKA.Pi_noApprox}
 H_{\rm wire}
&\approx
    \dfrac{1}{2}    \Big( \dfrac{\Phi_0}{2\pi} \Big)^2
    \dfrac{1}{\bar{L}}
    \int_0^l    d{z}\,
    \hat{\phi}'(z)^2,
\hspace{0.2cm}
    \text{for }\lambda_L \ll r_w.
\end{align}
Details of the integration are given in \cref{eq:Pi.HKA.Pi}.

\subsubsection{Lumped Element Inductive Energy}
We evaluate the integral in \cref{eq:Pi.HKA.Pi_noApprox} in the lumped element approximation in which $\hat{\phi}'(z) = (\hat{\phi}_B - \hat{\phi}_A) / l$ is independent of $z$, so that
$
    \int_0^l \dd{z}
    \hat{\phi}'^2
=
    (\hat{\phi}_{B}-\hat{\phi}_{A})^2 / l
$
and \cref{eq:Pi.HKA.Pi_noApprox} yields
\begin{align}
 H_{\rm wire}^{\rm lumped}
&=
    \dfrac{1}{2L}    \Big( \dfrac{\Phi_0}{2\pi} \Big)^2
  (\hat{\phi}_{B}-\hat{\phi}_{A})^2 
=
    \dfrac{E_L}{2}   (\hat{\phi}_{B}-\hat{\phi}_{A})^2, \nonumber\\
  &  \equiv  H_{L},\label{eqn:HL}
\end{align}
where  we define the inductive energy of the lumped element as  $E_L \equiv (\frac{\Phi_0}{2\pi})^2 / L$.

To make the lumped element approximation more explicit, we rederive the expression for $H_{L}$ in the segmented model for the wire.  As in \cref{eq:fieldprojector}, we divide the wire into $M$ segments, so that the finite difference approximation for the local phase gradient is $\hat \phi'(z_m)= (\hat \phi_{m+1}-\hat \phi_{m})M/\ell$, and then 
 \cref{eq:Pi.HKA.Pi_noApprox} becomes
\begin{align}
 H_{\rm wire}&=\frac{E_L}{2} M
    \sum_{m=0}^{M}
    ( \hat{\phi}_{m+1} - \hat{\phi}_m )^2,\label{eqn:Hwiresegment}
\end{align}
where $m=0$ and $m=M$ label the exterior segments at the ends of the wire, $z_A$ and $z_B$, respectively.
This Hamiltonian  acts on the Hilbert space $\mathcal{H}_A \otimes \cdots \otimes \mathcal{H}_B$, which is a large tensor-product of local BCS subspaces. 

To make the lumped element approximation, we assume that the phases at the interior segments \mbox{$0<m<M$} are constrained  to minimise the energy of the wire, given the exterior phases, $\hat{\phi}_{A}$ and $\hat{\phi}_B$.  It is straightforward to show that the partial minimisation of $ H_{\rm wire}$ over the values of the interior phases yields $ \hat{\phi}_{m+1} - \hat{\phi}_m = (\hat{\phi}_{B} - \hat{\phi}_A)/M$. Substituting this into \cref{eqn:Hwiresegment} yields \cref{eqn:HL}.

\subsubsection{Emergent Low-energy Hilbert Space of an Inductor}

While we have described a point-like superconducting element (e.g. a capacitor or tunnel junction) with a periodic coordinate $\phi\in[-\pi,\pi)$, which respects the $2\pi$ $\phi$-translation symmetries of these devices, the inductive Hamiltonian \cref{{eqn:HL}} breaks $\phi$-translation symmetry, and the Hilbert space of $\hat \phi_{AB}$ is substantially larger.  Here we provide a physically-motivated argument that provides an estimate on the size of the physically relevant Hilbert space for an inductive element.

As discussed above, the discretised 1D inductor is formed from $M$ segments, and recalling the definition in \cref{eq:Zset} of $\mathcal{Z}\subset[-\pi,\pi)$, the natural Hilbert space for the wire  is 
\begin{align}   \label{eq:emergent_space_range}
    \mathcal{H}_\text{wire} 
&= 
    \mathcal{H}_A  \otimes \cdots  \otimes  \mathcal{H}_B
=
    \mathcal{Z}^{\otimes M-1},
\end{align}
where the exponent $M-1$ indicates that the relevant internal degrees of freedom are the relative phases \mbox{$\delta\phi(z_m)=\phi(z_{m+1})-\phi(z_{m})\in\mathcal{Z}$} between neighbouring segments.

The subsequent ``lumped-element'' approximation treats the dynamics of the interior segments as being determined by the boundary  values $\phi_{A}$ and $\phi_{B}$,  and so the accessible states within $\mathcal{H}_\text{wire} $ satisfy \mbox{$\delta\phi(z_m) =(\phi_{B}-\phi_A)/M\in{\mathcal{Z}}$} for each segment $z_m$.  It follows that 
\begin{equation}
\phi_{B}-\phi_A\in M\mathcal{Z}\subset [-M \pi,M\pi), \label{eqn:InductiveHilbertSpace}
\end{equation} 
which determines the effective, lumped element Hilbert space of the discretised wire.

We next note that the number of segments, $M$, is bounded by the physical constraint that the supercurrent cannot exceed the critical current, $i_c$, of the wire.  In particular, the total current in the wire is \mbox{$\abs{i_s} = \Phi_0 |\phi_B-\phi_A|/L\leq \Phi_0 M\pi/L<i_c$}, and so 
\begin{equation}
   M \leq 
    {i_c L}/({\pi\,\Phi_0})\equiv M_\text{max}.\label{eqn:Mmax}
\end{equation}

An estimate for the critical current, 
\mbox{$
    i_c
=
    \frac{\Delta}{\lambda_L(T)}
    \sqrt{\frac{\rho_F}{2\mu_0}}
$} \cite{ref:annett}, indicates that the accessible Hilbert space for the lumped-element inductive wire depends on the physical properties of the wire. 
It follows that when $L$ is large $M_\text{max}\gg1$, and so the  lumped element phase difference operator $\hat{\phi}_B-\hat{\phi}_A$ has support over a large but finite interval given by \cref{eqn:InductiveHilbertSpace}, which is bounded above by \cref{eqn:Mmax}.  Unlike the Hilbert space of the point-like Josephson junction, the Hilbert space for the inductive element is no longer periodic, since lumped-element states with ${\phi}_B-{\phi}_A=\pm\pi M_\text{max}$ correspond to orthogonal states with opposite current flows. 

We comment in passing that this argument really asserts that the dissipationless, lumped-element effective model for an inductive element requires the phase drop across the inductor should be bounded in the interval $\phi_{B}-\phi_{A}\in [-M_\text{max}\pi,M_\text{max}\pi)$.  If  $\phi_{B}-\phi_{A}$ were to exceed this range, the supercurrent would exceed the critical current, and dissipative  quasiparticle formation would ensue.  

A more formal treatment of this dissipative limit on the accessible Hilbert space for a superconducting inductor would require the inclusion of quasiparticle states in a larger projective Hilbert space. We speculate that the dissipation would be described perturbatively by open-systems Lindblad superoperators whose effect would be to suppress large deviations in $\phi_{B}-\phi_{A}$.  This is beyond the goals of this work, and we leave this to future research.

\subsection{Effective LCJ Hamiltonian}
Combining the results given above, the  effective Hamiltonian  for the LCJ circuit shown in \cref{fig:system} is
\begin{align} \label{eq:Hamiltonian_LCJ}
	H_{\text{LCJ}}
&\approx
	E_C \hat{N}_{AB}^2- E_J	\cos\hat{\phi}_{AB} +\frac{E_L}{2} 
    \hat{\phi}_{AB}^2,
\end{align}
where $\hat{N}_{AB}=\hat{N}_{B}-\hat{N}_{A}$ and $\hat{\phi}_{AB}=\hat{\phi}_{B}-\hat{\phi}_{A}$.
This agrees with the standard expression for a quantised LCJ circuit derived by re-quantising GL mean field theory.  However, the microscopic derivation that we have undertaken here provides clarity on the low-energy Hilbert space for each component.

\begin{figure}[t]
	\centering
	\includegraphics[width=1.\columnwidth]{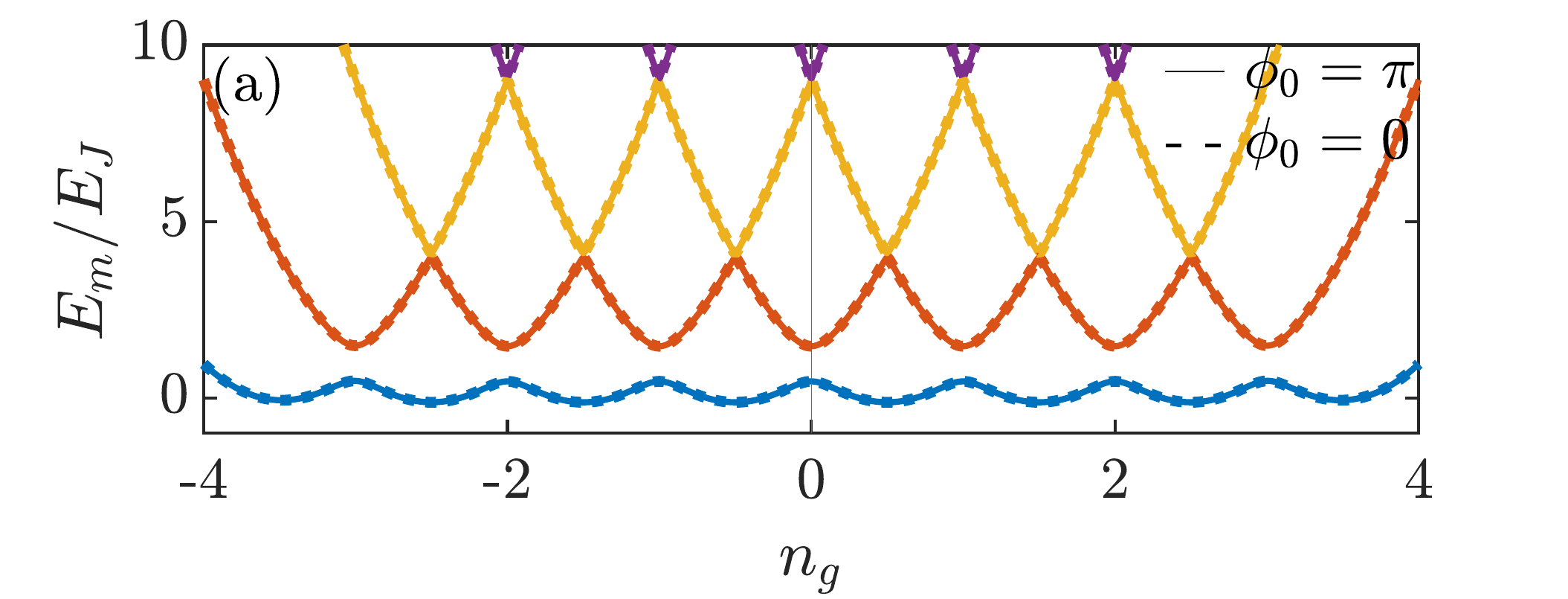}
    \includegraphics[width=1.\columnwidth]{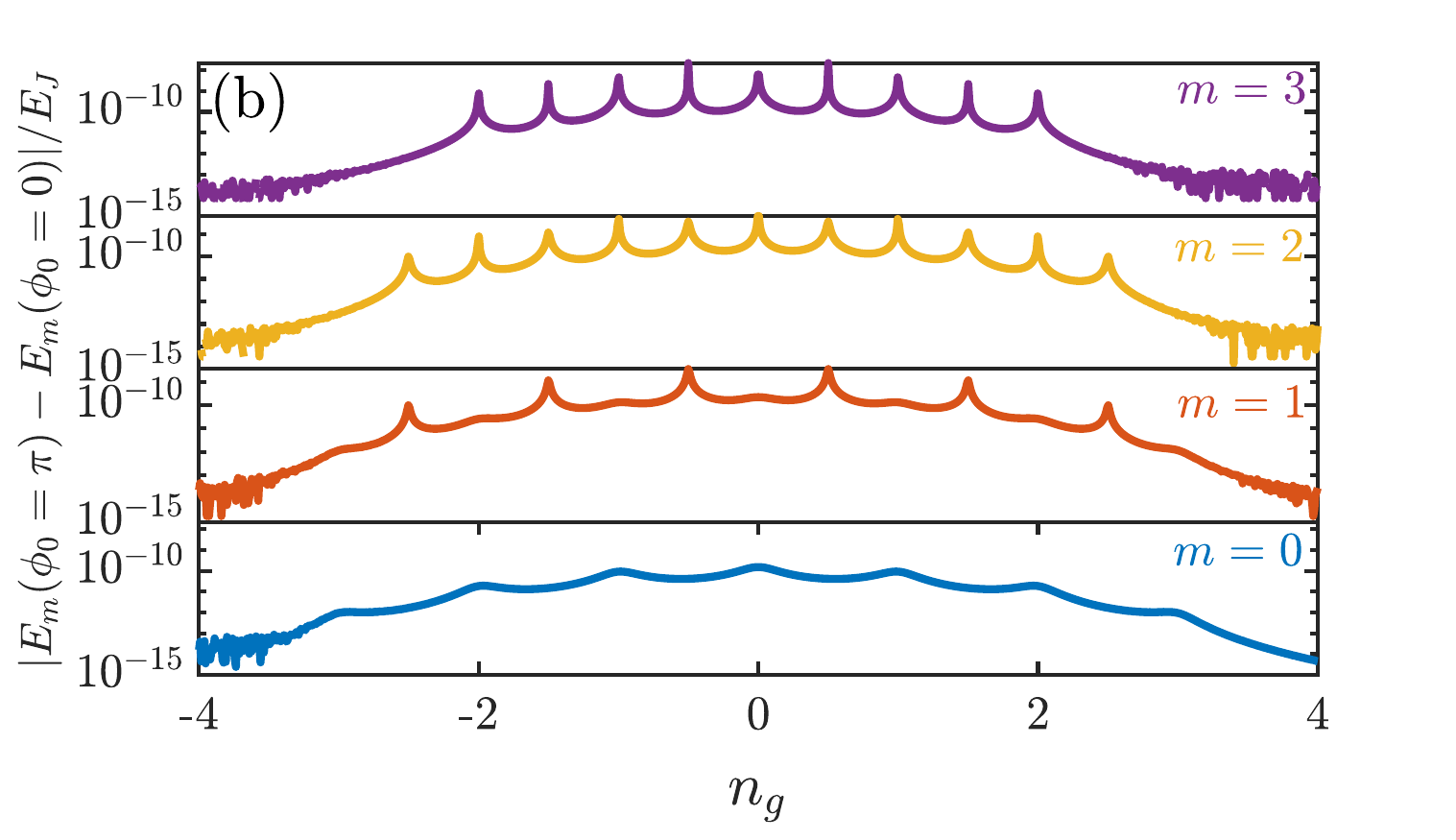}
	\caption{
        (a)
		The eigenenergies $E_m$ for the fist four levels of a transmon \cref{eq:Hamiltonian_CPB} for a $\jmax = 8$-electron system with $E_J = E_C$.
		The solid thin lines are with the reference phase $\phi_0=\pi$, while the dashed lines are with $\phi_0=0$.
		The maximal energy difference between each solid and dashed is of the order $\order{10^{-10}}$ and thus can be viewed as overlap.
		All energies are scaled by $E_J$.
        (b)
        Energy differences between two choices of reference phase ($\phi_0 = 0$ and $\pi$) for $m=0,1,2,3$ levels.
	}
	\label{fig:transmon_spectrum}
\end{figure}

Including external field and voltage bias terms is straightforward but somewhat messy, and results in additional linear terms in the effective Hamiltonian.  Different choices of the energy scales among the energy coefficients results in various  potential landscapes for well known superconducting qubits \cite{ref:geometricSuperinductanceQ}.

\section{Discussion}

\subsection{Effective Capacitor-Junction Hamiltonian}
Transmon and Cooper-pair box qubits are formed from a capacitor and a Josephson junction, so \cref{eq:Hamiltonian_LCJ} reduces to 
\begin{align} \label{eq:Hamiltonian_CPB}
	H_{\text{CJ}}&=
	E_C (\hat{N}-n_g)^2- E_J	\cos\hat{\phi},\nonumber\\
	&=
	E_C (\hat{N}-n_g)^2- E_J	(e^{i\hat{\phi}}	+	e^{-i\hat{\phi}})/2,
\end{align}
where we add a charge bias term by hand. 
Since the circuit is `point-like' with no extended inductive elements, the phase operator $\hat{\phi}$ is compact on the interval $\phi \in [-\pi,\pi)$, and the number operator is correspondingly discrete.

In the number basis, we represent the charge displacement operator as
\begin{align}	\label{eq:raisingOP_compact}
	e^{i\hat{\phi}}	
&={\sum}_{j=1}^{\jmax}\dyad{j-1}{j}
+e^{i(\jmax + 1)\phi_0}\dyad{\jmax}{0}.
\end{align}
Importantly, there is a residual dependence on $\phi_0$, which arises from the compactness of the Hilbert space \cite{ref:Pegg&Barnett}.  This parameter is  determined by convention, and so we address here the effect of this parameter choice.  \Cref{fig:transmon_spectrum}a shows the calculated eigenspectra of \cref{eq:Hamiltonian_CPB} as a function of the gate bias $n_g$, for $\phi_0=0$ (dashed) and $\pi$  (solid), which are almost identical, as shown by the difference plotted in  \Cref{fig:transmon_spectrum}b.  These plots were generated for the unphysically small value $\jmax=8$,  and the differences decreases as $\jmax$ increases.  Practically, this means that the choice of $\phi_0$ is essentially irrelevant for realistic systems.

\subsubsection{Experimental parameter estimates}\label{sec:experimental}

\begin{figure}[t]
    \includegraphics[width=0.6\columnwidth]{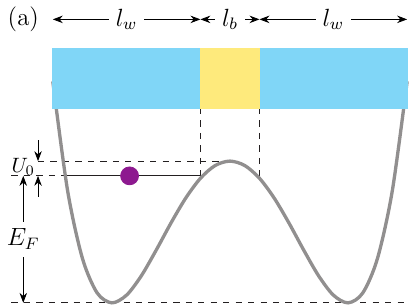}
	\caption{
    We estimate the tunnel matrix element $t$ using the WKB approximation in a double-well potential.
	The oxide layer acts as the barrier, whose energy is $U_0$-higher than the electron energy inside the superconductors.
    }
	\label{fig:double-well}
\end{figure}

Here we compare the microscopic expression for the junction energy in \cref{eq:2order} with physically measured values. 
Typical aluminium  parameters  are   $\Delta\approx340\,\si{\mu\electronvolt}$ \cite{ref:Al_gap}, 
electronic bandwidth $\mathcal{B}\approx10.6\,\si{\electronvolt}$ \cite{ref:Al_band}, so \mbox{$b\approx 3\times 10^{4}$}. 
\ycl{We adopt  device parameters in \cite{ref:exp_SJ}, and assume the metallic islands forming the junction  are identical to one another:
the thickness of the aluminium islands and the oxide barrier are approximately $l_w\approx30\,\si{\nano\meter}$ and $l_b\approx1\,\si{\nano\meter}$, respectively,  the area of the junction is $A_J\approx0.02\,\si{\micro\meter^2}$, and the  measured junction energy is \mbox{$E_J^{\text{\tiny{\cite{ref:exp_SJ}}}}\approx270\,\si{\mu\electronvolt}$}.  With this geometry, and for a face-centred cubic aluminium unit cell of side-length \mbox{$l_c=0.4\text{ nm}$} which contains 4 atoms per unit cell, 
each metallic island contains approximately 
\mbox{$N_{A,B}^e =N_e= 4l_w A_J/l_c^3 \approx 3 \times10^7$} electrons. 
}

\paragraph*{Microscopic tunnelling from WKB method:}
We model the junction barrier as one-dimensional double-well potential, shown  in \cref{fig:double-well}. 
Electrons within each metallic well travel with Fermi velocity $v_F\approx2\times10^6\,\si{\meter/\second}$  for aluminium \cite{ref:Ashcroft&Mermin}, 
colliding with the barrier with an ``attempt'' frequency
$f	=	v_F/ (2 l_w) \approx 30\,\si{\tera\hertz}$.

According to the WKB method, the tunnel matrix element $t$ is given by \cite{ref:Sakurai}
\begin{equation} \label{eq:tunnel_MtxElem}
	t = h\, f\, T_{\rm WKB},
\end{equation}
where 
\begin{equation}
    T_{\rm WKB} = e^{-2\xi \,l_b\sqrt{2m_e^* U_0}/\hbar}\label{eqn:ff}
\end{equation}
     is the WKB  tunnelling amplitude per collision. In this expression 
{$U_0\sim2\,\si{\electronvolt}$ is the barrier height  \cite{ref:Al-I-Al_barrier2014, ref:Al-I-Al_barrier}}, 
 the effective electron mass is $m_e^*=0.4m_e$ for aluminium oxide \cite{ref:Al2O3_effeciveMass}, and we include a dimensionless `fudge-factor' $\xi\sim1$ to account for relative uncertainties in the dimensional factors appearing in the exponent of $T_{\rm WKB}$. 

\begin{figure}[t]
    \includegraphics[width=0.8\columnwidth]{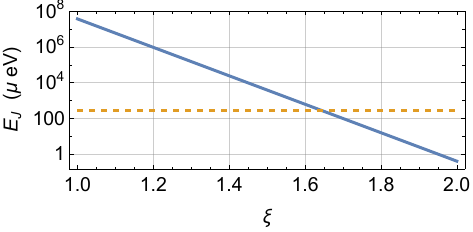}
	\caption{
    The estimated junction energy $E_J$, as a function of the ``fudge-factor''  $\xi$ in the tunnelling amplitude, given by  Equations (\ref{eqn:EJ}) and (\ref{eqn:ff}) (solid line).  The experimentally measured value of $E_J^{\text{\tiny{\cite{ref:exp_SJ}}}}\approx270\,\si{\mu\electronvolt}$ is also shown (dashed line), and matches the derived value of $E_J$ at $\xi\approx1.6$. 
    }
	\label{fig:FF}
\end{figure}

Assuming symmetric metal islands and adopting the geometric parameters in \cite{ref:exp_SJ}, the estimated junction energy is given by \cref{eqn:EJ},  $E_J \approx {t^2}{2\pi^2N_A^eN_B^e}\Delta/\mathcal{B}^2$, which depends exponentially strongly on $\xi$. 
\Cref{fig:FF} shows the  value of $E_J$ estimated from \cref{eqn:EJ}, and for $\xi\approx1.6$ we see that $E_J$ matches the observed value $E_J^{\text{\tiny{\cite{ref:exp_SJ}}}}$. This indicates that \cref{eqn:EJ} plausibly defines $E_J$.

\subsubsection{WKB in Non-Zero Magnetic Fields}
Consider a Josephson junction in an external field descried by $\bm{A}$. 
The tunnel matrix element is oscillatory in phase \cite{ref:JosephsonEffect_externalFields}
\begin{align}
	t'
&=
	t\,  e^{i\int_{\bm{l}} \bm{A} \vdot \dd{\bm{l}} / \Phi_0}\nonumber,\\
 &=
    t\, e^{i\Phi_b l_b / (2 l_w \Phi_0)},
\end{align}
where $t$ is the regular tunnel matrix element without fields, $\bm{l}$ is the tunneling path, and $\Phi_b$ is the bias magnetic flux threading the device.
The tunneling Hamiltonian gains a phase shift in $t$, becoming
\begin{align}
	E_J	\cos\hat{\phi}
&\mapsto
	E_J
	\cos(
		\hat{\phi}	
		-
\tfrac{l_b}{2l_w}{\Phi_b}/{\Phi_0}),\nonumber
\\
&\approx
	E_J	\cos\hat{\phi}, 
\end{align}
where the second line assumes the insulator is much thinner than the well width, $l_w\gg l_b$. 
We see that the bias offset ordinarily would not manifest itself in the constitutive relation for a point-like junction, and should instead be accounted for extended inductive loops  \cite{ref:circuitQuantisation_B_timeDep}.

\subsection{Higgs Modes}   \label{sec:Higgs_modes}
In this work we have treated the phase $\phi=\arg(\Delta)$ as a `position' coordinate, and used it to define a basis for the BCS subspace.  From this basis, we constructed the low-energy phase operator, $\hat \phi$ in \cref{eqn:phase_operator}, and via the Fourier transform, its conjugate $\hat n$.  In the foregoing analysis, we have treated $|\Delta|$ as a constant, determined by minimising the ground space energy of a metallic island, given by \cref{eq:Hl_diagElem}.  Here, we briefly comment on some further considerations of this energy landscape.

\begin{figure}[!]
		\centering
		\includegraphics[width=0.49\columnwidth]{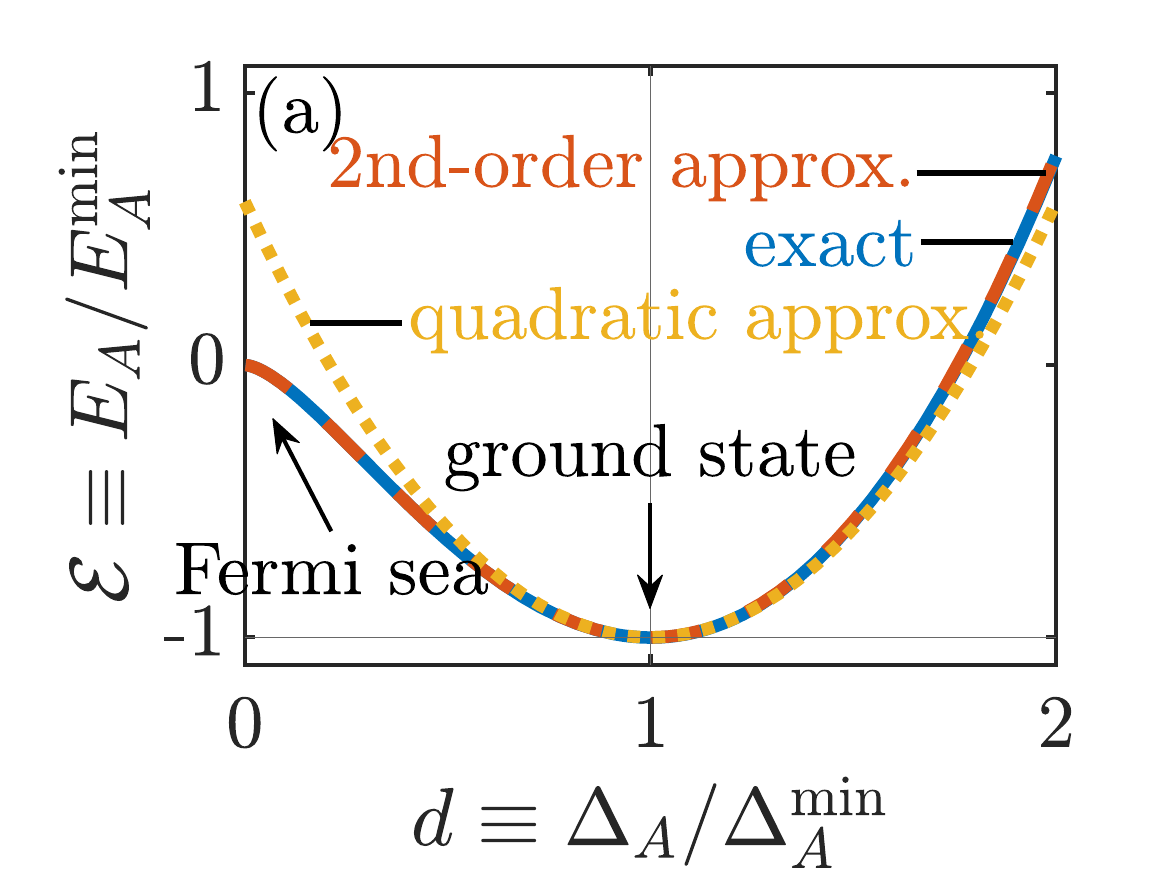}
		\centering
		\includegraphics[width=0.49\columnwidth]{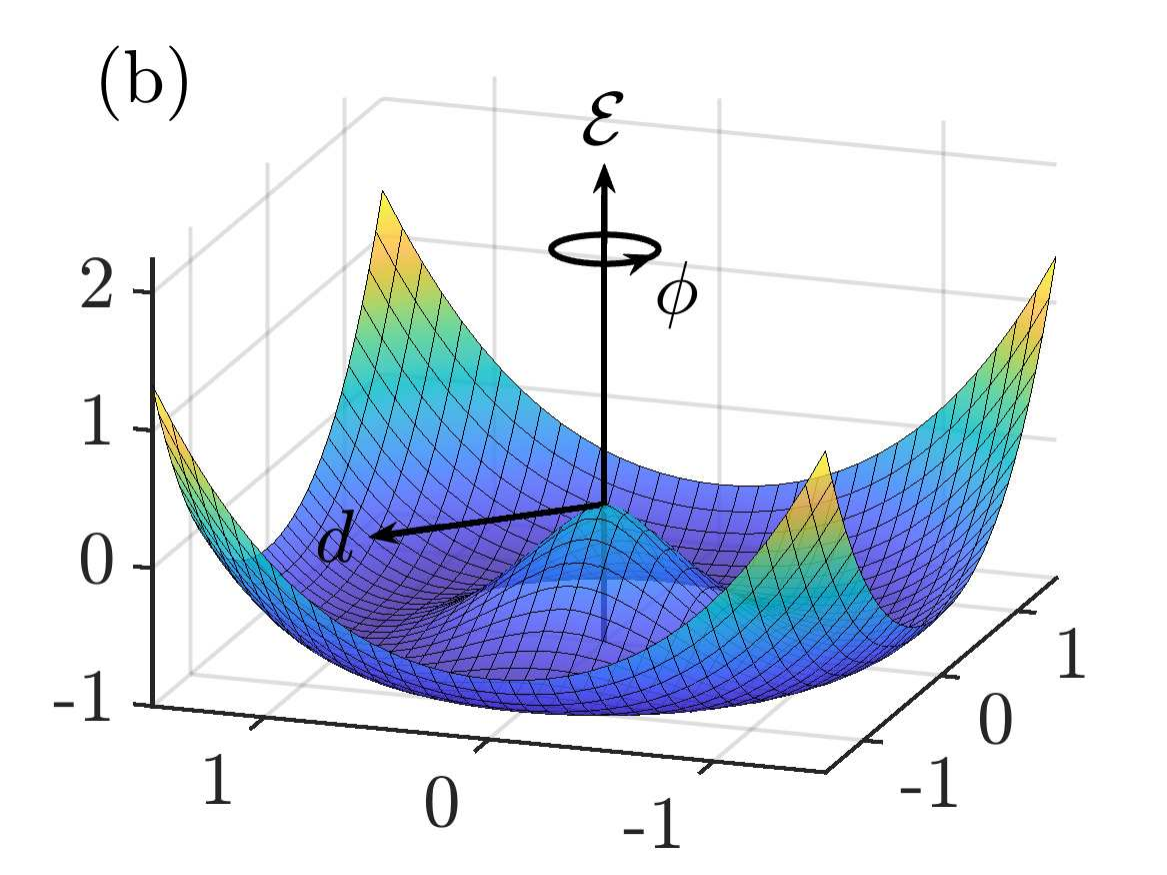}
	\caption{
		(a)
		The scaled energy spectrum $E_A / E_A^{\text{min}}$ with respect to the scaled order parameter $d \equiv \Delta_A / \Delta_A^{\text{min}}$.
		The second-order approximation \cref{eq:Hl_diagElem} (orange dashed) matches closely with the exact solution (blue solid) given in \cref{eqn:exactPHPappendix}, for large values of the dimensionless bandwidth $b$. 
		At $\Delta_A=\Delta_A^{\text{min}}$, we obtain the minimum of the energy spectrum which corresponds to the gap parameter at zero temperature.
		We use the quadratic expansion \cref{eq:Higgs_quadratic} (dotted yellow) to estimate the effective mass of the quasiparticle from the harmonic oscillator model.
		(b)
		The Mexican-hat-shaped potential-well of the superconductor with respect to the values of $\phi$, represented as an azimuthal angle, and $d$ as the radial coordinate.
	}
	\label{fig:energy_0orderCalc}
\end{figure}

The ground-state energy of a superconducting island given in \cref{eq:Hl_diagElem} can be conveniently rescaled relative to its minimum value $E^{\text{min}}=-\mathcal{B} e^{-2/\lambda} n $ and the location of the energetic minimum at $\Delta^{\text{min}}=2\mathcal{B}  e^{-1/\lambda}$.  We define $d \equiv \Delta / \Delta^{\text{min}}$, so the rescaled version of \cref{eq:Hl_diagElem} is
\begin{align}	
	\mathcal{E}
\nonumber
&\equiv 
    {E}/|{E^\text{min}}|,
\\
&\approx
	d^2
	\big(-1+2\ln d-2\lambda (\ln d)^2
	\big),
\nonumber
\\
&\approx
-1+	2 (1-\lambda)(d-1)^2+
	O(d-1)^3
    ,
\label{eq:Higgs_quadratic}
\end{align}
where the last line is a quadratic approximation for the energy around the minimum. 
\Cref{fig:energy_0orderCalc} shows $\mathcal{E}$, which corresponds to the radial coordinate of the  Mexican-hat potential.

For an isolated metallic island, $\mathcal{E}$ is independent of $\phi$, so $\mathcal{E}$ is symmetric in rotations of $\phi$. Plotting $\mathcal{E}$ as a function of the radial coordinate $|\Delta|$ and azimuthal angle $\phi$ yields the Mexican-hat potential, shown in \cref{fig:energy_0orderCalc}b.  In this potential, the low energy dynamics discussed in earlier sections are restricted to the minimum of this potential, and we have hitherto assumed that  $\Delta=\Delta^{\text{min}}$.  

However a real system can also undergo `Higgs'-like excitations of the radial coordinate.  Quantum mechanically there may be excitations of the Higgs mode, and so superconducting devices should more generally be analysed over a 2-dimensional  coordinate space $(\phi,\Delta_r)$, where $\phi$ is the azimuthal angle of the Mexican hat and $\Delta_r$ is the radial coordinates shown in \cref{fig:energy_0orderCalc} (b).
The phase-like mode is massless, while the radial-like gap mode (or Higgs mode) has some effective massive.

We briefly comment on the structure of this larger coordinate space. We suppose that the BCS ground states are now a function of both $\phi$ and $\Delta_r$, so that we would generalise the basis defined in \cref{eq:BCS_state} to a larger space spanned by the set of states $\{\ket{\Psi(\phi_j, \Delta_r)}\}_{j,r}$, where $\Delta_r$ is now part of the parameterisation of the larger basis.

As in \cref{sec:LowEspace}, we should analyse the overlap function $\mathcal{Y}(\phi_j,\phi_{j'};\Delta_r,\Delta_{r'})=\bra{\Psi(\phi_j, \Delta_r)}\Psi(\phi_{j'}, \Delta_{r'})$ in order to understand the effective dimension of this larger space.  As a start in this direction, for $j=j'$, we compute
\begin{align}	\label{eq:overlap_BCS_Δdiff}
	\mathcal{Y}(\phi_j,\phi_{j};\Delta_r,\Delta_{r'})
&=
	\ip{\Psi(\phi_j, \Delta_r)}{\Psi(\phi_j, \Delta_{r'})},\nonumber\\
&\approx
	e^{
		\frac{-3n\pi(\Delta_{r'} - \Delta_r)^2}{64\sqrt{2}b\Delta_r^2}
	},
\end{align}
where the second line is evaluated using a similar procedure as used to compute $\mathcal{W}(\varphi)$, given in \cref{eq:BCSoverlap_approx}. 
$\mathcal{Y}$ is approximately a Gaussian with width $\delta\Delta \sim \sqrt{\mathcal{B}\Delta/n}$, and so we would expect that a  discrete subset of radial modes would describe the space with high fidelity, similar to the emergent discrete set of phase states, described earlier. 
Namely, the number of nearly-orthogonal transverse modes that can be accommodated in the potential well $E_A(\mathcal{B},\Delta)$ for a given phase $\phi$ is roughly \mbox{$\Delta^{\text{min}}/\delta\Delta\sim \sqrt{ne^{-1/\lambda}}$}.  We leave more extensive analysis to future work.

The Higgs mode of a superconductor can be excited resonantly by a driving field with energy at $\Delta^{\text{min}}$, and oscillates at frequency $\omega_\text{Higgs} = 2\Delta^{\text{min}}/\hbar$ \cite{ref:Higgs_dynamics_Floquent}.
By treating the lowest-energy Higgs mode as a harmonic oscillator with the quadratic potential \cref{eq:Higgs_quadratic}, we obtain the effective mass of the oscillator associated with the potential $E(\mathcal{B},\Delta)$:
\begin{equation}
	m^*
\approx
    \frac{\hbar^2
    }{4}
    \cdot
    \frac{1}{\rho^2(E_F)|E^{\text{min}}|}.
\end{equation}
The depth of the Higgs potential in \cref{fig:energy_0orderCalc} can be expressed in terms of $\omega_\text{Higgs}$, and we estimate the number of bound Higgs-like excitations to be
\begin{equation}
    \text{\# bound states}
=
    {\abs{E^{\text{min}}}}/({\hbar\,\omega_\text{Higgs}})
\sim  
    {ne^{-1/\lambda}}.
\end{equation}
There is thus a non-trivial space of bound Higgs excitations that may be accessible in superconducting quantum devices, and these may provide useful or interesting for quantum technologies. We leave this to further work.

\section{Conclusion} \label{sec:conclusion}

We have  described here  a direct approach to the quantisation of superconducting circuits.
By projecting the microscopic Hamiltonian onto the low-energy BCS subspace, we derive the effective Hamiltonian of a superconducting LCJ circuit from first principles.

With this fully-quantum framework, we establish a direct pathway from the microscopic description to the macroscopic picture of a quantised circuit.  The low-energy BCS space has a natural discretisation that emerges from the structure of the BCS overlap function.  We have shown how to construct phase and charge operators on this subspace, and derived their commutation relation.

The projective approach we have described here reproduces the known results for capacitive, inductive and Josephson tunnelling elements.  One conclusion from our analysis is that since inductors are inherently extended objects, the Hilbert space associated to them is larger than that for  point-like capacitive and Josephson tunnelling elements, whose Hilbert space is periodic on the interval $[-\pi ,\pi)$.  We believe that this resolves some of the questions about the compactness or otherwise of superconducting circuits reviewed in \cite{ref:compact-noncompact}.

The approach we take here opens a number of directions for further work.  Firstly, we have given a physically-motivated argument to estimate  the support of the accessible Hilbert space of an inductive element, and we postulate that this will emerge dynamically from decoherence induced by quasi-particle formation at sufficiently high currents.  This could be formalised by extending the projective subspace to include few-quasiparticle states.  These additional modes will produce decoherence relative to the BCS subspace, and should dynamically constrain the accessible  states of extended wire-like inductive elements. 

Secondly, while there has been progress since \citet{Widom1981} in developing the theory of quantum-phase slip (QPS) devices (e.g.\ see \cite{ref:2nlinearSCqubit}), to our knowledge there is as yet no microscopic derivation analogous to that for Josephson junctions.  We expect that a detailed projective analysis of  superconducting wires whose thickness is comparable to or smaller than the penetration depth provides a viable pathway to realise such a derivation.

Lastly, while we have  focussed on the superconducting phase degree of freedom, $\phi$, the BCS family of states admits the possibility of quantising the gap amplitude, $|\Delta|$, which should yield Higgs-like excitations of superconducting islands.  We have provided a starting point for this analysis, and speculate that this may provide for a greatly expanded  family of superconducting quantum devices that could be built experimentally. 

\begin{acknowledgments}
We thank Abhijeet  Alase,  Benjamin Levitan, Ari Mizel,  Akshat Pandey, Barry Sanders,  and Tim Spiller for  helpful comments and discussions.  
TMS and YCL  acknowledge funding from the Australian Research Council Centre of Excellence for Engineered Quantum Systems (Project No.\ CE170100009).
\end{acknowledgments}

\bibliographystyle{apsrev4-2} 
\bibliography{reference.bib}

\begin{thebibliography}{50}%
\makeatletter
\providecommand \@ifxundefined [1]{%
 \@ifx{#1\undefined}
}%
\providecommand \@ifnum [1]{%
 \ifnum #1\expandafter \@firstoftwo
 \else \expandafter \@secondoftwo
 \fi
}%
\providecommand \@ifx [1]{%
 \ifx #1\expandafter \@firstoftwo
 \else \expandafter \@secondoftwo
 \fi
}%
\providecommand \natexlab [1]{#1}%
\providecommand \enquote  [1]{``#1''}%
\providecommand \bibnamefont  [1]{#1}%
\providecommand \bibfnamefont [1]{#1}%
\providecommand \citenamefont [1]{#1}%
\providecommand \href@noop [0]{\@secondoftwo}%
\providecommand \href [0]{\begingroup \@sanitize@url \@href}%
\providecommand \@href[1]{\@@startlink{#1}\@@href}%
\providecommand \@@href[1]{\endgroup#1\@@endlink}%
\providecommand \@sanitize@url [0]{\catcode `\\12\catcode `\$12\catcode
  `\&12\catcode `\#12\catcode `\^12\catcode `\_12\catcode `\%12\relax}%
\providecommand \@@startlink[1]{}%
\providecommand \@@endlink[0]{}%
\providecommand \url  [0]{\begingroup\@sanitize@url \@url }%
\providecommand \@url [1]{\endgroup\@href {#1}{\urlprefix }}%
\providecommand \urlprefix  [0]{URL }%
\providecommand \Eprint [0]{\href }%
\providecommand \doibase [0]{https://doi.org/}%
\providecommand \selectlanguage [0]{\@gobble}%
\providecommand \bibinfo  [0]{\@secondoftwo}%
\providecommand \bibfield  [0]{\@secondoftwo}%
\providecommand \translation [1]{[#1]}%
\providecommand \BibitemOpen [0]{}%
\providecommand \bibitemStop [0]{}%
\providecommand \bibitemNoStop [0]{.\EOS\space}%
\providecommand \EOS [0]{\spacefactor3000\relax}%
\providecommand \BibitemShut  [1]{\csname bibitem#1\endcsname}%
\let\auto@bib@innerbib\@empty
\bibitem [{\citenamefont {Devoret}\ and\ \citenamefont
  {Schoelkopf}(2013)}]{ref:SCcircuit_QI_outlook}%
  \BibitemOpen
  \bibfield  {author} {\bibinfo {author} {\bibfnamefont {M.~H.}\ \bibnamefont
  {Devoret}}\ and\ \bibinfo {author} {\bibfnamefont {R.~J.}\ \bibnamefont
  {Schoelkopf}},\ }\href {https://doi.org/10.1126/science.1231930} {\bibfield
  {journal} {\bibinfo  {journal} {Science}\ }\textbf {\bibinfo {volume}
  {339}},\ \bibinfo {pages} {1169} (\bibinfo {year} {2013})}\BibitemShut
  {NoStop}%
\bibitem [{\citenamefont {Clarke}\ and\ \citenamefont
  {Wilhelm}(2008)}]{ref:SCqubits_wilhelm}%
  \BibitemOpen
  \bibfield  {author} {\bibinfo {author} {\bibfnamefont {J.}~\bibnamefont
  {Clarke}}\ and\ \bibinfo {author} {\bibfnamefont {F.~K.}\ \bibnamefont
  {Wilhelm}},\ }\href {https://doi.org/10.1038/nature07128} {\bibfield
  {journal} {\bibinfo  {journal} {Nature}\ }\textbf {\bibinfo {volume} {453}},\
  \bibinfo {pages} {1031} (\bibinfo {year} {2008})}\BibitemShut {NoStop}%
\bibitem [{\citenamefont {Bocko}\ \emph {et~al.}(1997)\citenamefont {Bocko},
  \citenamefont {Herr},\ and\ \citenamefont {Feldman}}]{ref:SC_coherent}%
  \BibitemOpen
  \bibfield  {author} {\bibinfo {author} {\bibfnamefont {M.}~\bibnamefont
  {Bocko}}, \bibinfo {author} {\bibfnamefont {A.}~\bibnamefont {Herr}},\ and\
  \bibinfo {author} {\bibfnamefont {M.}~\bibnamefont {Feldman}},\ }\href
  {https://doi.org/10.1109/77.622206} {\bibfield  {journal} {\bibinfo
  {journal} {IEEE Transactions on Applied Superconductivity}\ }\textbf
  {\bibinfo {volume} {7}},\ \bibinfo {pages} {3638} (\bibinfo {year}
  {1997})}\BibitemShut {NoStop}%
\bibitem [{\citenamefont {Blais}\ \emph {et~al.}(2004)\citenamefont {Blais},
  \citenamefont {Huang}, \citenamefont {Wallraff}, \citenamefont {Girvin},\
  and\ \citenamefont {Schoelkopf}}]{ref:1Dtransmon_control}%
  \BibitemOpen
  \bibfield  {author} {\bibinfo {author} {\bibfnamefont {A.}~\bibnamefont
  {Blais}}, \bibinfo {author} {\bibfnamefont {R.-S.}\ \bibnamefont {Huang}},
  \bibinfo {author} {\bibfnamefont {A.}~\bibnamefont {Wallraff}}, \bibinfo
  {author} {\bibfnamefont {S.~M.}\ \bibnamefont {Girvin}},\ and\ \bibinfo
  {author} {\bibfnamefont {R.~J.}\ \bibnamefont {Schoelkopf}},\ }\href
  {https://doi.org/10.1103/PhysRevA.69.062320} {\bibfield  {journal} {\bibinfo
  {journal} {Phys. Rev. A}\ }\textbf {\bibinfo {volume} {69}},\ \bibinfo
  {pages} {062320} (\bibinfo {year} {2004})}\BibitemShut {NoStop}%
\bibitem [{\citenamefont {Burkard}\ \emph {et~al.}(2004)\citenamefont
  {Burkard}, \citenamefont {Koch},\ and\ \citenamefont
  {DiVincenzo}}]{ref:decoherence_SCqubits}%
  \BibitemOpen
  \bibfield  {author} {\bibinfo {author} {\bibfnamefont {G.}~\bibnamefont
  {Burkard}}, \bibinfo {author} {\bibfnamefont {R.~H.}\ \bibnamefont {Koch}},\
  and\ \bibinfo {author} {\bibfnamefont {D.~P.}\ \bibnamefont {DiVincenzo}},\
  }\href {https://doi.org/10.1103/PhysRevB.69.064503} {\bibfield  {journal}
  {\bibinfo  {journal} {Phys. Rev. B}\ }\textbf {\bibinfo {volume} {69}},\
  \bibinfo {pages} {064503} (\bibinfo {year} {2004})}\BibitemShut {NoStop}%
\bibitem [{\citenamefont {Osborne}\ \emph {et~al.}(2024)\citenamefont
  {Osborne}, \citenamefont {Larson}, \citenamefont {Jones}, \citenamefont
  {Simmonds}, \citenamefont {Gyenis},\ and\ \citenamefont
  {Lucas}}]{ref:circuit_quantization}%
  \BibitemOpen
  \bibfield  {author} {\bibinfo {author} {\bibfnamefont {A.}~\bibnamefont
  {Osborne}}, \bibinfo {author} {\bibfnamefont {T.}~\bibnamefont {Larson}},
  \bibinfo {author} {\bibfnamefont {S.~G.}\ \bibnamefont {Jones}}, \bibinfo
  {author} {\bibfnamefont {R.~W.}\ \bibnamefont {Simmonds}}, \bibinfo {author}
  {\bibfnamefont {A.}~\bibnamefont {Gyenis}},\ and\ \bibinfo {author}
  {\bibfnamefont {A.}~\bibnamefont {Lucas}},\ }\href
  {https://doi.org/10.1103/PRXQuantum.5.020309} {\bibfield  {journal} {\bibinfo
   {journal} {PRX Quantum}\ }\textbf {\bibinfo {volume} {5}},\ \bibinfo {pages}
  {020309} (\bibinfo {year} {2024})}\BibitemShut {NoStop}%
\bibitem [{\citenamefont {Milburn}(2003)}]{ref:quantumDot_Milburn}%
  \BibitemOpen
  \bibfield  {author} {\bibinfo {author} {\bibfnamefont {G.}~\bibnamefont
  {Milburn}},\ }\href {https://doi.org/10.1088/2058-7058/16/10/33} {\bibfield
  {journal} {\bibinfo  {journal} {Physics World}\ }\textbf {\bibinfo {volume}
  {16}},\ \bibinfo {pages} {24} (\bibinfo {year} {2003})}\BibitemShut {NoStop}%
\bibitem [{\citenamefont {Yamamoto}\ and\ \citenamefont
  {Imamoglu}(1999)}]{ref:mesoscopic_QO}%
  \BibitemOpen
  \bibfield  {author} {\bibinfo {author} {\bibfnamefont {Y.}~\bibnamefont
  {Yamamoto}}\ and\ \bibinfo {author} {\bibfnamefont {A.}~\bibnamefont
  {Imamoglu}},\ }\href
  {https://www.wiley.com/en-us/Mesoscopic+Quantum+Optics-p-9780471148746}
  {\emph {\bibinfo {title} {Mesoscopic Quantum Optics}}}\ (\bibinfo
  {publisher} {John Wiley and Sons},\ \bibinfo {year} {1999})\BibitemShut
  {NoStop}%
\bibitem [{\citenamefont {Stace}\ \emph {et~al.}(2004)\citenamefont {Stace},
  \citenamefont {Barnes},\ and\ \citenamefont
  {Milburn}}]{ref:mesoscopicChannel_QHE}%
  \BibitemOpen
  \bibfield  {author} {\bibinfo {author} {\bibfnamefont {T.~M.}\ \bibnamefont
  {Stace}}, \bibinfo {author} {\bibfnamefont {C.~H.~W.}\ \bibnamefont
  {Barnes}},\ and\ \bibinfo {author} {\bibfnamefont {G.~J.}\ \bibnamefont
  {Milburn}},\ }\href {https://doi.org/10.1103/PhysRevLett.93.126804}
  {\bibfield  {journal} {\bibinfo  {journal} {Phys. Rev. Lett.}\ }\textbf
  {\bibinfo {volume} {93}},\ \bibinfo {pages} {126804} (\bibinfo {year}
  {2004})}\BibitemShut {NoStop}%
\bibitem [{\citenamefont {Kuhn}\ \emph {et~al.}(2024)\citenamefont {Kuhn},
  \citenamefont {Sothmann},\ and\ \citenamefont
  {Cayao}}]{ref:Higgs_dynamics_Floquent}%
  \BibitemOpen
  \bibfield  {author} {\bibinfo {author} {\bibfnamefont {T.}~\bibnamefont
  {Kuhn}}, \bibinfo {author} {\bibfnamefont {B.}~\bibnamefont {Sothmann}},\
  and\ \bibinfo {author} {\bibfnamefont {J.}~\bibnamefont {Cayao}},\ }\href
  {https://doi.org/10.1103/PhysRevB.109.134517} {\bibfield  {journal} {\bibinfo
   {journal} {Phys. Rev. B}\ }\textbf {\bibinfo {volume} {109}},\ \bibinfo
  {pages} {134517} (\bibinfo {year} {2024})}\BibitemShut {NoStop}%
\bibitem [{\citenamefont {Glazman}\ and\ \citenamefont
  {Catelani}(2021)}]{ref:quasiparticles_SCqubits}%
  \BibitemOpen
  \bibfield  {author} {\bibinfo {author} {\bibfnamefont {L.~I.}\ \bibnamefont
  {Glazman}}\ and\ \bibinfo {author} {\bibfnamefont {G.}~\bibnamefont
  {Catelani}},\ }\href {https://doi.org/10.21468/SciPostPhysLectNotes.31}
  {\bibfield  {journal} {\bibinfo  {journal} {SciPost Phys. Lect. Notes}\ ,\
  \bibinfo {pages} {31}} (\bibinfo {year} {2021})}\BibitemShut {NoStop}%
\bibitem [{\citenamefont {Widom}(1979)}]{widom1979}%
  \BibitemOpen
  \bibfield  {author} {\bibinfo {author} {\bibfnamefont {A.}~\bibnamefont
  {Widom}},\ }\href {https://doi.org/10.1007/BF00119200} {\bibfield  {journal}
  {\bibinfo  {journal} {Journal of Low Temperature Physics}\ }\textbf {\bibinfo
  {volume} {37}},\ \bibinfo {pages} {449} (\bibinfo {year} {1979})}\BibitemShut
  {NoStop}%
\bibitem [{\citenamefont {Widom}\ \emph {et~al.}(1981)\citenamefont {Widom},
  \citenamefont {Megaloudis}, \citenamefont {Sacco},\ and\ \citenamefont
  {Clark}}]{Widom1981}%
  \BibitemOpen
  \bibfield  {author} {\bibinfo {author} {\bibfnamefont {A.}~\bibnamefont
  {Widom}}, \bibinfo {author} {\bibfnamefont {G.}~\bibnamefont {Megaloudis}},
  \bibinfo {author} {\bibfnamefont {J.~E.}\ \bibnamefont {Sacco}},\ and\
  \bibinfo {author} {\bibfnamefont {T.~D.}\ \bibnamefont {Clark}},\ }\href
  {https://doi.org/10.1007/BF02721707} {\bibfield  {journal} {\bibinfo
  {journal} {Il Nuovo Cimento B (1971-1996)}\ }\textbf {\bibinfo {volume}
  {61}},\ \bibinfo {pages} {112} (\bibinfo {year} {1981})}\BibitemShut
  {NoStop}%
\bibitem [{\citenamefont {Vool}\ and\ \citenamefont
  {Devoret}(2017)}]{https://doi.org/10.1002/cta.2359}%
  \BibitemOpen
  \bibfield  {author} {\bibinfo {author} {\bibfnamefont {U.}~\bibnamefont
  {Vool}}\ and\ \bibinfo {author} {\bibfnamefont {M.}~\bibnamefont {Devoret}},\
  }\href {https://doi.org/https://doi.org/10.1002/cta.2359} {\bibfield
  {journal} {\bibinfo  {journal} {International Journal of Circuit Theory and
  Applications}\ }\textbf {\bibinfo {volume} {45}},\ \bibinfo {pages} {897}
  (\bibinfo {year} {2017})},\ \Eprint
  {https://arxiv.org/abs/https://onlinelibrary.wiley.com/doi/pdf/10.1002/cta.2359}
  {https://onlinelibrary.wiley.com/doi/pdf/10.1002/cta.2359} \BibitemShut
  {NoStop}%
\bibitem [{\citenamefont {Pandey}\ and\ \citenamefont
  {Ghosh}(2024)}]{Pandey_2024}%
  \BibitemOpen
  \bibfield  {author} {\bibinfo {author} {\bibfnamefont {A.}~\bibnamefont
  {Pandey}}\ and\ \bibinfo {author} {\bibfnamefont {S.}~\bibnamefont {Ghosh}},\
  }\href {https://doi.org/10.1088/1402-4896/ad8842} {\bibfield  {journal}
  {\bibinfo  {journal} {Physica Scripta}\ }\textbf {\bibinfo {volume} {99}},\
  \bibinfo {pages} {125106} (\bibinfo {year} {2024})}\BibitemShut {NoStop}%
\bibitem [{\citenamefont {Mizel}(2024)}]{Mizel2024}%
  \BibitemOpen
  \bibfield  {author} {\bibinfo {author} {\bibfnamefont {A.}~\bibnamefont
  {Mizel}},\ }\href {https://doi.org/10.1103/PhysRevApplied.21.024030}
  {\bibfield  {journal} {\bibinfo  {journal} {Phys. Rev. Appl.}\ }\textbf
  {\bibinfo {volume} {21}},\ \bibinfo {pages} {024030} (\bibinfo {year}
  {2024})}\BibitemShut {NoStop}%
\bibitem [{\citenamefont {Thanh~Le}\ \emph {et~al.}(2020)\citenamefont
  {Thanh~Le}, \citenamefont {Cole},\ and\ \citenamefont
  {Stace}}]{ref:compact-noncompact}%
  \BibitemOpen
  \bibfield  {author} {\bibinfo {author} {\bibfnamefont {D.}~\bibnamefont
  {Thanh~Le}}, \bibinfo {author} {\bibfnamefont {J.~H.}\ \bibnamefont {Cole}},\
  and\ \bibinfo {author} {\bibfnamefont {T.~M.}\ \bibnamefont {Stace}},\ }\href
  {https://doi.org/10.1103/PhysRevResearch.2.013245} {\bibfield  {journal}
  {\bibinfo  {journal} {Phys. Rev. Res.}\ }\textbf {\bibinfo {volume} {2}},\
  \bibinfo {pages} {013245} (\bibinfo {year} {2020})}\BibitemShut {NoStop}%
\bibitem [{\citenamefont {Leggett}(1966)}]{Leggett1966}%
  \BibitemOpen
  \bibfield  {author} {\bibinfo {author} {\bibfnamefont {A.~J.}\ \bibnamefont
  {Leggett}},\ }\href {https://doi.org/10.1143/PTP.36.901} {\bibfield
  {journal} {\bibinfo  {journal} {Progress of Theoretical Physics}\ }\textbf
  {\bibinfo {volume} {36}},\ \bibinfo {pages} {901} (\bibinfo {year} {1966})},\
  \Eprint
  {https://arxiv.org/abs/https://academic.oup.com/ptp/article-pdf/36/5/901/5256693/36-5-901.pdf}
  {https://academic.oup.com/ptp/article-pdf/36/5/901/5256693/36-5-901.pdf}
  \BibitemShut {NoStop}%
\bibitem [{\citenamefont {Ambegaokar}\ \emph {et~al.}(1982)\citenamefont
  {Ambegaokar}, \citenamefont {Eckern},\ and\ \citenamefont
  {Sch\"on}}]{Ambegaokar1982}%
  \BibitemOpen
  \bibfield  {author} {\bibinfo {author} {\bibfnamefont {V.}~\bibnamefont
  {Ambegaokar}}, \bibinfo {author} {\bibfnamefont {U.}~\bibnamefont {Eckern}},\
  and\ \bibinfo {author} {\bibfnamefont {G.}~\bibnamefont {Sch\"on}},\ }\href
  {https://doi.org/10.1103/PhysRevLett.48.1745} {\bibfield  {journal} {\bibinfo
   {journal} {Phys. Rev. Lett.}\ }\textbf {\bibinfo {volume} {48}},\ \bibinfo
  {pages} {1745} (\bibinfo {year} {1982})}\BibitemShut {NoStop}%
\bibitem [{\citenamefont {Bardeen}\ \emph {et~al.}(1957)\citenamefont
  {Bardeen}, \citenamefont {Cooper},\ and\ \citenamefont
  {Schrieffer}}]{ref:BCS_superconductivity_theory}%
  \BibitemOpen
  \bibfield  {author} {\bibinfo {author} {\bibfnamefont {J.}~\bibnamefont
  {Bardeen}}, \bibinfo {author} {\bibfnamefont {L.~N.}\ \bibnamefont
  {Cooper}},\ and\ \bibinfo {author} {\bibfnamefont {J.~R.}\ \bibnamefont
  {Schrieffer}},\ }\href {https://doi.org/10.1103/PhysRev.108.1175} {\bibfield
  {journal} {\bibinfo  {journal} {Phys. Rev.}\ }\textbf {\bibinfo {volume}
  {108}},\ \bibinfo {pages} {1175} (\bibinfo {year} {1957})}\BibitemShut
  {NoStop}%
\bibitem [{\citenamefont {Bouchiat}\ \emph {et~al.}(1998)\citenamefont
  {Bouchiat}, \citenamefont {Vion}, \citenamefont {Joyez}, \citenamefont
  {Esteve},\ and\ \citenamefont {Devoret}}]{ref:CPB}%
  \BibitemOpen
  \bibfield  {author} {\bibinfo {author} {\bibfnamefont {V.}~\bibnamefont
  {Bouchiat}}, \bibinfo {author} {\bibfnamefont {D.}~\bibnamefont {Vion}},
  \bibinfo {author} {\bibfnamefont {P.}~\bibnamefont {Joyez}}, \bibinfo
  {author} {\bibfnamefont {D.}~\bibnamefont {Esteve}},\ and\ \bibinfo {author}
  {\bibfnamefont {M.~H.}\ \bibnamefont {Devoret}},\ }\href
  {https://doi.org/10.1238/Physica.Topical.076a00165} {\bibfield  {journal}
  {\bibinfo  {journal} {Physica Scripta}\ }\textbf {\bibinfo {volume} {1998}},\
  \bibinfo {pages} {165} (\bibinfo {year} {1998})}\BibitemShut {NoStop}%
\bibitem [{\citenamefont {Annett}(2004)}]{ref:annett}%
  \BibitemOpen
  \bibfield  {author} {\bibinfo {author} {\bibfnamefont {J.}~\bibnamefont
  {Annett}},\ }\href {https://books.google.com.au/books?id=-wiHrgEACAAJ} {\emph
  {\bibinfo {title} {Superconductivity, Superfluids and Condensates}}},\ Oxford
  Master Series in Physics\ (\bibinfo  {publisher} {OUP Oxford},\ \bibinfo
  {year} {2004})\BibitemShut {NoStop}%
\bibitem [{\citenamefont {Coleman}(2015)}]{ref:Coleman}%
  \BibitemOpen
  \bibfield  {author} {\bibinfo {author} {\bibfnamefont {P.}~\bibnamefont
  {Coleman}},\ }\href {https://doi.org/10.1017/CBO9781139020916} {\emph
  {\bibinfo {title} {Introduction to Many-Body Physics}}}\ (\bibinfo
  {publisher} {Cambridge University Press},\ \bibinfo {year}
  {2015})\BibitemShut {NoStop}%
\bibitem [{\citenamefont {Pegg}\ and\ \citenamefont
  {Barnett}(1989)}]{ref:Pegg&Barnett}%
  \BibitemOpen
  \bibfield  {author} {\bibinfo {author} {\bibfnamefont {D.~T.}\ \bibnamefont
  {Pegg}}\ and\ \bibinfo {author} {\bibfnamefont {S.~M.}\ \bibnamefont
  {Barnett}},\ }\href {https://doi.org/10.1103/PhysRevA.39.1665} {\bibfield
  {journal} {\bibinfo  {journal} {Phys. Rev. A}\ }\textbf {\bibinfo {volume}
  {39}},\ \bibinfo {pages} {1665} (\bibinfo {year} {1989})}\BibitemShut
  {NoStop}%
\bibitem [{\citenamefont {Mila}\ and\ \citenamefont {Schmidt}(2011)}]{ref:POM}%
  \BibitemOpen
  \bibfield  {author} {\bibinfo {author} {\bibfnamefont {F.}~\bibnamefont
  {Mila}}\ and\ \bibinfo {author} {\bibfnamefont {K.~P.}\ \bibnamefont
  {Schmidt}},\ }\bibinfo {title} {Strong-coupling expansion and effective
  hamiltonians},\ in\ \href {https://doi.org/10.1007/978-3-642-10589-0_20}
  {\emph {\bibinfo {booktitle} {Introduction to Frustrated Magnetism:
  Materials, Experiments, Theory}}},\ \bibinfo {editor} {edited by\ \bibinfo
  {editor} {\bibfnamefont {C.}~\bibnamefont {Lacroix}}, \bibinfo {editor}
  {\bibfnamefont {P.}~\bibnamefont {Mendels}},\ and\ \bibinfo {editor}
  {\bibfnamefont {F.}~\bibnamefont {Mila}}}\ (\bibinfo  {publisher} {Springer
  Berlin Heidelberg},\ \bibinfo {address} {Berlin, Heidelberg},\ \bibinfo
  {year} {2011})\ pp.\ \bibinfo {pages} {537--559}\BibitemShut {NoStop}%
\bibitem [{\citenamefont {Onishi}\ and\ \citenamefont
  {Yoshida}(1966)}]{ONISHI1966367}%
  \BibitemOpen
  \bibfield  {author} {\bibinfo {author} {\bibfnamefont {N.}~\bibnamefont
  {Onishi}}\ and\ \bibinfo {author} {\bibfnamefont {S.}~\bibnamefont
  {Yoshida}},\ }\href
  {https://doi.org/https://doi.org/10.1016/0029-5582(66)90096-4} {\bibfield
  {journal} {\bibinfo  {journal} {Nuclear Physics}\ }\textbf {\bibinfo {volume}
  {80}},\ \bibinfo {pages} {367} (\bibinfo {year} {1966})}\BibitemShut
  {NoStop}%
\bibitem [{\citenamefont {Gunnarsson}(1997)}]{ref:SC_bandwidth}%
  \BibitemOpen
  \bibfield  {author} {\bibinfo {author} {\bibfnamefont {O.}~\bibnamefont
  {Gunnarsson}},\ }\href {https://doi.org/10.1103/RevModPhys.69.575} {\bibfield
   {journal} {\bibinfo  {journal} {Rev. Mod. Phys.}\ }\textbf {\bibinfo
  {volume} {69}},\ \bibinfo {pages} {575} (\bibinfo {year} {1997})}\BibitemShut
  {NoStop}%
\bibitem [{\citenamefont {Poto{\v{c}}nik}\ \emph {et~al.}(2014)\citenamefont
  {Poto{\v{c}}nik}, \citenamefont {Krajnc}, \citenamefont {Jegli{\v{c}}},
  \citenamefont {Takabayashi}, \citenamefont {Ganin}, \citenamefont
  {Prassides}, \citenamefont {Rosseinsky},\ and\ \citenamefont
  {Ar{\v{c}}on}}]{ref:SC_gap_Cs3C60}%
  \BibitemOpen
  \bibfield  {author} {\bibinfo {author} {\bibfnamefont {A.}~\bibnamefont
  {Poto{\v{c}}nik}}, \bibinfo {author} {\bibfnamefont {A.}~\bibnamefont
  {Krajnc}}, \bibinfo {author} {\bibfnamefont {P.}~\bibnamefont
  {Jegli{\v{c}}}}, \bibinfo {author} {\bibfnamefont {Y.}~\bibnamefont
  {Takabayashi}}, \bibinfo {author} {\bibfnamefont {A.~Y.}\ \bibnamefont
  {Ganin}}, \bibinfo {author} {\bibfnamefont {K.}~\bibnamefont {Prassides}},
  \bibinfo {author} {\bibfnamefont {M.~J.}\ \bibnamefont {Rosseinsky}},\ and\
  \bibinfo {author} {\bibfnamefont {D.}~\bibnamefont {Ar{\v{c}}on}},\ }\href
  {https://doi.org/10.1038/srep04265} {\bibfield  {journal} {\bibinfo
  {journal} {Scientific Reports}\ }\textbf {\bibinfo {volume} {4}},\ \bibinfo
  {pages} {4265} (\bibinfo {year} {2014})}\BibitemShut {NoStop}%
\bibitem [{\citenamefont {Mahan}(1981)}]{ref:mahan}%
  \BibitemOpen
  \bibfield  {author} {\bibinfo {author} {\bibfnamefont {G.~D.}\ \bibnamefont
  {Mahan}},\ }\href {https://nla.gov.au/nla.cat-vn1437043} {\emph {\bibinfo
  {title} {Many-particle physics / Gerald D. Mahan}}}\ (\bibinfo  {publisher}
  {Plenum Press New York},\ \bibinfo {year} {1981})\BibitemShut {NoStop}%
\bibitem [{\citenamefont {Korsbakken}\ \emph {et~al.}(2009)\citenamefont
  {Korsbakken}, \citenamefont {Wilhelm},\ and\ \citenamefont
  {Whaley}}]{Korsbakken_2009}%
  \BibitemOpen
  \bibfield  {author} {\bibinfo {author} {\bibfnamefont {J.~I.}\ \bibnamefont
  {Korsbakken}}, \bibinfo {author} {\bibfnamefont {F.~K.}\ \bibnamefont
  {Wilhelm}},\ and\ \bibinfo {author} {\bibfnamefont {K.~B.}\ \bibnamefont
  {Whaley}},\ }\href {https://doi.org/10.1088/0031-8949/2009/T137/014022}
  {\bibfield  {journal} {\bibinfo  {journal} {Physica Scripta}\ }\textbf
  {\bibinfo {volume} {2009}},\ \bibinfo {pages} {014022} (\bibinfo {year}
  {2009})}\BibitemShut {NoStop}%
\bibitem [{\citenamefont {Bruus}\ \emph {et~al.}(2004)\citenamefont {Bruus},
  \citenamefont {Flensberg},\ and\ \citenamefont {Flensberg}}]{ref:bruus}%
  \BibitemOpen
  \bibfield  {author} {\bibinfo {author} {\bibfnamefont {H.}~\bibnamefont
  {Bruus}}, \bibinfo {author} {\bibfnamefont {K.}~\bibnamefont {Flensberg}},\
  and\ \bibinfo {author} {\bibfnamefont {{\O}.}~\bibnamefont {Flensberg}},\
  }\href {https://books.google.com.au/books?id=v5vhg1tYLC8C} {\emph {\bibinfo
  {title} {Many-Body Quantum Theory in Condensed Matter Physics: An
  Introduction}}},\ Oxford Graduate Texts\ (\bibinfo  {publisher} {OUP
  Oxford},\ \bibinfo {year} {2004})\BibitemShut {NoStop}%
\bibitem [{\citenamefont {Smith}\ \emph {et~al.}(2020)\citenamefont {Smith},
  \citenamefont {Kou}, \citenamefont {Xiao}, \citenamefont {Vool},\ and\
  \citenamefont {Devoret}}]{ref:2CPB_circuit}%
  \BibitemOpen
  \bibfield  {author} {\bibinfo {author} {\bibfnamefont {W.~C.}\ \bibnamefont
  {Smith}}, \bibinfo {author} {\bibfnamefont {A.}~\bibnamefont {Kou}}, \bibinfo
  {author} {\bibfnamefont {X.}~\bibnamefont {Xiao}}, \bibinfo {author}
  {\bibfnamefont {U.}~\bibnamefont {Vool}},\ and\ \bibinfo {author}
  {\bibfnamefont {M.~H.}\ \bibnamefont {Devoret}},\ }\href
  {https://doi.org/10.1038/s41534-019-0231-2} {\bibfield  {journal} {\bibinfo
  {journal} {npj Quantum Information}\ }\textbf {\bibinfo {volume} {6}},\
  \bibinfo {pages} {8} (\bibinfo {year} {2020})}\BibitemShut {NoStop}%
\bibitem [{\citenamefont {Jackson}(1999)}]{ref:Jackson_electrodynamics}%
  \BibitemOpen
  \bibfield  {author} {\bibinfo {author} {\bibfnamefont {J.~D.}\ \bibnamefont
  {Jackson}},\ }\href {http://cdsweb.cern.ch/record/490457} {\emph {\bibinfo
  {title} {Classical electrodynamics}}},\ \bibinfo {edition} {3rd}\ ed.\
  (\bibinfo  {publisher} {Wiley},\ \bibinfo {address} {New York, {NY}},\
  \bibinfo {year} {1999})\BibitemShut {NoStop}%
\bibitem [{\citenamefont {Tinkham}(1996)}]{ref:Tinkham_superconductivity}%
  \BibitemOpen
  \bibfield  {author} {\bibinfo {author} {\bibfnamefont {M.}~\bibnamefont
  {Tinkham}},\ }\href
  {https://app.knovel.com/hotlink/toc/id:kpISE00023/introduction-superconductivity/introduction-superconductivity}
  {\emph {\bibinfo {title} {Introduction to Superconductivity (2nd Edition)}}}\
  (\bibinfo  {publisher} {Dover Publications},\ \bibinfo {year}
  {1996})\BibitemShut {NoStop}%
\bibitem [{\citenamefont {Ketterson}\ and\ \citenamefont
  {Song}(1999)}]{ref:Kertterson}%
  \BibitemOpen
  \bibfield  {author} {\bibinfo {author} {\bibfnamefont {J.~B.}\ \bibnamefont
  {Ketterson}}\ and\ \bibinfo {author} {\bibfnamefont {S.~N.}\ \bibnamefont
  {Song}},\ }\href@noop {} {\emph {\bibinfo {title} {Superconductivity}}}\
  (\bibinfo  {publisher} {Cambridge University Press},\ \bibinfo {year}
  {1999})\BibitemShut {NoStop}%
\bibitem [{\citenamefont {Peruzzo}\ \emph {et~al.}(2021)\citenamefont
  {Peruzzo}, \citenamefont {Hassani}, \citenamefont {Szep}, \citenamefont
  {Trioni}, \citenamefont {Redchenko}, \citenamefont {\ifmmode \check{Z}\else
  \v{Z}\fi{}emli\ifmmode~\check{c}\else \v{c}\fi{}ka},\ and\ \citenamefont
  {Fink}}]{ref:geometricSuperinductanceQ}%
  \BibitemOpen
  \bibfield  {author} {\bibinfo {author} {\bibfnamefont {M.}~\bibnamefont
  {Peruzzo}}, \bibinfo {author} {\bibfnamefont {F.}~\bibnamefont {Hassani}},
  \bibinfo {author} {\bibfnamefont {G.}~\bibnamefont {Szep}}, \bibinfo {author}
  {\bibfnamefont {A.}~\bibnamefont {Trioni}}, \bibinfo {author} {\bibfnamefont
  {E.}~\bibnamefont {Redchenko}}, \bibinfo {author} {\bibfnamefont
  {M.}~\bibnamefont {\ifmmode \check{Z}\else
  \v{Z}\fi{}emli\ifmmode~\check{c}\else \v{c}\fi{}ka}},\ and\ \bibinfo {author}
  {\bibfnamefont {J.~M.}\ \bibnamefont {Fink}},\ }\href
  {https://doi.org/10.1103/PRXQuantum.2.040341} {\bibfield  {journal} {\bibinfo
   {journal} {PRX Quantum}\ }\textbf {\bibinfo {volume} {2}},\ \bibinfo {pages}
  {040341} (\bibinfo {year} {2021})}\BibitemShut {NoStop}%
\bibitem [{\citenamefont {Douglass}\ and\ \citenamefont
  {Meservey}(1964)}]{ref:Al_gap}%
  \BibitemOpen
  \bibfield  {author} {\bibinfo {author} {\bibfnamefont {D.~H.}\ \bibnamefont
  {Douglass}}\ and\ \bibinfo {author} {\bibfnamefont {R.}~\bibnamefont
  {Meservey}},\ }\href {https://doi.org/10.1103/PhysRev.135.A19} {\bibfield
  {journal} {\bibinfo  {journal} {Phys. Rev.}\ }\textbf {\bibinfo {volume}
  {135}},\ \bibinfo {pages} {A19} (\bibinfo {year} {1964})}\BibitemShut
  {NoStop}%
\bibitem [{\citenamefont {Levinson}\ \emph {et~al.}(1983)\citenamefont
  {Levinson}, \citenamefont {Greuter},\ and\ \citenamefont
  {Plummer}}]{ref:Al_band}%
  \BibitemOpen
  \bibfield  {author} {\bibinfo {author} {\bibfnamefont {H.~J.}\ \bibnamefont
  {Levinson}}, \bibinfo {author} {\bibfnamefont {F.}~\bibnamefont {Greuter}},\
  and\ \bibinfo {author} {\bibfnamefont {E.~W.}\ \bibnamefont {Plummer}},\
  }\href {https://doi.org/10.1103/PhysRevB.27.727} {\bibfield  {journal}
  {\bibinfo  {journal} {Phys. Rev. B}\ }\textbf {\bibinfo {volume} {27}},\
  \bibinfo {pages} {727} (\bibinfo {year} {1983})}\BibitemShut {NoStop}%
\bibitem [{\citenamefont {Corlevi}(2006)}]{ref:exp_SJ}%
  \BibitemOpen
  \bibfield  {author} {\bibinfo {author} {\bibfnamefont {S.}~\bibnamefont
  {Corlevi}},\ }\emph {\bibinfo {title} {Quantum effects in nanoscale Josephson
  junction circuits}},\ \href
  {https://api.semanticscholar.org/CorpusID:122570576} {Ph.D. thesis} (\bibinfo
  {year} {2006})\BibitemShut {NoStop}%
\bibitem [{\citenamefont {Ashcroft}\ and\ \citenamefont
  {Mermin}(1976)}]{ref:Ashcroft&Mermin}%
  \BibitemOpen
  \bibfield  {author} {\bibinfo {author} {\bibfnamefont {N.}~\bibnamefont
  {Ashcroft}}\ and\ \bibinfo {author} {\bibfnamefont {N.}~\bibnamefont
  {Mermin}},\ }\href@noop {} {\emph {\bibinfo {title} {Solid State Physics}}}\
  (\bibinfo  {publisher} {Saunders College Publishing},\ \bibinfo {address}
  {Fort Worth},\ \bibinfo {year} {1976})\BibitemShut {NoStop}%
\bibitem [{\citenamefont {Sakurai}(1993)}]{ref:Sakurai}%
  \BibitemOpen
  \bibfield  {author} {\bibinfo {author} {\bibfnamefont {J.~J.}\ \bibnamefont
  {Sakurai}},\ }\href {http://www.worldcat.org/isbn/0201539292} {\emph
  {\bibinfo {title} {Modern Quantum Mechanics (Revised Edition)}}},\ \bibinfo
  {edition} {1st}\ ed.\ (\bibinfo  {publisher} {Addison Wesley},\ \bibinfo
  {year} {1993})\BibitemShut {NoStop}%
\bibitem [{\citenamefont {Aref}\ \emph {et~al.}(2014)\citenamefont {Aref},
  \citenamefont {Averin}, \citenamefont {van Dijken}, \citenamefont {Ferring},
  \citenamefont {Koberidze}, \citenamefont {Maisi}, \citenamefont {Nguyend},
  \citenamefont {Nieminen}, \citenamefont {Pekola},\ and\ \citenamefont
  {Yao}}]{ref:Al-I-Al_barrier2014}%
  \BibitemOpen
  \bibfield  {author} {\bibinfo {author} {\bibfnamefont {T.}~\bibnamefont
  {Aref}}, \bibinfo {author} {\bibfnamefont {A.}~\bibnamefont {Averin}},
  \bibinfo {author} {\bibfnamefont {S.}~\bibnamefont {van Dijken}}, \bibinfo
  {author} {\bibfnamefont {A.}~\bibnamefont {Ferring}}, \bibinfo {author}
  {\bibfnamefont {M.}~\bibnamefont {Koberidze}}, \bibinfo {author}
  {\bibfnamefont {V.~F.}\ \bibnamefont {Maisi}}, \bibinfo {author}
  {\bibfnamefont {H.~Q.}\ \bibnamefont {Nguyend}}, \bibinfo {author}
  {\bibfnamefont {R.~M.}\ \bibnamefont {Nieminen}}, \bibinfo {author}
  {\bibfnamefont {J.~P.}\ \bibnamefont {Pekola}},\ and\ \bibinfo {author}
  {\bibfnamefont {L.~D.}\ \bibnamefont {Yao}},\ }\href
  {https://doi.org/10.1063/1.4893473} {\bibfield  {journal} {\bibinfo
  {journal} {Journal of Applied Physics}\ }\textbf {\bibinfo {volume} {116}},\
  \bibinfo {pages} {073702} (\bibinfo {year} {2014})}\BibitemShut {NoStop}%
\bibitem [{\citenamefont {Goodman}(2003)}]{ref:Al-I-Al_barrier}%
  \BibitemOpen
  \bibfield  {author} {\bibinfo {author} {\bibfnamefont {A.~M.}\ \bibnamefont
  {Goodman}},\ }\href {https://doi.org/10.1063/1.1659185} {\bibfield  {journal}
  {\bibinfo  {journal} {Journal of Applied Physics}\ }\textbf {\bibinfo
  {volume} {41}},\ \bibinfo {pages} {2176} (\bibinfo {year}
  {2003})}\BibitemShut {NoStop}%
\bibitem [{\citenamefont {Kulchin}\ \emph {et~al.}(2014)\citenamefont
  {Kulchin}, \citenamefont {Dzyuba},\ and\ \citenamefont
  {Amosov}}]{ref:Al2O3_effeciveMass}%
  \BibitemOpen
  \bibfield  {author} {\bibinfo {author} {\bibfnamefont {Y.}~\bibnamefont
  {Kulchin}}, \bibinfo {author} {\bibfnamefont {V.}~\bibnamefont {Dzyuba}},\
  and\ \bibinfo {author} {\bibfnamefont {A.}~\bibnamefont {Amosov}},\ }\href
  {https://doi.org/https://doi.org/10.1016/j.pscr.2015.02.003} {\bibfield
  {journal} {\bibinfo  {journal} {Pacific Science Review}\ }\textbf {\bibinfo
  {volume} {16}},\ \bibinfo {pages} {170} (\bibinfo {year} {2014})}\BibitemShut
  {NoStop}%
\bibitem [{\citenamefont {Yang}\ and\ \citenamefont
  {Agterberg}(2000)}]{ref:JosephsonEffect_externalFields}%
  \BibitemOpen
  \bibfield  {author} {\bibinfo {author} {\bibfnamefont {K.}~\bibnamefont
  {Yang}}\ and\ \bibinfo {author} {\bibfnamefont {D.~F.}\ \bibnamefont
  {Agterberg}},\ }\href {https://doi.org/10.1103/PhysRevLett.84.4970}
  {\bibfield  {journal} {\bibinfo  {journal} {Phys. Rev. Lett.}\ }\textbf
  {\bibinfo {volume} {84}},\ \bibinfo {pages} {4970} (\bibinfo {year}
  {2000})}\BibitemShut {NoStop}%
\bibitem [{\citenamefont {Riwar}\ and\ \citenamefont
  {DiVincenzo}(2022)}]{ref:circuitQuantisation_B_timeDep}%
  \BibitemOpen
  \bibfield  {author} {\bibinfo {author} {\bibfnamefont {R.-P.}\ \bibnamefont
  {Riwar}}\ and\ \bibinfo {author} {\bibfnamefont {D.~P.}\ \bibnamefont
  {DiVincenzo}},\ }\href {https://doi.org/10.1038/s41534-022-00539-x}
  {\bibfield  {journal} {\bibinfo  {journal} {npj Quantum Information}\
  }\textbf {\bibinfo {volume} {8}},\ \bibinfo {pages} {36} (\bibinfo {year}
  {2022})}\BibitemShut {NoStop}%
\bibitem [{\citenamefont {Le}\ \emph {et~al.}(2019)\citenamefont {Le},
  \citenamefont {Grimsmo}, \citenamefont {M\"uller},\ and\ \citenamefont
  {Stace}}]{ref:2nlinearSCqubit}%
  \BibitemOpen
  \bibfield  {author} {\bibinfo {author} {\bibfnamefont {D.~T.}\ \bibnamefont
  {Le}}, \bibinfo {author} {\bibfnamefont {A.}~\bibnamefont {Grimsmo}},
  \bibinfo {author} {\bibfnamefont {C.}~\bibnamefont {M\"uller}},\ and\
  \bibinfo {author} {\bibfnamefont {T.~M.}\ \bibnamefont {Stace}},\ }\href
  {https://doi.org/10.1103/PhysRevA.100.062321} {\bibfield  {journal} {\bibinfo
   {journal} {Phys. Rev. A}\ }\textbf {\bibinfo {volume} {100}},\ \bibinfo
  {pages} {062321} (\bibinfo {year} {2019})}\BibitemShut {NoStop}%
\bibitem [{\citenamefont {Flamant}(2016)}]{ref:Prob5}%
  \BibitemOpen
  \bibfield  {author} {\bibinfo {author} {\bibfnamefont {C.}~\bibnamefont
  {Flamant}},\ }\href
  {https://canvas.harvard.edu/files/3287579/download?download_frd=1&verifier=IlUbgxJBGPLjVzbGcmtMlsgyiQW2HsBqMcN9VCuk}
  {\bibinfo {title} {Ph 295b, many-body physics: Set 5}} (\bibinfo {year}
  {2016})\BibitemShut {NoStop}%
\bibitem [{\citenamefont {Uchino}(2021)}]{ref:tunneling_JJ}%
  \BibitemOpen
  \bibfield  {author} {\bibinfo {author} {\bibfnamefont {S.}~\bibnamefont
  {Uchino}},\ }\href {https://doi.org/10.1103/PhysRevResearch.3.043058}
  {\bibfield  {journal} {\bibinfo  {journal} {Phys. Rev. Res.}\ }\textbf
  {\bibinfo {volume} {3}},\ \bibinfo {pages} {043058} (\bibinfo {year}
  {2021})}\BibitemShut {NoStop}%
\bibitem [{\citenamefont {Gor'kov}(1958)}]{ref:Gorkov}%
  \BibitemOpen
  \bibfield  {author} {\bibinfo {author} {\bibfnamefont {L.~P.}\ \bibnamefont
  {Gor'kov}},\ }\href {http://jetp.ras.ru/cgi-bin/e/index/e/7/3/p505?a=list}
  {\bibfield  {journal} {\bibinfo  {journal} {Soviet Physics Journal of
  Experimental and Theoretical Physics}\ }\textbf {\bibinfo {volume} {7}},\
  \bibinfo {pages} {505} (\bibinfo {year} {1958})}\BibitemShut {NoStop}%
\end{thebibliography}%

\clearpage

\appendix
\section{Eigenstate of Electron Number Operator} \label{append:N_eigenstate}
In this appendix, we show that the number state defined in \cref{eq:|N>_def} is the eigenstate of both electron and Cooper pair number operators, $\hat{N}^e$, $\hat{N}$ respectively.
\begin{proof}
	The BCS state in $N$-representation is described by
	\begin{align}	\label{eq:|N>_def_BCS}
		\ket{\Psi(N)}
	\nonumber
	&=
		\left(
			\dfrac{\alpha}{\sqrt{n-1}+\alpha}
		\right)^{\sfrac{1}{2}} 
	\times
	\\
	&\quad
		\sum_{j=0}^{\sqrt{n-1}/\alpha}
		e^{-iN\phi_j}
		\prod_{\bm{k}=\bm{k}_1}^{\bm{k}_n}
		\left(
			u_{\bm{k}} + v_{\bm{k}} e^{i\phi_j}
			c_{\bm{k}\uparrow}^{\dagger}
			c_{-\bm{k}\downarrow}^{\dagger}
		\right)
		\ket{0}.
	\end{align}
	We can expand the product of binomials in \cref{eq:|N>_def_BCS} into a sum of product by the lemma \cite{ref:Prob5}:
	\begin{lemma} \label{lemma:product2sum+product}
		$\forall$ $A_k$, $B_k$ such that $[{A_k},{B_k}]=0$, we have an identity
		\[
			\prod_{k=1}^n
			\left( A_k + B_k \right)
		=
			\sum_{m=0}^n
			\sum_{\chi\in\mathcal{S}_m}
			\prod_{i\in\chi^c}
			A_i
			\prod_{j\in\chi}
			B_j,
		\]
		where $\mathcal{S}_m$ is the set of all subsets of $\{1,2,.\cdots,n\}$ with exactly $m$ elements, i.e.,
		\[
			\mathcal{S}_m
		\equiv
			\left\{
				\chi \subseteq
				\{1,2,.\cdots,n\},
				\abs{\chi} = m,
				\right\},
		\quad
			\abs{\mathcal{S}_m} = \binom{n}{m},
		\]
		and $\chi^c$ is the complement of $\chi$.
	\end{lemma}
	With the substitution $u_{\bm{k}} \rightarrow A_k$, $v_{\bm{k}} e^{i\phi_j} c_{\bm{k}\uparrow}^{\dagger} c_{-\bm{k}\downarrow}^{\dagger} \rightarrow B_k$, \cref{eq:|N>_def_BCS} becomes
    \begin{widetext}
	\begin{subequations}\begin{align} 
			\ket{\Psi(N)} 
		&=
			\left(
				\dfrac{\alpha}{\sqrt{n-1}+\alpha}
			\right)^{\sfrac{1}{2}} 
			\sum_{j=0}^{\sqrt{n-1}/\alpha}
			e^{-iN\phi_j}
			\sum_{m=0}^n \sum_{\chi\in\mathcal{S}_m}
			\prod_{\bm{k}_i\in\chi^c}
			u_{\bm{k}_i}
			\prod_{\bm{k}_j\in\chi}
			v_{\bm{k}_j}  e^{i\phi_j}
			c_{\bm{k}_j\uparrow}^{\dagger} c_{-\bm{k}_j\downarrow}^{\dagger}
			\ket{0},
		\label{eq:|N>_expand_1st}
		\\\nonumber
		&=
			\left(
				\dfrac{\alpha}{\sqrt{n-1}+\alpha}
			\right)^{\sfrac{-1}{2}} 
			\dfrac{\alpha}{\sqrt{n-1}+\alpha}
			\sum_{j=0}^{\sqrt{n-1}/\alpha}
			e^{-iN\phi_j}
			\sum_{m=0}^n 
			e^{im\phi_j}
			\sum_{\chi\in\mathcal{S}_m}
			\prod_{\bm{k}_i\in\chi^c}
			u_{\bm{k}_i}
			\prod_{\bm{k}_j\in\chi}
			v_{\bm{k}_j}  
			c_{\bm{k}_j\uparrow}^{\dagger} c_{-\bm{k}_j\downarrow}^{\dagger}
			\ket{0},
		\\
		&=
			\bigg(
				\dfrac{\alpha}{\sqrt{n-1}+\alpha}
			\bigg)^{\sfrac{-1}{2}} 
			\sum_{m=0}^n 
			\left(
				\dfrac{\alpha}{\sqrt{n-1}+\alpha}
				\sum_{j=0}^{\sqrt{n-1}/\alpha}
				e^{i(m-N)\phi_j}
			\right)
			\sum_{\chi\in\mathcal{S}_m}
			\prod_{\bm{k}_i\in\chi^c}
			u_{\bm{k}_i}
			\prod_{\bm{k}_j\in\chi}
			v_{\bm{k}_j}  
			c_{\bm{k}_j\uparrow}^{\dagger} c_{-\bm{k}_j\downarrow}^{\dagger}
			\ket{0}.
		\label{eq:|N>_expand}
	\end{align}\end{subequations}
    \end{widetext}
	For large $n$ such that $\sqrt{n}\gg\alpha$, we can define a new variable $n' = \sqrt{n}/\alpha \approx \sqrt{n-1}/\alpha$, so the term inside the bracket of \cref{eq:|N>_expand} can be rewritten as
	\begin{align*}
		&\dfrac{\alpha}{\sqrt{n-1}+\alpha}
		\sum_{j=0}^{\sqrt{n-1}/\alpha}
		e^{i(m-N)\phi_j}
		\\
		&\quad\quad\quad=
		\dfrac{\alpha}{\sqrt{n-1}+\alpha}
		\sum_{j=0}^{\sqrt{n-1}/\alpha}
		e^{i(m-N)
			(-\pi + 2\pi\alpha j/\sqrt{n})
		},
		\\
		&\quad\quad\quad\approx
		\dfrac{1}{n'+1}
		\sum_{j=0}^{n'}
		e^{
			i(m-N)
			(-\pi + 2\pi j/n')
		},
		\\
		&\quad\quad\quad=
		e^{-\pi i (m-N)}
		\cdot
		\dfrac{1}{n'+1}
		\sum_{j=0}^{n'}
		e^{2\pi i(m-N)j/n'}.
	\end{align*}
Taking the limit $n\rightarrow\infty$ and  $n'\rightarrow\infty$, we have
	\begin{align*}
		\lim\limits_{n\rightarrow\infty}
		&\dfrac{\alpha}{\sqrt{n-1}+\alpha}
		\sum_{j=0}^{\sqrt{n-1}/\alpha}
		e^{i(m-N)\phi_j}
		\\
		&=
		e^{-\pi i (m-N)}
		\lim\limits_{n'\rightarrow\infty}
		\dfrac{1}{n'+1}
		\sum_{j=0}^{n'}
		e^{2\pi i(m-N)j/n'},
		\\
		&\approx
		e^{-\pi i (m-N)}
		\cdot
		\underbrace{
			\lim\limits_{n'\rightarrow\infty}
			\dfrac{1}{n'}
			\sum_{j=0}^{n'}
			e^{2\pi i(m-N)j/n'}
		}_{=\delta_{mN}},
		\\
		&=
		e^{-\pi i (m-N)}
		\delta_{m,N},
	\end{align*}
	where $\delta_{m,N}$ is the Kronecker delta  function. 
	Now \cref{eq:|N>_expand} is 
	\begin{align} \label{eq:|N>_sum_prod}
		\ket{\Psi(N)} \nonumber
		&=
		\left(
		\dfrac{\alpha}{\sqrt{n-1}+\alpha}
		\right)^{\sfrac{-1}{2}} 
		\sum_{m=0}^n
		e^{-\pi i (m-N)}
		\delta_{m,N}
		\\\nonumber
		&\hspace{1cm}\times
        \sum_{\chi\in\mathcal{S}_m}
		\prod_{\bm{k}_i\in\chi^c}
		u_{\bm{k}_i}
		\prod_{\bm{k}_j\in\chi}
		v_{\bm{k}_j}  
		c_{\bm{k}_j\uparrow}^{\dagger} c_{-\bm{k}_j\downarrow}^{\dagger}
		\ket{0},
		\\
        &=
		\tfrac{1}{{\sqrt{\jmax}}}
        \sum_{\chi\in\mathcal{S}_N}
		\prod_{\bm{k}_i\in\chi^c}
		\!\!u_{\bm{k}_i}\!\!
		\prod_{\bm{k}_j\in\chi}\!
		v_{\bm{k}_j}  
		c_{\bm{k}_j\uparrow}^{\dagger} c_{-\bm{k}_j\downarrow}^{\dagger}
		\ket{0}.
	\end{align}
	
	Next, with \cref{eq:|N>_sum_prod}, the electron number operator $\hat{N}^e$ of a single superconductor acting on this number state $\ket{\Psi(N)}$ gives
     \ycl{
	\begin{widetext}\begin{align}
		\hat{N}^e \ket{\Psi(N)} \nonumber
	&=
		\sum_{\bm{k}'}
		\big(
			c_{\bm{k}'\uparrow}^{\dagger}   c_{\bm{k}'\uparrow}
			+
			c_{-\bm{k}'\downarrow}^{\dagger}   c_{-\bm{k}'\downarrow}
		\big)
        \dfrac{1}{\sqrt{\jmax}}
		\sum_{\chi\in\mathcal{S}_N}
		\prod_{\bm{k}_i\in\chi^c}
		u_{\bm{k}_i}
		\prod_{\bm{k}_j\in\chi}
		v_{\bm{k}_j}  
		c_{\bm{k}_j\uparrow}^{\dagger} c_{-\bm{k}_j\downarrow}^{\dagger}
		\ket{0},
	\\\nonumber
	&=
        \dfrac{1}{\sqrt{\jmax}}
		\sum_{\chi\in\mathcal{S}_N}
        \sum_{\bm{k}'\in\chi}
		\prod_{\bm{k}_i\in\chi^c}
		u_{\bm{k}_i}
		\big(
			c_{\bm{k}'\uparrow}^{\dagger}   c_{\bm{k}'\uparrow}
			+
			c_{-\bm{k}'\downarrow}^{\dagger}   c_{-\bm{k}'\downarrow}
		\big)
		\prod_{\bm{k}_j\in\chi}
		v_{\bm{k}_j}  
		c_{\bm{k}_j\uparrow}^{\dagger} c_{-\bm{k}_j\downarrow}^{\dagger}
		\ket{0},
	\\\nonumber
	&=
		\dfrac{1}{\sqrt{\jmax}}
		\sum_{\chi\in\mathcal{S}_N}
        \sum_{\bm{k}'\in\chi}   2
		\prod_{\bm{k}_i\in\chi^c}
		u_{\bm{k}_i}  
		\prod_{\bm{k}_j\in\chi}
		v_{\bm{k}_j}  
		c_{\bm{k}_j\uparrow}^{\dagger} c_{-\bm{k}_j\downarrow}^{\dagger}
		\ket{0},
	\\\nonumber
	&=
		2 N
        \Big(   
        \dfrac{1}{\sqrt{\jmax}}
		\sum_{\chi\in\mathcal{S}_N}
		\prod_{\bm{k}_i\in\chi^c}
		u_{\bm{k}_i}
		\prod_{\bm{k}_j\in\chi}
		v_{\bm{k}_j}  
		c_{\bm{k}_j\uparrow}^{\dagger} c_{-\bm{k}_j\downarrow}^{\dagger}
		\ket{0}
        \Big),
        \quad \because \abs{\chi}=N,
	\\
	&=
		2N \ket{\Psi(N)},
	\end{align}\end{widetext}
    where $\chi$ is a subset of the single-particle modes.}
	The eigenvalue equation $\hat{N}^e\ket{\Psi(N)} = 2N \ket{\Psi(N)}$ implies that we have exactly $2N$ electron in the state $\ket{\Psi(N)}$.
	This number state is actually that of the Cooper pairs: $\hat{N}^e = 2\hat{N}$.
\end{proof}

\section{Matrix Elements of the Superconducting Island Hamiltonian} \label{append:Melem_islandHamiltonain}
The matrix elements of the island Hamiltonian projected onto the low-energy Hilbert space in \cref{eqn:PHAP} is given by three momentum sums, which can be approximated by the integrals: $\sum_{\bm{k}=-\infty}^\infty  \rightarrow 2\int_\Delta^{b\Delta} d{E} \rho(E)$, where $\rho(E) = \frac{n}{2\Delta\sqrt{b^2 - 1}}\frac{E}{\sqrt{E^2 - \Delta^2}}$. 
{Recalling $u_{\bf k}$ and $v_{\bf k}$ from \cref{eq:BogoCoeff} and using the results} 
\begin{widetext}
\ycl{
\begin{align}
    \mel{\Psi(\phi_j)}{c_{\bm{k}\uparrow}^\dagger c_{\bm{k}\uparrow}}{\Psi(\phi_{j'})}
\nonumber
&=
    \bra{0} \prod_{\bm{k}' \neq \bm{k}}
    (u_{\bm{k}'} + v_{\bm{k}'} e^{-i\phi_j} c_{-\bm{k}'\downarrow } c_{\bm{k}'\uparrow})
    (u_{\bm{k}} + v_{\bm{k}} e^{-i\phi_j} c_{-\bm{k}\downarrow } c_{\bm{k}\uparrow})
    c_{\bm{k}\uparrow}^\dagger c_{\bm{k}\uparrow}
    (u_{\bm{k}} + v_{\bm{k}} e^{i\phi_{j'}} c_{\bm{k}\uparrow}^\dagger c_{-\bm{k}\downarrow}^\dagger)
\\\nonumber 
&\quad\times
    \prod_{\bm{k}'' \neq \bm{k}}
    (u_{\bm{k}''} + v_{\bm{k}''} e^{i\phi_{j'}} c_{\bm{k}''\uparrow}^\dagger c_{-\bm{k}''\downarrow}^\dagger) \ket{0},
\\\nonumber
&=
    \bra{0} \prod_{\bm{k}' \neq \bm{k}}
    (u_{\bm{k}'} + v_{\bm{k}'} e^{-i\phi_j} c_{-\bm{k}'\downarrow } c_{\bm{k}'\uparrow})
    \cdot
    v_{\bm{k}}^2    e^{i\varphi}    
    \cdot
    \prod_{\bm{k}'' \neq \bm{k}}
    (u_{\bm{k}''} + v_{\bm{k}''} e^{i\phi_{j'}} c_{\bm{k}''\uparrow}^\dagger c_{-\bm{k}''\downarrow}^\dagger) \ket{0},
\\\nonumber
&=
    e^{i\varphi} v_{\bm{k}}^2
    \dfrac{
            \ip{\Psi(\phi_j)}{\Psi(\phi_{j'})}
        }
        {
            \bra{0}
            (u_{\bm{k}} + v_{\bm{k}} e^{-i\phi_j} c_{-\bm{k}\downarrow} c_{\bm{k}\uparrow})
            (u_{\bm{k}} + v_{\bm{k}} e^{i\phi_{j'}} c_{\bm{k}\uparrow}^\dagger c_{-\bm{k}\downarrow}^\dagger)
            \ket{0}
        },
\\
&=
    \dfrac{v_{\bm{k}}^2}{u_{\bm{k}}^2 + v_{\bm{k}}^2 e^{i\varphi}}
    e^{i\varphi}   \mathcal{W}(\varphi),
\end{align}
and
\begin{align}
    \mel{\Psi(\phi_j)}
        {
            c_{\bm{k}\uparrow}^\dagger c_{-\bm{k}\downarrow}^\dagger
            c_{-\bm{k}'\downarrow}  c_{\bm{k}'\uparrow}
        }
        {\Psi(\phi_{j'})}
&=
    \bra{0} \prod_{\bm{\kappa} \neq \bm{k}, \bm{k}'}
    (u_{\bm{\kappa}} + v_{\bm{\kappa}} e^{-i\phi_j} c_{-\bm{\kappa}\downarrow } c_{\bm{\kappa}\uparrow})
\\\nonumber
&\quad\times
    (u_{\bm{k}} + v_{\bm{k}} e^{-i\phi_j} c_{-\bm{k}\downarrow } c_{\bm{k}\uparrow})
    c_{\bm{k}\uparrow}^\dagger c_{-\bm{k}\downarrow}^\dagger
    (u_{\bm{k}} + v_{\bm{k}} e^{i\phi_{j'}} c_{\bm{k}\uparrow}^\dagger c_{-\bm{k}\downarrow}^\dagger)
\\\nonumber
&\quad\times
    (u_{\bm{k}'} + v_{\bm{k}'} e^{-i\phi_j} c_{-\bm{k}'\downarrow } c_{\bm{k}'\uparrow})   
    c_{-\bm{k}'\downarrow}  c_{\bm{k}'\uparrow}
    (u_{\bm{k}'} + v_{\bm{k}'} e^{i\phi_{j'}} c_{\bm{k}'\uparrow}^\dagger c_{-\bm{k}'\downarrow}^\dagger)
\\\nonumber
&\quad\times
    \prod_{\bm{\kappa}'' \neq \bm{k}, \bm{k}'}
    (u_{\bm{\kappa}''} + v_{\bm{\kappa}''} e^{i\phi_{j'}} c_{\bm{\kappa}''\uparrow}^\dagger c_{-\bm{\kappa}''\downarrow}^\dagger) \ket{0},
\\\nonumber
&=
    u_{\bm{k}}  v_{\bm{k}}  u_{\bm{k}'}  v_{\bm{k}'}
    e^{i\varphi}
\\\nonumber
&\quad
    \tfrac{
            \ip{\Psi(\phi_j)}{\Psi(\phi_{j'})}
        }
        {
            \bra{0}
            (u_{\bm{k}} + v_{\bm{k}} e^{-i\phi_j} c_{-\bm{k}\downarrow} c_{\bm{k}\uparrow})
            (u_{\bm{k}} + v_{\bm{k}} e^{i\phi_{j'}} c_{\bm{k}\uparrow}^\dagger c_{-\bm{k}\downarrow}^\dagger)
            (u_{\bm{k}'} + v_{\bm{k}'} e^{-i\phi_j} c_{-\bm{k}'\downarrow} c_{\bm{k}'\uparrow})
            (u_{\bm{k}'} + v_{\bm{k}'} e^{i\phi_{j'}} c_{\bm{k}'\uparrow}^\dagger c_{-\bm{k}'\downarrow}^\dagger)
            \ket{0}
        },
\\\nonumber
&=
    \Big(
        \dfrac{u_{\bm{k}}  v_{\bm{k}}}{u_{\bm{k}}^2 + v_{\bm{k}}^2 e^{i\varphi}}
    \Big)^2
    e^{i\varphi}    \mathcal{W}(\varphi),
\end{align}
where $\varphi = \phi_{j'} - \phi_j$,} we compute the matrix elements of $PH_AP$ as follows 
\begin{align}
	\mel{\Psi(\phi_j)}{H_A}{\Psi(\phi_{j'})}
\nonumber
&=
	2\sum_{\bm{k}}
	\dfrac{\left(	\epsilon_{\bm{k}}	-	\mu_A	\right)	v_{\bm{k}}^2}
	{u_{\bm{k}}^2	+	v_{\bm{k}}^2	e^{i\varphi}}
    e^{i\varphi} \mathcal{W}(\varphi)
	-
	\abs{g}^2
	\left(
		\sum_{\bm{k}}
		\dfrac{u_{\bm{k}}	v_{\bm{k}}}
		{u_{\bm{k}}^2	+	v_{\bm{k}}^2	e^{i\varphi}}
	\right)^2
    e^{i\varphi} \mathcal{W}(\varphi),
\\\nonumber
&=
	2    e^{i\varphi} \mathcal{W}(\varphi)\sum_{\bm{k}}
	\dfrac{
		\sqrt{E_{\bm{k}}^2 - \Delta^2}
		\left(
		E_{\bm{k}}
		-
		\sqrt{E_{\bm{k}}^2 - \Delta^2}
		\right)
	}
	{
		E_{\bm{k}}
		(
		1	+	e^{i\varphi}
		)
		+
		\sqrt{E_{\bm{k}}^2 - \Delta^2}
		(
		1	-	e^{i\varphi}
		)
	}
    \nonumber\\
    &\quad
	-
	\abs{g}^2 e^{i\varphi} \mathcal{W}(\varphi)
	\Big(
		\Delta\sum_{\bm{k}}
		\dfrac{1}
		{
			E_{\bm{k}}
			\left(
			1	+	e^{i\varphi}
			\right)
			+
			\sqrt{E_{\bm{k}}^2 - \Delta^2}
			\left(
			1	-	e^{i\varphi}
			\right)
		}
	\Big)^2
    ,\nonumber
\\
\nonumber
&\approx
		e^{i\varphi} \mathcal{W}(\varphi)
		\int_\Delta^{b\Delta}	dE 
		\rho(E) \frac{4 e^{-i \varphi } E (\Delta^2 -E^2)}{2 E^2-\Delta^2(1-\cos (\varphi ))}
\nonumber\\
&\quad
-
	\abs{g}^2e^{i\varphi} \mathcal{W}(\varphi)
		\left(
			\int_\Delta^{b\Delta} dE \rho(E) \frac{\Delta  E \left(1+e^{-i \varphi }\right)}{2 E^2-\Delta ^2 (1-\cos (\varphi ))}
		\right)^2,
\\
&\equiv
	\mathcal{K}(\varphi)
	-
	\abs{g}^2 \mathcal{V}(\varphi).\label{eqn:exactPHPappendix}
\end{align}
The complex functions $\mathcal{K}(\varphi)$ and $\mathcal{V}(\varphi)$ are
\begin{align}
    \mathcal{K}(\varphi)
&={n\,\Delta\,\mathcal{W}(\varphi) \left( 2\,{\ln} (2 b) \cos (\varphi )+\varphi  \sin (\varphi
   )-2 b^2\right)}/({4 b}) +O(b^{-2})
   \\
\mathcal{V} (\varphi)
&={n^2 \mathcal{W}(\varphi) \left(2\,{\ln}(2 b) \cos \left({\varphi
   }/{2}\right)+\varphi  \sin \left({\varphi }/{2}\right)\right)^2}/({4 b})^2
   +O(b^{-3}).
\end{align}
\end{widetext}
In addition, the diagonal elements of $ \mathcal{K}$ and $ \mathcal{V}$ evaluate to the simple expressions
\begin{align}
 \mathcal{K} (0)&=  {n\,\Delta  \big({\ln} (2 b)-b^2\big)}/({2 \,b}),\label{eqn:K0}\\
   \mathcal{V} (0)&=  ({n \ln(2 b)})^2/({2 \,b})^2,\label{eqn:V0}
\end{align}
where we retain only terms at leading orders in $1/b$.

\section{Josephson Effect \& Greens' Functions}	\label{append:GreensFun}
We use Greens' functions to calculate  the second-order perturbation term. 
For a given diagonalisable Hamiltonian, the Fourier transform of its propagator relates to its Hamiltonian inverse.
That is,
\[
	H_T
	\dfrac{1}{E_0\bar{P} - \bar{P}H_0\bar{P}}
	H_TD_{\varphi}
=
	\mathscr{F}_{\tau\rightarrow E_0}
	\left[
	H_T U_0(\tau) H_TD_{\varphi}
	\right],
\]
where $U_0(\tau)$ is the propagator for $H_0$ at time $\tau$.
The matrix elements of the projection in \cref{eq:time_ordering_2FF} is then the Fourier transformation of the expectation value:
\begin{align}
	\bra{\Psi(\phi_j)}
	H_T
	(E_0\bar{P} - &\bar{P}H_0\bar{P})	^{-1}
	H_TD_\varphi
	\ket{\Psi(\phi_j)}
\nonumber
\\
&=
	\mathscr{F}_{\tau\rightarrow E_0}
	\ev{H_T U_0(\tau) H_TD_{\varphi}}.
\end{align}
The expectation value on the right-hand side is taken with respect to the BCS ground state:
\begin{align*}
	\ev{H_T(\tau)H_TD_\varphi}
&=
	\ev{H_T(\tau)H_TD_\varphi}{\Psi(\phi_j)},
\\
&=
	\ev{U_0^\dagger(\tau)H_TU_0(\tau)H_TD_\varphi}
	{\Psi(\phi_j)},
\\
&=
	e^{iE_0\tau}
	\ev{H_TU_0(\tau)H_TD_\varphi}
	{\Psi(\phi_j)},
\end{align*}
so
\begin{equation}
	\ev{H_T U_0(\tau) H_T D_\varphi}
=
	e^{-iE_0\tau}
	\ev{H_T(\tau) H_T D_{\varphi}}{\Psi(\phi_j)}.
\end{equation}
Taking the Fourier transform on the both sides,
\begin{align} \label{eq:goal_FT}
	\mathscr{F}_{\tau\rightarrow E_0}
	&\left[
		\mel{\Psi(\phi_j)}
			{H_T U_0(\tau) H_T }
			{\Psi(\phi_{j'})}
	\right]
\nonumber
\\\nonumber
&=
	\mathscr{F}_{\tau\rightarrow E_0}
	\left[
		e^{-iE_0\tau}
		\mel{\Psi(\phi_j)}
			{H_T(\tau)  H_T D_{\varphi}}
			{\Psi(\phi_{j})}
	\right],
\\\nonumber
&=
	\mathscr{F}_{\tau\rightarrow E_0}
	\left[
		\mel{\Psi(\phi_j)}
			{U_0(\tau)H_T(\tau)  H_T D_{\varphi}}
			{\Psi(\phi_{j})}
	\right],
\\\nonumber
&=
	\mathscr{F}_{\tau\rightarrow E_0}
	\left[
		\mel{\Psi(\phi_j)}
			{\mathcal{T}_{\tau}
				\left( H_T(\tau)  H_T D_{\varphi} \right)
			}
			{\Psi(\phi_{j})}
	\right],
\\
&=
	\mathscr{F}_{\tau\rightarrow E_0}
	\left[
		\mathcal{T}_{\tau}
		\ev{ H_T(\tau)  H_T D_{\varphi} }
	\right],
\end{align}
i.e.\ the matrix elements are the Fourier transform of the Greens' functions 
\(
\mathscr{F}_{\tau\rightarrow E_0}
\ev{\mathcal{T}_{\tau}
	\left( H_T(\tau)  H_T D_{\varphi} \right)}.
\)
\begin{proof}
The propagator $U_0(\tau)$ can be expanded as the time ordered series
\begin{align*}
	U_0(\tau)
&=
	\mathcal{T}_{\tau}
	\exp(
	\dfrac{-i}{\hbar}
	\int_0^\tau
	\dd{t} H_0(t)
	),
\\
&=
	\mathcal{T}_{\tau}
	\left(
	\mathbb{I}
	-
	\dfrac{i}{\hbar}
	\int_0^\tau
	\dd{t} H_0(t)
	+
	\cdots
	\right),
\end{align*}
such that
\begin{align*}
	U_0(\tau) H_T(\tau) H_TD_{\varphi}
&=
	\mathcal{T}_{\tau}
	\big(
		H_T(\tau) H_TD_{\varphi}
\\
&\quad-
		\dfrac{i}{\hbar}
		\int_0^\tau
		\dd{t} H_0(t)
		H_T(\tau) H_TD_{\varphi}
\\
&\quad+
		\cdots
	\big),
\\
&\approx
	\mathcal{T}_{\tau}
	\left(
		H_T(\tau) H_TD_{\varphi}
	\right).
\end{align*}
\end{proof}

Expanding the time-ordered product in \cref{eq:goal_FT} and applying Wick's theorem to factorise the correlation functions, we have
\begin{widetext}
\begin{subequations}\label{eq:time_ordering_2FF_NoExpand}\begin{align} 
		\ev{ \mathcal{T}_{\tau}
			c_{\bm{k}_{A}s}^{\dagger}(\tau)
			c_{\bm{k}_{B}s}(\tau)
			c_{\bm{k}_{A}\sigma}^{\dagger}
			c_{\bm{k}_{B}\sigma}
			D_{\varphi_A}D_{\varphi_B}
		}
& =
		\theta(\tau)
		\ev{ c_{\bm{k}_{A}s}^{\dagger}(\tau)
			c_{\bm{k}_{B}s}(\tau)
			c_{\bm{k}_{A}\sigma}^{\dagger}
			c_{\bm{k}_{B}\sigma}
			D_{\varphi_A}D_{\varphi_B}
		}
\nonumber
\\\nonumber
&\quad
		-\theta(-\tau)
		\ev{c_{\bm{k}_{A}\sigma}^{\dagger}
			c_{\bm{k}_{B}\sigma}
			D_{\varphi_A}D_{\varphi_B}
			c_{\bm{k}_{A}s}^{\dagger}(\tau)
			c_{\bm{k}_{B}s}(\tau)
		},
\\
&=
		2\theta(\tau)
		\ev{ c_{\bm{k}_{A}s}^{\dagger}(\tau)
			c_{\bm{k}_{A}\sigma}^{\dagger}
			D_{\varphi_A}
		}
		\ev{
			c_{\bm{k}_{B}s}(\tau)
			c_{\bm{k}_{B}\sigma}
			D_{\varphi_B}
		}
\\\nonumber
&\quad
		-2\theta(-\tau)
		\ev{
			c_{\bm{k}_{A}\sigma}^{\dagger}
			D_{\varphi_A}
			c_{\bm{k}_{A}s}^{\dagger}(\tau)
		}
		\ev{
			c_{\bm{k}_{B}\sigma}
			D_{\varphi_B}
			c_{\bm{k}_{B}s}(\tau)
		},
\\\nonumber
& =
		2\ev{ \mathcal{T}_{\tau}
			c_{\bm{k}_{A}s}^{\dagger}(\tau)
			c_{\bm{k}_{A}\sigma}^{\dagger}
			D_{\varphi_A}
		}
		\ev{ \mathcal{T}_{\tau}
			c_{\bm{k}_{B}\sigma}
			D_{\varphi_B}
			c_{\bm{k}_{B}s}(\tau)
		},
\\
& =
		2\ev{ \mathcal{T}_{\tau}
			c_{\bm{k}_{A}s}^{\dagger}(\tau)
			c_{\bm{k}_{A}\sigma}^{\dagger}
			D_{\varphi_A}
		}
		\ev{ \mathcal{T}_{\tau}
			c_{\bm{k}_{B}\sigma}(-\tau)
			c_{\bm{k}_{B}s}
			D_{\varphi_B}
		},
\label{eq:factorised_corrFun}
\end{align}\end{subequations}
\end{widetext}
where subscripts $s$, $\sigma$ denote the spin indices, and the expectation values are taken with respect to $\ket{\Psi(\phi_j)}$.
With the factorisation of the expectation value (proved in Section~\ref{append:factorize_<c+.c+.N^n>})
\begin{equation} \label{eq:factorise_<c+.c+.N^n>}
	\ev{
		c_{\bm{k}_{A}s}^{\dagger}(\tau)
		c_{\bm{k}_{A}\sigma}^{\dagger}(N_A^e)^n
	}
=
	\ev{ c_{\bm{k}_{A}s}^{\dagger}(\tau)
		c_{\bm{k}_{A}\sigma}^{\dagger}
	}
	\ev{
		(N_A^e)^n
	},
\end{equation}
and the BCS overlap expressed with the phase displacement operator
\begin{equation} \label{eq:series_BCSoverlap}
	\sum_{n}
	\dfrac{\left(i\varphi\right)^{n}}{n!}
	\ev{
		\left(N^{e}\right)^n
	}
=
	\mathcal{W}(\varphi)
\quad
	\varphi = \phi' - \phi,
\end{equation}
we expand the phase displacement operator in the first term in \cref{eq:factorised_corrFun} by Taylor's series as
\small\begin{subequations}\begin{align} \label{eq:<c+.c+.D>_forA_1}
	\langle
		\mathcal{T}_{\tau} 
		&c_{\bm{k}_{A}s}^{\dagger}(\tau) 
		c_{\bm{k}_{A}\sigma}^{\dagger} 
		D_{\varphi_A}
	\rangle
\nonumber
\\\nonumber
& =
	\ev{
		c_{\bm{k}_{A}s}^{\dagger}(\tau)
		c_{\bm{k}_{A}\sigma}^{\dagger}
	}
	+
	(i\varphi_{A}/2)
	\ev{ c_{\bm{k}_{A}s}^{\dagger}(\tau)
		c_{\bm{k}_{A}\sigma}^{\dagger}(2N_{A}^{e})
	}
\\\nonumber
& \quad+
	\dfrac{(i\varphi_A/2)^2}{2!}
	\ev{ c_{\bm{k}_{A}s}^{\dagger}(\tau)
		c_{\bm{k}_{A}\sigma}^{\dagger}
		(2N_{A}^e)^2
	}
	+
	\cdots,
\\\nonumber
& =
	\ev{
		c_{\bm{k}_{A}s}^{\dagger}(\tau)
		c_{\bm{k}_{A}\sigma}^{\dagger}
	} 
	+
	(i\varphi_{A})
	\ev{
		c_{\bm{k}_{A}s}^{\dagger}(\tau)
		c_{\bm{k}_{A}\sigma}^{\dagger}N_A^e
	}
\\
&\quad+
	\dfrac{(i\varphi_{A})^2}{2!}
	\ev{
		c_{\bm{k}_{A}s}^{\dagger}(\tau)
		c_{\bm{k}_{A}\sigma}^{\dagger}
		(N_A^e)^2
	}
	+
	\cdots,
\\\nonumber
&=
	\ev{
		c_{\bm{k}_{A}s}^{\dagger}(\tau)
		c_{\bm{k}_{A}\sigma}^{\dagger}N_A^e
	}
\\
&\quad\cdot
	\left(
		1 
		+
		(i\varphi_A) \ev{N_A^e} 
		+
		\dfrac{(i\varphi_A)^2}{2!} 	 \ev{(N_A^e)^2} 
		+ 
		\cdots
	\right),
\\
&=
	\ev{
		c_{\bm{k}_As}^{\dagger}(\tau) 
		c_{\bm{k}_A\sigma}^{\dagger}
	}
	\mathcal{W}(\varphi_A),
\end{align}\end{subequations}\normalsize
where $\mathcal{W}(\varphi_A) = \ip{\Psi_A(\phi_j)}{\Psi_A(\phi_{j'})}$ is the overlap function of the BCS ground state for island $A$.
Similarly for island $B$:
\[
	\ev{
		\mathcal{T}_{\tau}
		c_{\bm{k}_B\sigma}(-\tau)
		c_{\bm{k}_Bs}D_{\varphi_B}
	}
=
	\ev{
		c_{\bm{k}_B\sigma}(-\tau) 
		c_{\bm{k}_Bs}
	}
	\mathcal{W}(\varphi_B).
\]
\begin{proof}
We have computed the BCS overlap function $\mathcal{W}(\varphi)$ directly from the definition in Section~\ref{sec:H0_space}.
Here with the introduction of the phase displacement operator $D_\varphi$, we will do the inner product again to obtain additional insight.
The overlap now becomes the expectation value of $D_\varphi$:
\begin{subequations}\begin{align}
	\mathcal{W}(\varphi)
&=
	\ip{\Psi(\phi_j)}{\Psi(\phi_{j'})}
=
	\ev{D_\varphi}{\Psi(\phi_j)}
\\\nonumber
&=
	\bra{0}
	\prod_{\bm{q}'}
	\left(
		u_{\bm{q}'}	+	v_{\bm{q}'}	e^{-i\phi_j}	c_{-\bm{q}'\downarrow}	c_{\bm{q}'\uparrow}
	\right)
\\\nonumber
&\quad\times
	\left[
		\mathbb{I}
		+
		(i\varphi/2)
		\left(
			\sum_{\bm{k}}
			c_{\bm{k}\uparrow}^\dagger	c_{\bm{k}\uparrow}
			+
			c_{-\bm{k}\downarrow}^\dagger	c_{-\bm{k}\downarrow}
		\right)
		+
		\cdots
	\right]
\\\nonumber
&\quad\times
	\prod_{\bm{q}'}
	\left(
		u_{\bm{q}}	+	v_{\bm{q}}	e^{i\phi_{j'}}	c_{\bm{q}\uparrow}^\dagger	c_{-\bm{q}\downarrow}^\dagger
	\right)
	\ket{0},
\\\nonumber
&=
	1
	+
	(i\varphi/2)	\ev{N}
	+
	\dfrac{(i\varphi/2)^2}{2!}	\ev{N^2}
	+
	\cdots,
\\
&=
	1
	+
	(i\varphi)	\ev{N^e}
	+
	\dfrac{(i\varphi)^2}{2!}	\ev{(N^e)^2}
	+
	\cdots,
\\
&=
	\sum_{n=0}^\infty
	\dfrac{(i\varphi)^n}{n!}	\ev{(N^e)^n},
\\
&=
	\langle{e^{i\varphi\hat{N}^e}}\rangle,
\end{align}\end{subequations}
i.e., we can express $\mathcal{W}(\varphi)$ as the expectation value of $D_\varphi$.
\end{proof}

Hence, by taking the tunneling matrix element as a constant $t_{\bm{k}_A\bm{k}_B} \approx t$ \cite{ref:tunneling_JJ}, a convenient approximation for a symmetric junction, we obtain 
\begin{widetext}\begin{align} 
	\ev{
		\mathcal{T}_{\tau}H_T (\tau)H_T
		D_{\varphi}
	}\nonumber
&=
	2\sum_{\bm{k}_{A}\bm{k}_{B}s\sigma}
	e^{i\phi_j}
	t_{\bm{k}_{A}\bm{k}_{B}}^{2}
	\ev{
		\mathcal{T}_{\tau}c_{\bm{k}_{A}s}^{\dagger}\left(\tau\right)c_{\bm{k}_{A}\sigma}^{\dagger}
	}
	\ev{
		\mathcal{T}_{\tau}c_{\bm{k}_{B}\sigma}\left(-\tau\right)c_{\bm{k}_{B}s}
	}
	\mathcal{W}(\varphi_A) \mathcal{W}(\varphi_B)
	+
	c.c.,
\\\nonumber
&=
	2\sum_{\bm{k}_{A}\bm{k}_{B}s\sigma}
	e^{i\phi_j}
	t_{\bm{k}_A\bm{k}_B}^2
	\ev{\mathcal{T}_{\tau}  c_{\bm{k}_As}^{\dagger} (\tau)  c_{\bm{k}_A{\sigma}}^{\dagger} }
	\ev{\mathcal{T}_{\tau}  c_{\bm{k}_B\sigma} (-\tau)  c_{\bm{k}_As} }
	\mathcal{W}(\varphi)
	+
	c.c.,
\\\nonumber
&=
	2e^{i\phi_j}
	\sum_{\bm{k}_{A}\bm{k}_{B}s\sigma}
	t_{\bm{k}_{A}\bm{k}_{B}}^{2}
	\mathcal{F}_{s\sigma}\left(\bm{k}_{A},\tau\right)
	\mathcal{F}_{s\sigma}^{*}\left(\bm{k}_{B},-\tau\right)
	\ip{\Psi(\phi_j)}{\Psi(\phi_{j'})}
	+
	c.c.,
\\\nonumber
&=
	2e^{i\phi_j}
	\sum_{\bm{k}_{A}\bm{k}_{B}}
	t_{\bm{k}_{A}\bm{k}_{B}}^{2}
	\mathcal{F}_{\uparrow\downarrow}\left(\bm{k}_{A},\tau\right)
	\mathcal{F}_{\uparrow\downarrow}^{*}\left(\bm{k}_{B},-\tau\right)
	\ip{\Psi(\phi_j)}{\Psi(\phi_{j'})}
	+
	c.c,
\\
&\approx
	2e^{i\phi_j} t^2
	\sum_{\bm{k}_{A}\bm{k}_{B}}
	\mathcal{F}_{\uparrow\downarrow} (\bm{k}_A,\tau)
	\mathcal{F}_{\uparrow\downarrow}^* (\bm{k}_B,-\tau)
	\ip{\Psi(\phi_j)}{\Psi(\phi_{j'})}
	+
	c.c.,
\end{align}\end{widetext}
which is provided in \cref{eq:time_ordering_2FF}.
Note that the anomalous Greens' functions 
\(
\mathcal{F}_{\uparrow\uparrow}
=
\mathcal{F}_{\downarrow\downarrow}
=
0.
\)
\begin{proof}
Here we show that the anomalous Greens' functions $\mathcal{F}_{ss} = 0$.
For simplicity, we consider only a single mode existing on the given superconductor.
By Bogoliubov transformation 
\begin{equation} \label{eq:bogolon_fieldOP}
	\left(\begin{array}{c}
		b_{\bm{k}\uparrow}^{\dagger}\\
		b_{-\bm{k}\downarrow}
	\end{array}\right)
	=
	\left(\begin{array}{cc}
		u_{\bm{k}} & -v_{\bm{k}}e^{-i\phi}\\
		v_{\bm{k}}e^{i\phi} & u_{\bm{k}}
	\end{array}\right)
	\left(\begin{array}{c}
		c_{\bm{k}\uparrow}^{\dagger}\\
		c_{-\bm{k}\downarrow}
	\end{array}\right).
\end{equation}
the time-dependent fermionic annihilation operator is given by
\begin{subequations}\begin{align}
	c_{\bm{k}_l\uparrow}(\tau)
&=
	e^{H_l\tau}
	\left(
		u_{\bm{k}_l}	b_{\bm{k}_l\uparrow}
		+
		v_{\bm{k}_l}	e^{i\phi}	b_{-\bm{k}_l\downarrow}^\dagger
	\right)
	e^{-H_l\tau},
\\\nonumber
&=
	u_{\bm{k}_l}	b_{\bm{k}_l\uparrow}	e^{-E_{\bm{k}_l}\tau}
	+
	v_{\bm{k}_l}	e^{i\phi}	b_{-\bm{k}_l\downarrow}^\dagger	e^{E_{\bm{k}_l}\tau},
\\\nonumber
&=
	u_{\bm{k}_l}	
	\left(
		u_{\bm{k}_l}	c_{\bm{k}_l\uparrow}
		-
		v_{\bm{k}_l}	e^{i\phi}	c_{-\bm{k}_l\downarrow}^\dagger	
	\right)
	e^{-E_{\bm{k}_l}\tau}
\\\nonumber
&\quad+
	v_{\bm{k}_l}	e^{i\phi}
	\left(
		v_{\bm{k}_l}	e^{-i\phi}	c_{\bm{k}_l\uparrow}
		+
		u_{\bm{k}_l}	c_{-\bm{k}_l\downarrow}^\dagger
	\right)
	e^{E_{\bm{k}_l}\tau},
\\\nonumber
&=
	\left(
		u_{\bm{k}_l}^2	e^{-E_{\bm{k}_l}\tau}
		+
		v_{\bm{k}_l}^2	e^{E_{\bm{k}_l}\tau}
	\right)
	c_{\bm{k}_l\uparrow}
\\
&\quad+
	u_{\bm{k}_l}	v_{\bm{k}_l}	e^{i\phi}
	\left(
		e^{E_{\bm{k}_l}\tau}
		-
		e^{-E_{\bm{k}_l}\tau}
	\right)
	c_{-\bm{k}_l\downarrow}^\dagger.
\end{align}\end{subequations}
$H_l$ is the superconductor Hamiltonian for the island $l$.
The Greens' functions are then 
\small\begin{align}
	\ev{c_{\bm{k}_l\uparrow}(\tau)	c_{\bm{k}_l\uparrow}}
\nonumber
&=
	\left(
		u_{\bm{k}_l}^2	e^{-E_{\bm{k}_l}\tau}
		+
		v_{\bm{k}_l}^2	e^{E_{\bm{k}_l}\tau}
	\right)
	\cancelto{0}{
		\ev{c_{\bm{k}_l\uparrow}	c_{\bm{k}_l\uparrow}}
	}
\\\nonumber
&\quad+
	u_{\bm{k}_l}	v_{\bm{k}_l}	e^{i\phi}
	\left(
		e^{E_{\bm{k}_l}\tau}
		-
		e^{-E_{\bm{k}_l}\tau}
	\right)
	\cancelto{0}{
		\ev{c_{-\bm{k}_l\downarrow}	c_{-\bm{k}_l\downarrow}^\dagger }
	},
\\
&=
	0,
\\
	\ev{c_{\bm{k}_l\uparrow}	c_{\bm{k}_l\uparrow}(\tau)}
&=
	0.
\end{align}\normalsize
The result can be generalised to the spin-down case:
\[
	\ev{c_{-\bm{k}_l\downarrow}(\tau)	c_{-\bm{k}_l\downarrow}}
=
	\ev{c_{-\bm{k}_l\downarrow}	c_{-\bm{k}_l\downarrow}(\tau)}
=
	0.
\]
The electron mode $\bm{k}$ can be arbitrary.
Therefore, 
\begin{equation}
	\mathcal{F}_{ss}
\equiv
	\theta(\tau)	\ev{c_{\bm{k}s}(\tau)	c_{\bm{k}s}}
	-
	\theta(-\tau)	\ev{c_{\bm{k}s}	c_{\bm{k}s}(\tau)}
=
	0,
\end{equation}
the anomalous Greens' functions with the identical spin indices vanish:
\(
	\mathcal{F}_{\uparrow\uparrow}
=
	\mathcal{F}_{\downarrow\downarrow}
=
	0.
\)
\end{proof}

\subsection{Factorisation in Greens' Functions}	\label{append:factorize_<c+.c+.N^n>}
We prove \cref{eq:factorise_<c+.c+.N^n>} here.
The expectation value $\ev{(N^e)^n}$ is given by
\begin{widetext}\begin{align}	\label{eq:<Ne^n>}
	\ev{(N^e)^n}
\nonumber
&=
	\bra{0}
	\prod_{\bm{q}'}
	\left(
		u_{\bm{q}'}	+	v_{\bm{q}'}	e^{-i\phi_j}	c_{-\bm{q}'\downarrow}	c_{\bm{q}'\uparrow}
	\right)
	\left(
		\sum_{\bm{k}}
		c_{\bm{k}\uparrow}^\dagger	c_{\bm{k}\uparrow}
		+
		c_{-\bm{k}\downarrow}^\dagger	c_{-\bm{k}\downarrow}
	\right)^n
	\prod_{\bm{q}'}
	\left(
		u_{\bm{q}}	+	v_{\bm{q}}	e^{i\phi_{j'}}	c_{\bm{q}\uparrow}^\dagger	c_{-\bm{q}\downarrow}^\dagger
	\right)
	\ket{0},
\\\nonumber
&=
	2^n
	\sum_{\bm{k}}
	\bra{0}	\prod_{\bm{q}\bm{q}'}
	\left(
		u_{\bm{q}'}	+	v_{\bm{q}'}	e^{-i\phi_j}	c_{-\bm{q}'\downarrow}	c_{\bm{q}'\uparrow}
	\right)
\\\nonumber
&\quad\times
	\left\{
		\left(
			c_{\bm{k}\uparrow}^\dagger	c_{\bm{k}\uparrow}
		\right)^n
	+
		\left(
			\sum_{\bm{k}_1	\neq\bm{k}}
			c_{\bm{k}_1\uparrow}^\dagger	c_{\bm{k}_1\uparrow}
			c_{\bm{k}\uparrow}^\dagger	c_{\bm{k}\uparrow}
			\cdots
			c_{\bm{k}\uparrow}^\dagger	c_{\bm{k}\uparrow}
		\right)
	+
		\left(
			\sum_{\bm{k}_2	\neq\bm{k}}
			c_{\bm{k}_2\uparrow}^\dagger	c_{\bm{k}_2\uparrow}
			c_{\bm{k}_2\uparrow}^\dagger	c_{\bm{k}_2\uparrow}
			\cdots
			c_{\bm{k}\uparrow}^\dagger	c_{\bm{k}\uparrow}
		\right)
		+
		\cdots
	\right\}
\\\nonumber
&\quad\times
	\left(
		u_{\bm{q}}	+	v_{\bm{q}}	e^{i\phi_{j'}}	c_{\bm{q}\uparrow}^\dagger	c_{-\bm{q}\downarrow}^\dagger
	\right)
	\ket{0},
\\
&=
	2^n	\sum_{\bm{k}}
	\left(
		v_{\bm{k}}^{2n}
		+
		\sum_{\bm{k}_1	\neq\bm{k}}
		v_{\bm{k}_1}^2
		\underbrace{
			v_{\bm{k}}^2	\cdots	v_{\bm{k}}^2
		}_{n-1}
		+
		\sum_{\bm{k}_2	\neq\bm{k}}
		v_{\bm{k}_2}^2	v_{\bm{k}_2}^2
		\underbrace{
			\cdots	v_{\bm{k}}^2
		}_{n-2}
		+
		\cdots
	\right).
\end{align}
On the other hand, the correlation functions are
\begin{subequations}\label{eq:anomalousF_all}\begin{align}
	\ev{c_{\bm{k}\uparrow}^\dagger(\tau)	c_{\bm{k}\uparrow}^\dagger }
&=
	\mathcal{F}_{\uparrow\uparrow}^*
=
	0,
\\
	\ev{c_{-\bm{k}\downarrow}^\dagger(\tau)	c_{-\bm{k}\downarrow}^\dagger }
&=
	\mathcal{F}_{\downarrow\downarrow}^*
=
	0,
\\
	\ev{c_{\bm{k}\uparrow}^\dagger(\tau)	c_{-\bm{k}\downarrow}^\dagger }
&=
	\mathcal{F}_{\uparrow\downarrow}
=
	u_{\bm{k}}	v_{\bm{k}}^3
	\left(
		e^{E_{\bm{k}}\tau}	-	e^{-E_{\bm{k}}\tau}
	\right),
\\
	\ev{c_{-\bm{k}\downarrow}^\dagger(\tau)	c_{\bm{k}\uparrow}^\dagger }
&=
	\mathcal{F}_{\downarrow\uparrow}
=
	u_{\bm{k}}	v_{\bm{k}}^3
	\left(
		e^{E_{\bm{k}}\tau}	-	e^{-E_{\bm{k}}\tau}
	\right).
\end{align}\end{subequations}

The left-hand side of \cref{eq:factorise_<c+.c+.N^n>} is discussed separately.
For $s\neq\sigma$, we have
\begin{align}\label{eq:<c+.c+.N^n>_sσ}
	\ev{
		c_{\bm{k}s}^\dagger(\tau)
		c_{\bm{k}\sigma}^\dagger
		(N^e)^n
	}
\nonumber
&=
	2^n\sum_{\bm{k}}
	\ev{
		c_{\bm{k}s}^\dagger(\tau)	c_{\bm{k}\sigma}^\dagger
		\left(
			c_{\bm{k}\uparrow}^\dagger	c_{\bm{k}\uparrow}
		\right)^n
	}
	+
	\ev{
		c_{\bm{k}s}^\dagger(\tau)	c_{\bm{k}\sigma}^\dagger
		\left(
			\sum_{\bm{k}_1	\neq\bm{k}}
			c_{\bm{k}_1\uparrow}^\dagger	c_{\bm{k}_1\uparrow}
			c_{\bm{k}\uparrow}^\dagger	c_{\bm{k}\uparrow}
			\cdots
			c_{\bm{k}\uparrow}^\dagger	c_{\bm{k}\uparrow}
		\right)
	}
	+
	\cdots,
\\\nonumber
&=
	2^n\sum_{\bm{k}}
	\left(
		u_{\bm{k}}^2	e^{-E_{\bm{k}\tau}}
		+
		v_{\bm{k}}^2	e^{E_{\bm{k}\tau}}
	\right)
	\cancelto{0}{\ev{
		c_{\bm{k}s}^\dagger	c_{\bm{k}\sigma}^\dagger
		\left(
		c_{\bm{k}\uparrow}^\dagger	c_{\bm{k}\uparrow}
		\right)^n
	}}
\\\nonumber
&\quad+
	u_{\bm{k}}	v_{\bm{k}}^3
	\left(
		e^{E_{\bm{k}}\tau}	-	e^{-E_{\bm{k}}\tau}
	\right)
	\ev{
		c_{\bm{k}\sigma}^\dagger	c_{\bm{k}\sigma}^\dagger
		\left(
		c_{\bm{k}\uparrow}^\dagger	c_{\bm{k}\uparrow}
		\right)^n
	}
\\\nonumber
&\quad+
	\left(
		u_{\bm{k}}^2	e^{-E_{\bm{k}\tau}}
		+
		v_{\bm{k}}^2	e^{E_{\bm{k}\tau}}
	\right)
	\cancelto{0}{
		\ev{
			c_{\bm{k}s}^\dagger	c_{\bm{k}\sigma}^\dagger
			\sum_{\bm{k}_1	\neq\bm{k}}
			c_{\bm{k}_1\uparrow}^\dagger	c_{\bm{k}_1\uparrow}
			c_{\bm{k}\uparrow}^\dagger	c_{\bm{k}\uparrow}
			\cdots
			c_{\bm{k}\uparrow}^\dagger	c_{\bm{k}\uparrow}
		}
	}
\\\nonumber
&\quad+
	u_{\bm{k}}	v_{\bm{k}}^3
	\left(
		e^{E_{\bm{k}}\tau}	-	e^{-E_{\bm{k}}\tau}
	\right)
	\ev{
		c_{\bm{k}s}^\dagger	c_{\bm{k}\sigma}^\dagger
		\sum_{\bm{k}_1	\neq\bm{k}}
		c_{\bm{k}_1\uparrow}^\dagger	c_{\bm{k}_1\uparrow}
		c_{\bm{k}\uparrow}^\dagger	c_{\bm{k}\uparrow}
		\cdots
		c_{\bm{k}\uparrow}^\dagger	c_{\bm{k}\uparrow}
	}
\\\nonumber
&\quad+\cdots,
\\
&=
	2^n
	u_{\bm{k}'}	v_{\bm{k}'}^3
	\left(
		e^{E_{\bm{k}'}\tau}	-	e^{-E_{\bm{k}'}\tau}
	\right)
	\sum_{\bm{k}}
	\left(
		v_{\bm{k}}^{2n}
		+
		\sum_{\bm{k}_1	\neq\bm{k}}
		v_{\bm{k}_1}^2
			v_{\bm{k}}^2	\cdots	v_{\bm{k}}^2
		+
		\cdots
	\right).
\end{align}
However if $s=\sigma$, all the expectation values vanish according to the exclusion principle,
\begin{equation}	\label{eq:<c+.c+.N^n>_ss}
	\ev{
		c_{\bm{k}s}^\dagger(\tau)
		c_{\bm{k}s}^\dagger
		(N^e)^n
	}
=
	0.
\end{equation}

Combining the results in Eqs.~\eqref{eq:<Ne^n>},~\eqref{eq:anomalousF_all},~\eqref{eq:<c+.c+.N^n>_sσ},~\eqref{eq:<c+.c+.N^n>_ss}, we obtain
\small\begin{align}
	\ev{
		c_{\bm{k}s}^\dagger(\tau)
		c_{\bm{k}\sigma}^\dagger
	}
	\left( (N^e)^n \right)
\nonumber
&=
	\ev{
		c_{\bm{k}s}^\dagger(\tau)
		c_{\bm{k}\sigma}^\dagger
		(N^e)^n
	},
\\
&=
	\begin{cases}
		0,	&s=\sigma	\\
		2^n	
		u_{\bm{k}'}	v_{\bm{k}'}^3
		\left(
			e^{E_{\bm{k}'}\tau}	-	e^{-E_{\bm{k}'}\tau}
		\right)
		\displaystyle\sum_{\bm{k}}
		\left(
			v_{\bm{k}}^{2n}
			+
			\displaystyle\sum_{\bm{k}_1	\neq\bm{k}}
			v_{\bm{k}_1}^2
			v_{\bm{k}}^2	\cdots	v_{\bm{k}}^2
			+
			\displaystyle\sum_{\bm{k}_2	\neq\bm{k}}
			v_{\bm{k}_2}^2	v_{\bm{k}_2}^2
            \cdots	v_{\bm{k}}^2
			+
			\cdots
		\right)
		&	s \neq \sigma
	\end{cases}.
\end{align}\normalsize

\subsection{Finite-Temperature Fourier Transform}   \label{append:FT&sums}
In this section, we compute the matrix elements 
\(
    \mel{\Psi(\phi_j)}
    {H_T (E_0 \bar{P} - \bar{P}H_0\bar{P})^{-1} H_T}
    {\Psi(\phi_{j'})}.
\)
Firstly, we take the finite-temperature Fourier transform of \cref{eq:time_ordering_2FF} \cite{ref:bruus} , to get
\begin{align}
	\mathscr{F}_{\tau\rightarrow E_0}
	\ev{H_T(\tau) H_T D_\varphi}
\nonumber
&=
	2e^{i\phi_j}t^2	\dfrac{1}{\beta}
	\sum_{i\mathcal{E}_n i\mathcal{E}_{n'}}
	\sum_{\bm{k}_{A}\bm{k}_{B}}
	\mathcal{F}_{\uparrow\downarrow} (\bm{k}_A, i\mathcal{E}_n)
	\mathcal{F}_{\uparrow\downarrow}^* (\bm{k}_B, i\mathcal{E}_{n'})
	\underbrace{
		\dfrac{1}{\beta}
		\int_0^{\beta} \dd{\tau}
		e^{i\tau(\mathcal{E}_{n'} - \mathcal{E}_n + E_0)}
	}_{=\delta_{\mathcal{E}_{n'}, \mathcal{E}_n-E_0}}
	\ip{\Psi(\phi_j)}{\Psi(\phi_{j'})},
\\
&=
	2e^{i\phi_j}t^2	
	\sum_{\bm{k}_{A}\bm{k}_{B}}
	\dfrac{1}{\beta}
	\sum_{i\mathcal{E}_n}
		\mathcal{F}_{\uparrow\downarrow} (\bm{k}_A, i\mathcal{E}_n)
		\mathcal{F}_{\uparrow\downarrow}^* (\bm{k}_B, i(\mathcal{E}_n-E_0))
	\ip{\Psi(\phi_j)}{\Psi(\phi_{j'})}
	+
	\text{c.c.},
\end{align}
where $\beta$ is the Boltzmann temperature and $E_0$ is the eigenenergy of $H_0$.
We then compute the Matsubara sum by a contour integral \cite{ref:bruus}:
\begin{subequations}\begin{align}
	\dfrac{1}{\beta}
	\sum_{i\mathcal{E}_n}
	\mathcal{F}_{\uparrow\downarrow} (\bm{k}_A, i\mathcal{E}_n)
	\mathcal{F}_{\uparrow\downarrow}^* (\bm{k}_B, i(\mathcal{E}_n-E_0))
\nonumber
&=
	\dfrac{1}{\beta}
	\sum_{i\mathcal{E}_n}
	\dfrac{-\Delta}
	{(i\mathcal{E}_n)^2
		-
		\left(
			(\epsilon_{\bm{k}_A}-\mu_A)^2
			+
			\abs{\Delta}^2
		\right)
	}
	\dfrac{-\Delta}
	{(i\mathcal{E}_n-iE_0)^2
		-
		\left(
			(\epsilon_{\bm{k}_B}-\mu_B)^2
			+
			\abs{\Delta}^2
		\right)
	},
\\
&\approx
	\dfrac{\Delta^2}{4E_{\bm{k}_A}E_{\bm{k}_B}}
	\Big(
		\dfrac{1}{iE_0 + E_{\bm{k}_A} + E_{\bm{k}_B}}
		-
		\dfrac{1}{iE_0 - E_{\bm{k}_A} - E_{\bm{k}_B}}
	\Big),
\;\;\text{for }
	\beta\rightarrow\infty,
\\
&=
	\dfrac{\Delta^2}{2E_{\bm{k}_A}E_{\bm{k}_B}}
	\cdot
	\dfrac{E_{\bm{k}_A} + E_{\bm{k}_B}}
	{E_0^2 + (E_{\bm{k}_A} + E_{\bm{k}_B})^2}.
\label{eq:Matsubara_sum}
\end{align}\end{subequations}
The residues occurs at the poles 
\(
i\mathcal{E}_r
=
\pm E_{\bm{k}_A}, iE_0\pm E_{\bm{k}_B}.
\)
The number of excitation is chosen to be zero as we assume zero temperature $\beta\rightarrow\infty$ in our case.
We compute the momentum sum by integral approximation
\(
\sum_{\bm{k}=-\infty}^\infty
\rightarrow
2
\int_\Delta^{\infty}
\dd{E_{\bm{k}}}
\rho(E_{\bm{k}})
\).
After some algebra given in \cref{append:double_integral}, we obtain
\begin{align}	\label{eq:double_sum}
	\sum_{\bm{k}_A\bm{k}_B}
	\dfrac{1}{\beta}
	\sum_{i\mathcal{E}_n}
	\mathcal{F}_{\uparrow\downarrow} (\bm{k}_A, i\mathcal{E}_n)
	\mathcal{F}_{\uparrow\downarrow}^* (\bm{k}_B, i(\mathcal{E}_n-E_0))
=
	\dfrac{2N_A^eN_B^e}{b^2-1}	
	\mathcal{I}(E_0,\Delta).
\end{align}
We evaluate the double integral $\mathcal{I} (E_0, \Delta)$ in \cref{append:double_integral}.

\subsection{Double Integral $\mathcal{I} (E_0,\Delta)$}	\label{append:double_integral}
\ycl{
Given the Matsubara sum \cref{eq:Matsubara_sum} and applying the integral approximation to the momentum sums, the \cref{eq:double_sum} becomes
\begin{align}
    \sum_{\bm{k}_A\bm{k}_B}
	\dfrac{1}{\beta}
	\sum_{i\mathcal{E}_n}
	\mathcal{F}_{\uparrow\downarrow} (\bm{k}_A, i\mathcal{E}_n)
	\mathcal{F}_{\uparrow\downarrow}^* (\bm{k}_B, i(\mathcal{E}_n-E_0))
\nonumber
&=
    \int_\Delta^{\infty}    \dd{E_{\bm{k}_A}}
    \int_\Delta^{\infty}    \dd{E_{\bm{k}_B}}
    \rho(E_{\bm{k}_A})  \rho(E_{\bm{k}_B})
    \dfrac{\Delta^2}{2E_{\bm{k}_A}E_{\bm{k}_B}}
	\dfrac{E_{\bm{k}_A} + E_{\bm{k}_B}}
	{E_0^2 + (E_{\bm{k}_A} + E_{\bm{k}_B})^2},
\\\nonumber
&\approx
    \dfrac{N_A^e N_B^e}{2(b^2 - 1)}
    \int_\Delta^{\infty}   
    \int_\Delta^{\infty}    
    \dfrac{(E_{\bm{k}_A} + E_{\bm{k}_B})  \dd{E_{\bm{k}_A}} \dd{E_{\bm{k}_B}}}
        {
            \sqrt{
				(E_{\bm{k}_A}^2 - \Delta^2)
				(E_{\bm{k}_B}^2 - \Delta^2)
                }
            \big( E_0^2 + (E_{\bm{k}_A} + E_{\bm{k}_B})^2 \big)
        }
\\\nonumber
&\quad+
    \order{E_0/\Delta},
\\
&\equiv
    \dfrac{N_A^e N_B^e}{2(b^2 - 1)}
    \mathcal{I}(E_0,\Delta),
\end{align}
in which we approximate the momentum sums by a double integral $\mathcal{I} (E_0, \Delta)$.
Here we compute $\mathcal{I} (E_0, \Delta)$ in two limits: (1) $E_0 \ll \Delta$ and (2) $E_0 \gg \Delta$.
}

\subsubsection{$E_0 \ll \Delta$}
We expand the integrand around $E_0=0$ :
\small\begin{align*}
	\mathcal{I}
&\approx
	\int_\Delta^{\infty} \dd{E_{\bm{k}_A}} \hspace{-0.1cm}
	\int_\Delta^{\infty} \dd{E_{\bm{k}_B}} \hspace{-0.1cm}
	\Big(
		\dfrac{\ycl{1}}{
			\left(E_{\bm{k}_A} + E_{\bm{k}_B}\right)
			\sqrt{
				(E_{\bm{k}_A}^2 - \Delta^2)
				(E_{\bm{k}_B}^2 - \Delta^2)
		}}
	-
		\dfrac{\ycl{E_0^2}}{
			\left(E_{\bm{k}_A} + E_{\bm{k}_B}\right)^3
			\sqrt{
				(E_{\bm{k}_A}^2 - \Delta^2)
				(E_{\bm{k}_B}^2 - \Delta^2)
		}}
	\Big)
    +
	\order{\tfrac{E_0}{\Delta}}^3,
\nonumber
\\
&=
	\ycl{
    \dfrac{-\pi(E_0^2 - 16\Delta^2)}{64\Delta^3}
    }
	+
	\order{\tfrac{E_0}{\Delta}}^4,
	\quad 
	\text{for }E_0 \ll\Delta.
\end{align*}\normalsize
\end{widetext}
So,
\begin{equation} \label{eq:DBsum_small_E0}
	\mathcal{I}(E_0,\Delta)
\approx
	\dfrac{\pi^2}{\ycl{4}\Delta}
	\left(
		1
		-
		\dfrac{E_0^2}{16\Delta^2}
	\right),
\quad
	\text{for }E_0\ll\Delta.
\end{equation}

\subsubsection{$E_0 \gg \Delta$}
\begin{figure}[!]
	\centering
 	\includegraphics[width=.49\columnwidth]{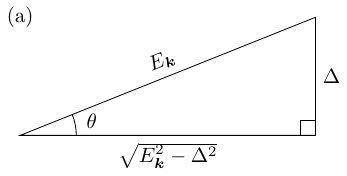}
    \includegraphics[width=.49\columnwidth]{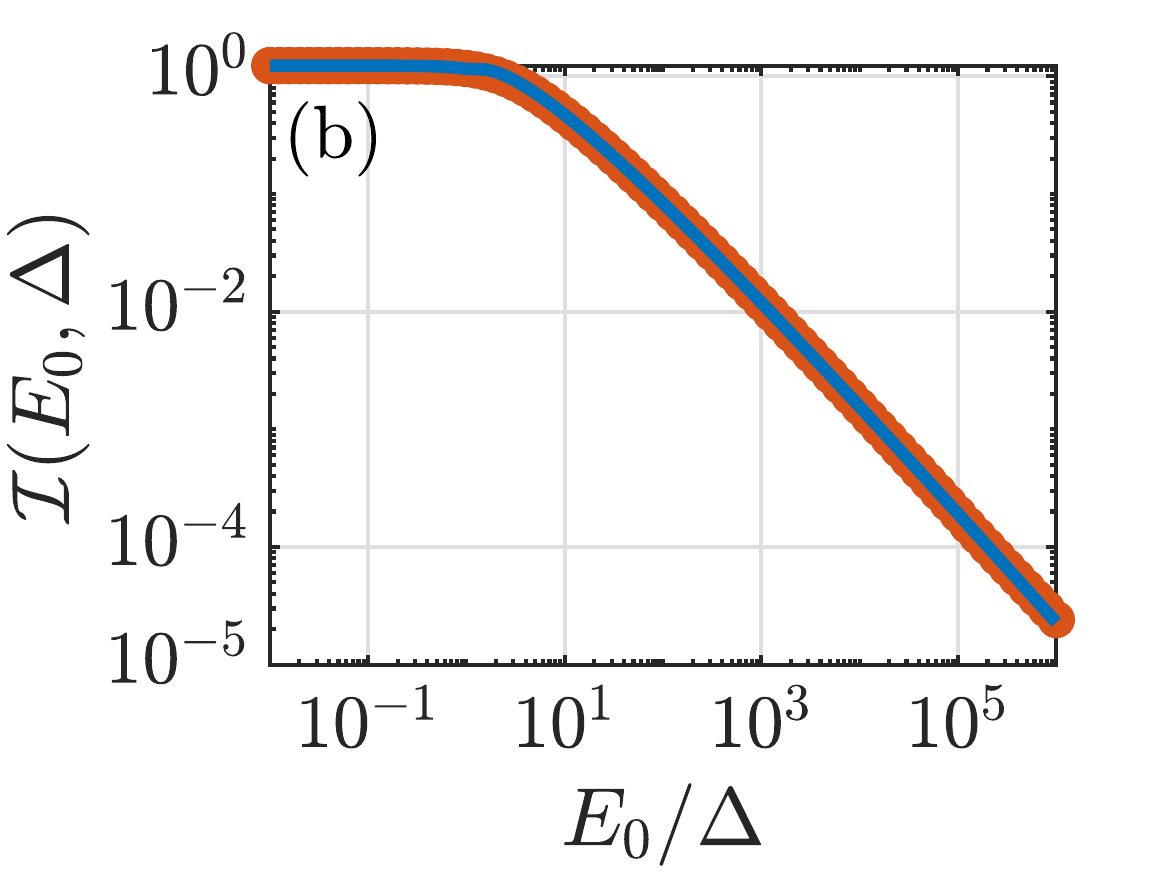}
	\caption{
        (a)
		Triangle transformation for calculating the momentum sum in $\mathcal{I}(E_0,\Delta)$.
		With this transformation, we change the integral 
		\(
		\int_0^\infty
		\dd{E_{\bm{k}}}
		\rightarrow
		\int_0^{\pi/2}
		\dd{\theta}
		(-\Delta\cot\theta\csc\theta).
		\)
        (b)
        Numerical (circles) and analytical approximation (solid line) of the double sum $\mathcal{I}(E_0,\Delta)$.
     	}
	\label{fig:triangle_transformation}
\end{figure}

We expand the integral around the limit  $E_0 \rightarrow \infty$.
Introducing the triangle transformation in \cref{fig:triangle_transformation}a with 
\mbox{$E_{\bm{k}_A} \equiv \Delta / \sin\theta_A$} and
\mbox{$E_{\bm{k}_B} \equiv \Delta / \sin\theta_B$},
the integral changes:
\(
\int_\Delta^\infty \dd{E_{\bm{k}_A}} 
\int_\Delta^\infty \dd{E_{\bm{k}_B}} 
\rightarrow
\int_0^{\pi/2} \dd{\theta_A}
	(-\Delta\cot\theta_A\csc\theta_A)
\int_0^{\pi/2} \dd{\theta_B}
	(-\Delta\cot\theta_B\csc\theta_B).
\)
Rescaling the energy by $\Delta$, we obtain
\small\begin{align}	\label{eq:DBsum_large_E0_2approx}
	\mathcal{I}(E_0,\Delta)
&=
	\int_0^{\pi/2} \int_0^{\pi/2}
	\dd\theta_A \dd\theta_B
	\dfrac{\csc^2\theta_A \csc\theta_B}
	{E_0^2 +  \Delta^2(\csc\theta_A + \csc\theta_B)^2},
\nonumber
\\
&\approx
	\dfrac{1}{\Delta}\cdot
\begin{cases}
		\int \dd\theta_B 
		\cdot1
	&\text{for }\theta_B\ll1
	\\
		\int \dd\theta_B
		\dfrac{\csc\theta_B(\pi E_0 - 2\csc\theta_A)}{2E_0^2}
		&\text{for }E_0\gg\Delta
\end{cases}
\end{align}\normalsize
A crossover approximation around $\theta_B = 1/E_0$ is dealt with by integrating the two approximations in \cref{eq:DBsum_large_E0_2approx} up to $\order{E_0}$:
\begin{align} \label{eq:DBsum_large_E0}
	\mathcal{I}(E_0,\Delta)
&\approx
	\dfrac{1}{\Delta}
	\left(
		\int_0^{1/E_0} \dd{\theta_B}
	\right.
\nonumber
\\
&\quad\quad\left.+
		\int_{1/E_0}^{\pi/2} \dd{\theta_B}
		\dfrac{
			\csc\theta_B(\pi E_0 - 2\csc\theta_B)
		}
		{2E_0^2}
	\right),
\nonumber
\\
&=
	\dfrac{\pi\ln(2E_0/\Delta)}
	{2E_0}
	+
	\order{\tfrac{E_0}{\Delta}}^3,
	\quad
	\text{for }E_0\gg\Delta.
\end{align}

Combining \cref{eq:DBsum_small_E0} and \cref{eq:DBsum_large_E0}, we obtain the double integral approximation for the double sum in \cref{eq:double_sum}:
\[
	\mathcal{I} (E_0, \Delta) =
	\begin{cases}
		\dfrac{\pi^2}{2\Delta}
		\left(
			1
			-
			\dfrac{E_0^2}{16\Delta^2}
		\right),
		&
		\text{for }E_0\ll\Delta
		\\
		\dfrac{\pi\ln(2E_0/\Delta)}{2E_0},
		&
		\text{for }E_0\gg\Delta
	\end{cases}.
\]
\Cref{fig:triangle_transformation}b shows the analytical solution $\mathcal{I} (E_0,\Delta)$ (solid line) and the numerical double sum (circle markers), indicating good agreement.

\section{Spatially-Varying BCS Theory}    \label{append:spatialBCS}
The inductance of an inductor depends on its geometry and  material properties. 
Consider a superconductor volume ``SC" in the presence of a vector potential $\bm{A}$. The Hamiltonian is given by \cite{ref:Gorkov, ref:Coleman}
\begin{align} \label{eq:Hamiltonian_SCspatial}
	H_{K_A} + H_{I_A}
\nonumber
&=
	\int_\text{SC} \dd{V}
	\big\{
		c_{\bm{r}s}^\dagger
		\big[
			\tfrac{1}{2m}
			(
				\bm{p} - e \bm{A}^\text{int}
			)^2
			-
			\mu
		\big]
		c_{\bm{r}s}
\\
&\quad-
		\abs{g}^2
		c_{\bm{r}\uparrow}^\dagger	c_{-\bm{r}\downarrow}^\dagger
		c_{-\bm{r}'\downarrow}		c_{\bm{r}'\uparrow}
	\big\}.
\end{align}
In the Hamiltonian, the electron momentum operator $\bm{p}$ acts on the fermionic Hilbert space, while the vector potential operator $\bm{A}^\text{int}$ acts on the electromagnetic Hilbert space.  However, in the low-frequency limit, the $\bm{A}^\text{int}$ field will be determined by the local current density, according to London's equation. 
The fermionic field operators in the spatial representation are the Fourier transform of those in the momentum space \cite{ref:Coleman},
\begin{equation} \label{eq:FourierTrans_fieldOPelectron}
	c_{\bm{r}s}
=
	\dfrac{1}{\sqrt{V}}
	\sum_{\bm{k}}
	e^{i\bm{k} \cdot \bm{r}}
	c_{\bm{k}s}.
\end{equation}
The matter phase of the superconductor now has position dependence, so the basis state depends on the position as well. The system projector onto the low-energy subspace becomes
\begin{equation}
    \Pi
=
    \bigotimes_{\bm{r}}    P_{\bm{r}},
\end{equation}
where $P_{\bm{r}} = \sum_{\phi(\bm{r})} \dyad{\Psi_{\bm{r}} [\phi(\bm{r})]}$ is the low-energy projector on the coordinate $\bm{r}$.
For simplicity, we consider the quasi-1D wire model,
so the projector reduces to \cref{eq:projector_inductor}, depending only on the axial coordinate $z$.
The $z$-dependent basis state can be expressed with the use of \cref{eq:FourierTrans_fieldOPelectron} and \cref{lemma:product2sum+product}:
\begin{equation}	\label{eq:BCS_position}
	\ket{\Psi_z(\phi[z])}
=
	\prod_{\bm{k}}
	\big[
		u_{\bm{k}}
		+
		v_{\bm{k}}
		\big(
			e^{i\phi(z)/2}
			c_{\bm{k}\uparrow}^\dagger
		\big)
		\big(
			e^{i\phi(z)/2}
			c_{-\bm{k}\downarrow}^\dagger
		\big)
	\big]
	\ket{0}.
\end{equation}
The derivation is given in \cref{append:spatial_BCS}.
The inclusion of the spatial-dependence gives rise to an additional phase in fermionic field operators.
After transforming the fermionic Hamiltonian \cref{eq:Hamiltonian_SCwoMFA} to the spatial representation \cref{eq:Hamiltonian_SCspatial} and constructing the new distributed projector $\Pi$ in \cref{eq:projector_inductor}, we can evaluate the inductance Hamiltonian by the projection:
\begin{align}   \label{eq:projection_inductance_mel}
    H_L 
\nonumber
&=   
    \Pi (H_{K_A} + H_{I_A}) \Pi  ,
\\\nonumber
&=
    \bigotimes_z
    P_z H P_z,
\\\nonumber
&=    
    \bigotimes_{z}    \sum_{\phi(z) \bar{\phi}(z)}
    \mel{\Psi_z [\phi(z)]}
        {(H_{K_A} + H_{I_A})}
        {\Psi_z [\bar{\phi}(z)]}
\\\nonumber
&\hspace{2.5cm}\times  
    \dyad{\Psi_z [\phi(z)]}{\Psi_z [\bar{\phi}(z)]},
\\\nonumber
&\approx
    \bigotimes_{z}    \sum_{\phi(z)}    
    \ev{(H_{K_A} + H_{I_A})}{\Psi_z [\phi(z)]}
\\
&\hspace{2.5cm}\times
    \dyad{\Psi_z [\phi(z)]}.
\end{align}
In the last line we apply the approximation
\begin{align*}
	&\mel{\Psi_z[\phi(z)]}{(H_{K_A} + H_{I_A})}{\Psi_z[\bar{\phi}(z)]}
\\
&\approx
    \mel{\Psi_z[\phi(z)]}{(H_{K_A} + H_{I_A})}{\Psi_z[\phi(z)]}
	\delta_{\phi(z),\bar{\phi}(z)},
\end{align*}
according to orthogonality of BCS basis states at each point along $z$. 
Namely, the off-diagonal matrix elements are negligible in the infinite limit $n\rightarrow\infty$.

For simplicity, we choose the gauge 
\begin{equation}
    \mathcal{G}_{\phi(z)}
= 
    e^{i\phi(z)\hat{N} / 2},
\end{equation}
such that   \mbox{
\(
	c_{zs}^\dagger
\mapsto
	\tilde{c}_{zs}^\dagger
=
	e^{-i\phi(z)/2}	c_{zs}^\dagger
\)
}
and \mbox{
\(
	c_{zs}
\mapsto
	\tilde{c}_{zs}
=
	e^{i\phi(z)/2}	c_{zs}.
\)
}
This gauge removes the $\phi(z)$-dependence on the bra and ket, i.e., 
\begin{align} 	\label{eq:kineticH_spatial_GT}
	&\ev{(H_{K_A} + H_{I_A})
	}
	{\Psi_z[\phi(z)]} \nonumber
\\
&=
    \ev{e^{-i\phi(z)/2} (H_{K_A} + H_{I_A}) e^{i\phi(z)/2}}{\Psi_z}.
\end{align}
In our quasi-1D model, the electron momentum operator is given by \mbox{$\bm{p} \mapsto -i\hbar \partial_z \bm{e}_z$} and the vector potential \mbox{$\bm{A}^\text{int} = A^\text{int}_z \bm{e}_z$}.
The derivative operators acting on the gauge transformed state are
\begin{equation}
    \partial_z  e^{i\phi(z)/2}
    c_{zs}	\ket{\Psi_z}
=
	e^{i\phi(z)/2}
		\dfrac{i}{2}  \phi'(z)
	c_{zs}	\ket{\Psi_z},
\end{equation}
and
\begin{align}
\partial_z^2 &e^{i\phi(z)/2}
 	c_{zs}	\ket{\Psi_z}
\nonumber
\\
&=
	e^{i\phi(z)/2}
	\left[
		\left(
			\dfrac{i}{2}	\phi'(z)
		\right)^2
	\hspace{-0.1cm}+
		\dfrac{i}{2}
	\right]
	c_{zs}	\ket{\Psi_z}.
\end{align}
Assuming the matter phase is adiabatic, $\phi''(z) = 0$, and choosing Coulomb gauge $\div{\bm{A}^\text{int}} = 0$, \cref{eq:kineticH_spatial_GT} reduces to
\begin{align}   \label{eq:HL_elem}
    &\ev{(H_{K_A} + H_{I_A})}{\Psi_z[\phi(z)]} 
\nonumber
\\
&=
    \dfrac{1}{2}
    \dfrac{De^2}{m}
    \int_\text{SC} \hspace{-0.1cm}  \dd{V}
    \Big(
        \dfrac{\Phi_0}{2\pi}    \phi'(z)
        -
        A^\text{int}_z
    \Big)^2
    +
    \mathcal{V}(\varphi),
\end{align}
where the electron density is defined by 
\begin{equation}
	D \equiv \ev{c_{zs}^\dagger c_{zs}}{\Psi_z}.\label{eqn:electrondensity}
\end{equation}
 and $\varphi = \bar{\phi} - \phi$.
As the projection of the interaction term gives a constant, we can choose the energy baseline to be $\Pi H_{I_A} \Pi = \mathcal{V}(\varphi) = 0$. 
The projection then becomes
\begin{align}   \label{eq:HL_integral}
    \Pi H_{K_A} \Pi
\nonumber
&\approx
    \bigotimes_z    \sum_{\phi(z)}
    \dfrac{1}{2}
    \dfrac{De^2}{m}
    \int_\text{SC}  \dd{V}
    \Big(
        \dfrac{\Phi_0}{2\pi}    \phi'(z)
        -
        A^\text{int}_z
    \Big)^2
\\\nonumber
&\hspace{3cm}\times
    \dyad{\Psi_z [\phi(z)]},
\\
&=
    \dfrac{1}{2}
    \dfrac{1}{\mu_0\lambda_L^2}
    \int_\text{SC}  \dd{V}
    \Big(
        \dfrac{\Phi_0}{2\pi}    \hat{\phi}'(z)
        -
        A^\text{int}_z (\hat{i}_s)
    \Big)^2,
\end{align}
in which we define the London penetration depth $\lambda_L$ via $\frac{De^2}{m} = \frac{1}{\mu_0\lambda_L^2}$.
Considering a cylindrical superconducting wire, the volume integral over the superconducting device is $\int_\text{SC} \dd{V} = \int_0^{2\pi} \dd{\theta} \int_0^l \dd{z} \int_0^{r_w} r\dd{r}$, and $A^\text{int}_z (\hat{i}_s)$ is given in \cref{eq:Afield_int}.
With the current operator \mbox{$\hat{i}_s = \frac{\Phi_0}{2\pi}\hat{\phi}'(z) / \bar{L}$} adopted from \cref{eq:supercurrentOP}, and applying the lumped element approximation, \mbox{$\hat{\phi}'(z) = (\hat{\phi}_B - \hat{\phi}_A) / l = -\hat{\phi} / l$}, \cref{eq:HL_integral} becomes
\begin{widetext}\begin{align}   \label{eq:Pi.HKA.Pi}
    \Pi H_{K_A} \Pi
\nonumber
&=
    \dfrac{1}{2}    \Big(\dfrac{\Phi_0}{2\pi}\Big)^2
    \dfrac{1}{\mu_0\lambda_L^2} 
    2\pi
    \underbrace{
        \int_0^l \dd{z} \hat{\phi}'^2(z)
    }_{=\hat{\phi}^2 / l}
    \int_0^{r_w}    r\dd{r}
    \Big(
        1
        -
        \dfrac{\mu_0 \lambda_L}{2\pi r_w \bar{L}}
        \dfrac{1 - I_0(r/\lambda_L)}{I_1(r_w/\lambda_L)}
    \Big)^2,
\\\nonumber
&=
    \dfrac{1}{2}    \Big(\dfrac{\Phi_0}{2\pi}\Big)^2
    \dfrac{2\pi}{\mu_0 l}
    \dfrac{\hat{\phi}^2}{\lambda_L^2}
    \Bigg\{
        \dfrac{r_w^2}{2}
        +
        \dfrac{\mu_0l}{2\pi L}
        \Bigg[
            \lambda_L^2
            \Big(
                \dfrac{\mu_0l}{4\pi L}
                \Big(
                    \dfrac{1 + I_0^2(r_w/\lambda_L)}{I_1^2(r_w/\lambda_L)}
                    -
                    1
                \Big)
                +
                2
            \Big)
            -
            \dfrac{1}{I_1(r_w/\lambda_L)}
            \Big(
                r_w \lambda_L
                -
                \dfrac{\mu_0 l}{\pi L}
                \dfrac{\lambda_L^3}{r_w}
            \Big)
        \Bigg]
    \Bigg\},
\\
&\approx
    \dfrac{1}{2}    \Big(\dfrac{\Phi_0}{2\pi}\Big)^2
    \hat{\phi}^2
    \dfrac{1}{L},
\quad
    \text{for }\lambda_L \ll r_w,
\end{align}\end{widetext}
where the $I_1^{-1}(r_w/\lambda_L)$ is exponentially suppressed and $(1 + I_0^2(r_w/\lambda_L)) / I_1^2(r_w/\lambda_L) \approx 1$ in the limit $r_w \gg \lambda_L$.

Combining these results we find
\begin{equation}
    H_L
=
    \Pi (H_{K_A} + H_{I_A}) \Pi
\approx
    \dfrac{E_L}{2}  \hat{\phi}^2,
\end{equation}
where the inductance energy of the system $E_L$ consists of the kinetic and geometric contributions:
$E_L = (\frac{\Phi_0}{2\pi})^2 / L$, where $L = L_K + L_G$.

\subsection{Spatial-Dependent BCS State \& Projections}	\label{append:spatial_BCS}
The BCS state changes when we consider a spatial-varying matter phase $\phi(\bm{r})$.
Here we consider the case in which the matter phase depends on all three coordinates of the space. 
To derive the BCS state in position representation, we require the Fourier's transform \cref{eq:FourierTrans_fieldOPelectron} and \cref{lemma:product2sum+product}.
With the substitution
\(
	u_{\bm{k}}
\rightarrow
	A_k,
\)
\(
	v_{\bm{k}} e^{i\phi}
	\int \dd{\bm{r}_{\bm{k}}}	\int \dd{\bm{r}'_{\bm{k}}}
	e^{i \bm{k} \cdot (\bm{r}_{\bm{k}} - \bm{r}'_{\bm{k}})}
	c_{\bm{r}_{\bm{k}}\uparrow}^\dagger	c_{-\bm{r}'_{\bm{k}}\downarrow}^\dagger
\rightarrow
	B_k,
\)
we obtain
\begin{widetext}
\begin{subequations}\begin{align}
	\ket{\Psi(\phi)}
\nonumber
&=
	\prod_{\bm{k}}
	\left(
		u_{\bm{k}}
		+
		v_{\bm{k}}	e^{i\phi}
		c_{\bm{k}\uparrow}^\dagger	c_{-\bm{k}\downarrow}^\dagger
	\right)	\ket{0},
\\\nonumber
&=
	\prod_{\bm{k}}
	\left(
		u_{\bm{k}}
		+
		v_{\bm{k}}	e^{i\phi}
		\int \dd{\bm{r}_{\bm{k}}}	\int \dd{\bm{r}'_{\bm{k}}}
		e^{i \bm{k} \cdot (\bm{r}_{\bm{k}} - \bm{r}'_{\bm{k}})}
		c_{\bm{r}_{\bm{k}}\uparrow}^\dagger	c_{-\bm{r}'_{\bm{k}}\downarrow}^\dagger
	\right)	\ket{0},
\\\nonumber
&=
	\sum_{m=0}^n
	\sum_{\chi\in\mathcal{S}_m}
	\prod_{\bm{k}_i\in\chi^c}	u_{\bm{k}_i}
	\prod_{\bm{k}_j\in\chi}
	v_{\bm{k}_j}	e^{i\phi}
	\int \dd{\bm{r}_{\bm{k}_j}}	\int \dd{\bm{r}'_{\bm{k}_j}}
	e^{i \bm{k}_j \cdot (\bm{r}_{\bm{k}_j} - \bm{r}'_{\bm{k}_j})}
	c_{\bm{r}_{\bm{k}_j}\uparrow}^\dagger	c_{-\bm{r}'_{\bm{k}_j}\downarrow}^\dagger
	\ket{0},
\\\nonumber
&=
	\sum_{m=0}^n
	\sum_{\chi\in\mathcal{S}_m}
	\prod_{\bm{k}_i\in\chi^c}	u_{\bm{k}_i}
	\prod_{\bm{k}_j\in\chi}
	v_{\bm{k}_j}	
	\int \dd{\bm{r}_{\bm{k}_j}}	\int \dd{\bm{r}'_{\bm{k}_j}}
	e^{i \bm{k}_j \cdot (\bm{r}_{\bm{k}_j} - \bm{r}'_{\bm{k}_j})}
	\left(
		e^{i\phi/2}	c_{\bm{r}_{\bm{k}_j}\uparrow}^\dagger	
	\right)
	\left(
		e^{i\phi/2}	c_{-\bm{r}'_{\bm{k}_j}\downarrow}^\dagger
	\right)
	\ket{0},
\\
&\rightarrow
	\sum_{m=0}^n
	\sum_{\chi\in\mathcal{S}_m}
	\prod_{\bm{k}_i\in\chi^c}	u_{\bm{k}_i}
	\prod_{\bm{k}_j\in\chi}
	v_{\bm{k}_j}	
	\int \dd{\bm{r}_{\bm{k}_j}}	\int \dd{\bm{r}'_{\bm{k}_j}}
	e^{i \bm{k}_j \cdot (\bm{r}_{\bm{k}_j} - \bm{r}'_{\bm{k}_j})}
	\left(
		e^{i\phi(\bm{r})/2}	c_{\bm{r}_{\bm{k}_j}\uparrow}^\dagger	
	\right)
	\left(
		e^{i\phi(\bm{r})/2}	c_{-\bm{r}'_{\bm{k}_j}\downarrow}^\dagger
	\right)
	\ket{0},
\label{eq:BCS_spatial_phase}
\\
&=
	\prod_{\bm{k}}
	\left[
		u_{\bm{k}}
		+
		v_{\bm{k}}
		\left(
			e^{i\phi(\bm{r})/2}
			c_{\bm{k}\uparrow}^\dagger
		\right)
		\left(
			e^{i\phi(\bm{r})/2}
			c_{-\bm{k}\downarrow}^\dagger
		\right)
	\right]
	\ket{0},
\\\nonumber
&\equiv
	\ket{\Psi[\phi(\bm{r})]}.
\end{align}\end{subequations}
In Eq.~\ref{eq:BCS_spatial_phase}, we generalise the matter phase from constant to position-dependent,
\(
	\phi
\rightarrow
	\phi(\bm{r}).
\)
Hence we express the BCS ground state in spatial representation.

The additional phase terms in \cref{eq:BCS_spatial_phase} vanish under the gauge transformation $\mathcal{G}_{\phi(\bm{r})} = e^{i\phi(\bm{r})\hat{N} / 2}$.
Therefore, the electron kinetic energy with respect to the ground state reads
\begin{subequations}\label{eq:kineticH_spatial_GT_detail}\begin{align} 
		\Big\langle
			\int \dd{\bm{r}}
			&c_{\bm{r}s}^\dagger
			\left[
				\dfrac{1}{2m}
				\left(
					i\hbar\grad{} - q \bm{A}
				\right)^2
				-
				\mu
			\right]
			c_{\bm{r}s}
		\Big\rangle
\nonumber
\\
&=
	\ev{
		\int \dd{\bm{r}}
		c_{\bm{r}s}^\dagger
		\left[
			\dfrac{1}{2m}
			\left(
				i\hbar\grad{} - q \bm{A}
			\right)^2
			-
			\mu
		\right]
		c_{\bm{r}s}
	}
	{\Psi[\phi(\bm{r})]},
\\\nonumber
&=
	\bra{0}
	\sum_{m=0}^n
	\sum_{\chi\in\mathcal{S}_m}
	\prod_{\bm{k}_i\in\chi^c}	u_{\bm{k}_i}
	\prod_{\bm{k}_j\in\chi}
	v_{\bm{k}_j}	
	\int \dd{\bm{r}_{\bm{k}_j}}	\int \dd{\bm{r}'_{\bm{k}_j}}
	e^{-i \bm{k}_j \cdot (\bm{r}_{\bm{k}_j} - \bm{r}'_{\bm{k}_j})}
	\left(
		e^{-i\phi(\bm{r})/2}	c_{-\bm{r}_{\bm{k}_j}\downarrow}
	\right)
	\left(
		e^{-i\phi(\bm{r})/2}	c_{\bm{r}'_{\bm{k}_j}\uparrow}
	\right)
\\\nonumber
&\quad\times
	\int\dd{\bm{r}}
	c_{\bm{r}s}^\dagger
	\left[
	\dfrac{1}{2m}	
	\left(
	i\hbar\grad{} - q \bm{A}
	\right)^2
	-
	\mu
	\right]
	c_{\bm{r}s}
	\sum_{m'=0}^n
	\sum_{\chi'\in\mathcal{S}'_m}
	\prod_{\bm{k}'_i\in\chi^c}	u_{\bm{k}'_i}
	\prod_{\bm{k}'_j\in\chi}
	v_{\bm{k}'_j}	
	\int \dd{\bm{r}''_{\bm{k}'_j}}	\int \dd{\bm{r}'''_{\bm{k}'_j}}
\\\nonumber
&\quad\times
	e^{i \bm{k}'_j \cdot (\bm{r}''_{\bm{k}'_j} - \bm{r}'''_{\bm{k}'_j})}
	\left(
	e^{i\phi(\bm{r})/2}	c_{\bm{r}''_{\bm{k}'_j}\uparrow}^\dagger
	\right)
	\left(
	e^{i\phi(\bm{r})/2}	c_{-\bm{r}'''_{\bm{k}'_j}\downarrow}^\dagger
	\right)
	\ket{0},
\\\nonumber
&=
	\bra{0}
	\sum_{m=0}^n	\sum_{\chi\in\mathcal{S}_m}
	\prod_{\bm{k}_i\in\chi^c}	u_{\bm{k}_i}
	\prod_{\bm{k}_j\in\chi}		v_{\bm{k}_j}	
	\int \dd{\bm{r}_{\bm{k}_j}}	\int \dd{\bm{r}'_{\bm{k}_j}}
	e^{-i \bm{k}_j \cdot (\bm{r}_{\bm{k}_j} - \bm{r}'_{\bm{k}_j})}
\\\nonumber
&\quad\times
	\mathcal{G}_{\phi(\bm{r})}^\dagger
	\underbrace{
		\mathcal{G}_{\phi(\bm{r})}
		\left(
		e^{-i\phi(\bm{r})/2}	c_{-\bm{r}_{\bm{k}_j}\downarrow}
		\right)
		\mathcal{G}_{\phi(\bm{r})}^\dagger
	}_{= c_{-\bm{r}_{\bm{k}_j}\downarrow}}
	\underbrace{\mathcal{G}_{\phi(\bm{r})}
		\left(
		e^{-i\phi(\bm{r})/2}	c_{\bm{r}'_{\bm{k}_j}\uparrow}
		\right)
		\mathcal{G}_{\phi(\bm{r})}^\dagger
	}_{= c_{\bm{r}'_{\bm{k}_j}\uparrow}}	
	\mathcal{G}_{\phi(\bm{r})}
\\\nonumber
&\quad\times
	\int\dd{\bm{r}}
	\mathcal{G}_{\phi(\bm{r})}^\dagger
	\underbrace{
		\mathcal{G}_{\phi(\bm{r})}	c_{\bm{r}s}^\dagger
		\mathcal{G}_{\phi(\bm{r})}^\dagger
	}_{= e^{i\phi(\bm{r})/2} c_{\bm{r}s}^\dagger}
	\mathcal{G}_{\phi(\bm{r})}
	\left[
	\dfrac{1}{2m}	
	\left(
	i\hbar\grad{} - q \bm{A}
	\right)^2
	-
	\mu
	\right]
	\mathcal{G}_{\phi(\bm{r})}^\dagger
	\underbrace{
		\mathcal{G}_{\phi(\bm{r})} c_{\bm{r}s}
		\mathcal{G}_{\phi(\bm{r})}^\dagger
	}_{= e^{i\phi(\bm{r})/2} c_{\bm{r}s}}
	\mathcal{G}_{\phi(\bm{r})}
\\\nonumber
&\quad\times
	\sum_{m'=0}^n	\sum_{\chi'\in\mathcal{S}'_m}
	\prod_{\bm{k}'_i\in\chi^c}	u_{\bm{k}'_i}
	\prod_{\bm{k}'_j\in\chi}	v_{\bm{k}'_j}	
	\int \dd{\bm{r}''_{\bm{k}'_j}}	\int \dd{\bm{r}'''_{\bm{k}'_j}}
	e^{i \bm{k}'_j \cdot (\bm{r}''_{\bm{k}'_j} - \bm{r}'''_{\bm{k}'_j})}
\\\nonumber
&\quad\times
	\mathcal{G}_{\phi(\bm{r})}^\dagger	
	\underbrace{\mathcal{G}_{\phi(\bm{r})}
		\left(
		e^{i\phi(\bm{r})/2}	c_{\bm{r}''_{\bm{k}'_j}\uparrow}^\dagger
		\right)
		\mathcal{G}_{\phi(\bm{r})}^\dagger
	}_{= c_{\bm{r}''_{\bm{k}'_j}\uparrow}^\dagger}	
	\underbrace{\mathcal{G}_{\phi(\bm{r})}
		\left(
		e^{i\phi(\bm{r})/2}	c_{-\bm{r}'''_{\bm{k}'_j}\downarrow}^\dagger
		\right)
		\mathcal{G}_{\phi(\bm{r})}^\dagger
	}_{c_{-\bm{r}'''_{\bm{k}'_j}\downarrow}^\dagger}	
	\mathcal{G}_{\phi(\bm{r})}
	\ket{0},
\\\nonumber
&=
	\bra{0}
	\sum_{m=0}^n	\sum_{\chi\in\mathcal{S}_m}
	\prod_{\bm{k}_i\in\chi^c}	u_{\bm{k}_i}
	\prod_{\bm{k}_j\in\chi}		v_{\bm{k}_j}	
	\int \dd{\bm{r}_{\bm{k}_j}}	\int \dd{\bm{r}'_{\bm{k}_j}}
	e^{-i \bm{k}_j \cdot (\bm{r}_{\bm{k}_j} - \bm{r}'_{\bm{k}_j})}
	\mathcal{G}_{\phi(\bm{r})}^\dagger	c_{-\bm{r}_{\bm{k}_j}\downarrow}	c_{\bm{r}'_{\bm{k}_j}\uparrow}
\\\nonumber
&\quad\times
	\dfrac{1}{2m}	\int\dd{\bm{r}}
	e^{-i\phi(\bm{r})/2}
	c_{\bm{r}s}^\dagger
	\left[
	\left(
	i\hbar\grad{} - q \bm{A}
	\right)^2
	-
	\mu
	\right]
	e^{i\phi(\bm{r})/2}
	c_{\bm{r}s}
\\\nonumber
&\quad\times
	\sum_{m'=0}^n
	\sum_{\chi'\in\mathcal{S}'_m}
	\prod_{\bm{k}'_i\in\chi^c}	u_{\bm{k}'_i}
	\prod_{\bm{k}'_j\in\chi}
	v_{\bm{k}'_j}	
	\int \dd{\bm{r}''_{\bm{k}'_j}}	\int \dd{\bm{r}'''_{\bm{k}'_j}}
	e^{i \bm{k}'_j \cdot (\bm{r}''_{\bm{k}'_j} - \bm{r}'''_{\bm{k}'_j})}
	c_{\bm{r}''_{\bm{k}'_j}\uparrow}^\dagger
	c_{-\bm{r}'''_{\bm{k}'_j}\downarrow}^\dagger
	\mathcal{G}_{\phi(\bm{r})}\ket{0},
\\\nonumber
&=
	\ev{
		\dfrac{1}{2m}	\int\dd{\bm{r}}
		e^{-i\phi(\bm{r})/2}
		c_{\bm{r}s}^\dagger
		\left[
		\left(
		i\hbar\grad{} - q \bm{A}
		\right)^2
		-t
		\mu
		\right]
		e^{i\phi(\bm{r})/2}
		c_{\bm{r}s}
	}
	{\Psi(\phi=0)},
\\
&=
	\ev{
		\dfrac{1}{2m}	\int\dd{\bm{r}}
		e^{-i\phi(\bm{r})/2}
		c_{\bm{r}s}^\dagger
		\left[
		\left(
		i\hbar\grad{} - q \bm{A}
		\right)^2
		-
		\mu
		\right]
		e^{i\phi(\bm{r})/2}
		c_{\bm{r}s}
	}
	{\Psi}.
\end{align}\end{subequations}\end{widetext}\normalsize
Note that the derivative operators $\gradient$, $\laplacian$ act on the position space, and the field operators $c_{\bm{r}s}^\dagger$, $c_{\bm{r}s}$ act on the fermionic Fock state.

\section{Vector potential of a Superconducting Wire} \label{append:Afield_SCwire}
In this appendix, we solve the London equation for a superconducting wire in cylindrical coordinates and compute the magnetic field inside the wire. The system can be referred to \cref{fig:SC_straightWire}.
We then integrate the flux around the wire out to a cutoff distance, and calculate the geometric inductance as the magnetic field is linked to the current.
In addition, we also compute the $\bm{A}^\text{int}$-field inside the superconductor induced by the supercurrent.
This field associates to the change of the electromotive force inside the metal and gives rise to the kinetic inductance.

With Ampére's law, the London equation is expressed as
\(
    \curl{\curl{\bm{H}^\text{int}}} + \bm{H}^\text{int} / \lambda_L^2
    =
    0,
\)
where $\bm{B}^\text{int} = \mu_0 \bm{H}^\text{int}$ and $\lambda_L$ is London penetration depth.
Considering a circular superconducting wire with radius $r_w$, we have $\bm{H}^\text{int} (r) = H_\theta(r) \bm{e}_\theta$ according to the cylindrical symmetry.
The fields inside the metal must be finite and the boundary condition guarantees $\bm{H}^\text{int} (r=r_w) = H_0 \bm{e}_\theta$, so the solution reads
\begin{equation}    \label{eq:Hfield_SCwire}
    \bm{H}^\text{int} (r)
=
    H_0
    \dfrac{I_1 (r/\lambda_L)}{I_1 (r_w/\lambda_L)}
    \bm{e}_\theta,
\end{equation}
where $I_\alpha (z)$ is the modified Bessel function of the first kind, and $H_0$ is the magnetic field at the surface of the cylinder. 
From Maxwell's equation (with zero electric field) $\curl{\bm{H}}^\text{int} = \bm{J}$, the supercurrent density is along the axial direction:
\begin{equation}    \label{eq:supercurrent_density}
    \bm{J} (r)
=
    \dfrac{H_0}{\lambda_L}
    \dfrac{I_0 (r/\lambda_L)}{I_1 (r_w/\lambda_L)}
    \bm{e}_z = J_z(r) \bm{e}_z.
\end{equation}
The supercurrent $\bm{J}$ in \cref{eq:supercurrent_density} is exponentially suppressed at the centre, but is not identically zero.
The maximal values is at the surface $\bm{J}_{\text{max}} = \bm{J}(r_w) = \frac{H_0}{\lambda_L}\bm{e}_z$.
The total current flowing through the wire is the surface integral over the cross section,
\begin{equation}    \label{eq:supercurrent}
    i_s
=
    \int_{\bm{\sigma}}
    \bm{J}  \vdot \dd{\bm{\sigma}}
=
    2\pi \int_0^{r_w} dr
    \,r \, J_z(r)
=
    2\pi r_w H_0,
\end{equation}
such that the $\bm{B}^\text{ext}$-field outside the wire is given by
\begin{equation}
    \bm{B}^\text{ext} (r)
=
    \frac{\mu_0 i_s}{2\pi r} \bm{e}_\theta
=
    \frac{\mu_0 r_w H_0}{r}  \bm{e}_\theta.
\end{equation}
The corresponding vector potential outside the metal is 
\begin{equation}
    \bm{A}^\text{ext} (r)
=
    \frac{\mu_0 i_s}{2\pi}
    \ln{
        \frac{\sqrt{l^2 + r^2} + l}{r}
    }
    \bm{e}_z.
\end{equation}

\paragraph*{Gauge Invariant Vector Potential inside a Superconductor}
We choose a gauge in which the internal vector potential is parallel to the supercurrent \cite{ref:Tinkham_superconductivity} and satisfies $\bm{H}^\text{int} = \curl{\bm{A}}^\text{int} / \mu_0$.
Solving for the current density from Maxwell's equation $\bm{J} = \curl{\curl{\bm{A}^\text{int}}} / \mu_0$, gives
\begin{align}   \label{eq:Afield_int}
    \bm{A}^\text{int} (r)
&=
        H_0 \mu_0 \lambda_L 
        \dfrac{1 - I_0(r / \lambda_L)}{I_1(r_w/\lambda_L)}
    \bm{e}_z,\nonumber
\\
&=
       i_s \frac{ \mu_0 \lambda_L }{2\pi r_w} 
        \dfrac{1 - I_0(r / \lambda_L)}{I_1(r_w/\lambda_L)}
    \bm{e}_z,\nonumber
\\
&\approx
       -i_s\frac{\mu_0 \lambda_L  e^{({r-r_w})/{\lambda_L }}}{2 \pi  \sqrt{r_w r}} 
       \bm{e}_z
\quad
    \text{ for } \lambda_L \ll r_w. 
\end{align} 

We obtain the gauge invariant vector potential inside the superconductor:
\begin{equation}
    \tilde{\bm{A}}^{\text{int}} (r)
=
    \bm{A}^{\text{int}} (r) 
    +
    \dfrac{\Phi_0}{2\pi}
    \grad{\hat{\phi}},
\end{equation}
where the scalar $\phi$-field is self-generated by the superconductor.

\subsection{Classical Flux}
The total inductance of the system consists of kinetic and geometric contributions. Below we compute the two terms from the viewpoint of flux.

\subsubsection{Flux outside the Device}
We can obtain $L_G$ by integrating the  flux outside the wire, induced by the current $\Phi_\text{ext}$, which is given by
\begin{equation}    \label{eq:flux_induced}
    \Phi_\text{ext}
=
    l \int_{r_w}^R  \dd{r}
    \abs{\bm{B}^{\rm ext} (r)}
=
    \underbrace{
        \dfrac{\mu_0 l \ln(R/r_w)}{2\pi}
    }_{=L_G}
    \cdot
    i_s.
\end{equation}

\subsubsection{Flux inside the superconductor}
Here we consider the flux inside the superconductor, which gives $\Phi_\text{int} = i_sL_K$.
By placing a superconductor in an magnetic field, the field penetrates the metal to a small depth, called the London penetration depth $\lambda_L$. 
The electrons  within this region move to induce the Meissner effect, and giving  rise to a kinetic inductance.

\paragraph*{2D model}
For the cylindrical wire model, the internal flux passing the effective surface within the wire is
\begin{align}
    \Phi_\text{int}
\nonumber
&=
    \int_{\bm{S}}   \bm{B}^{\text{int}} \vdot \dd{\bm{S}},
\\\nonumber
&=
    \dfrac{\mu_0 i_s l}{2\pi r_w}
    \dfrac{1}{I_1(r_w/\lambda_L)}
    \int_{\lambda_L}^{r_w}  \dd{r} I_1(r/\lambda_L),
\\\nonumber
&=
    \dfrac{\mu_0 i_s l}{2\pi r_w}
    \dfrac{\lambda_L \left(I_0(r_w/\lambda_L) - I_0(1) \right)}{I_1(r_w/\lambda_L)},
\\
&\approx
    \dfrac{\mu_0 i_s l \lambda_L}{2\pi r_w},
\quad
    \text{for }\lambda_L \ll r_w.
\end{align}
Given the London penetration depth $\lambda_L = \sqrt{m/(\mu_0 De^2)}$ and the effective cross-section area $\sigma_{\text{eff}} \approx 2\pi r_w \lambda_L$, the $\Phi_\text{int}$ can be written as
\begin{equation}    \label{eq:flux_internal2D}
    \Phi_\text{int}
=
    i_s \cdot
    \underbrace{
        \left(
            \dfrac{m}{De^2} 
            \dfrac{l}{\sigma_{\text{eff}}}
        \right)
    }_{\equiv L_K}.
\end{equation}

\paragraph*{1D model}
Approximating the superconducting wire to a point-like wire, the internal magnetic field is given by
 $
    \bm{B}^{\text{int}}
=
    \bm{B}^{\text{int}} (r=r_w)
=
    \frac{\mu_0 i_s}{2\pi r_w}
    \bm{e}_\theta,
$
so the flux enclosed reads
\begin{equation}    \label{eq:flux_internal1D}
    \Phi_\text{int}
=
    \int_{\bm{\sigma}_{\text{eff}}}
    \bm{B}^{\text{int}}
    \vdot   \dd{\bm{\sigma}}
=
    \dfrac{\mu_0 i_s l \lambda_L}{2\pi r_w}
=  
    L_K
    i_s.
\end{equation}
Results in both \eqref{eq:flux_total} and \eqref{eq:inductance_total} give the magnetic flux inside the superconducting wire, which suggests that we can link the two models by approximating the cross-section $\sigma \mapsto \sigma_{\text{eff}}$.

\subsubsection{Total Flux}
The total flux passing through a  surface from the wire core out to the distance $R$ is the sum
\begin{equation} \label{eq:flux_total}
    \Phi_{\text{tot}}
=
    \Phi_{\text{int}}   +   \Phi_{\text{ext}}
=
    \left( L_K + L_G \right) i_s,
\end{equation}
which is the bias magnetic flux of the device: $\Phi_\text{tot} = \Phi_b$.
\cref{eq:flux_total} implies that the total inductance of the system is 
\begin{equation}    \label{eq:inductance_total}
    L
=
    L_K + L_G,
\end{equation}
i.e., the inductors in series.
Results \cref{eq:flux_total} and \cref{eq:inductance_total} can be generalised to any shapes of the superconducting device.

\subsection{Classical Energies}
Here we compute the inductance from the energy and compare the results to the viewpoint of flux.

\subsubsection{Energy of the Field outside the Superconductor}
The electromagnetic energy stored in $\\bm{B}^\text{ext}$-field is given by
\begin{align}   \label{eq:EMenergy_Bind}
    E_{\bm{B}_\text{ext}}
=
    \dfrac{1}{2\mu_0}
    \int  \dd{V}
    \abs{\bm{B}^\text{ext}}^2
=  
    \dfrac{i_s^2}{2}
    \dfrac{\mu_0 l\ln(R/r_w)}{2\pi}.
\end{align}
As the field can transfer energy to the inductor by the current, we obtain the geometric inductance as
\begin{equation}    \label{eq:inductance_geometric}
    L_G
=
    \dfrac{\mu_0 l\ln(R/r_w)}{2\pi}.
\end{equation}

\subsubsection{Energies of the inside Superconductor}
The energy inside the wire, including the electron kinetic energy and the field energy,  contributes to the kinetic inductance.
\paragraph*{2D model}
In the presence of $\bm{A}^\text{int}$, 
the kinetic energy of all electrons is described by the Hamiltonian
\begin{equation}
    H_{\text{el}}^{(T)}
=
    \int_{\text{SC}}    \dd{V}
    c_{\bm{r}s}^\dagger \dfrac{\hat{\bm{p}}_{\text{tot}}^2}{2m} c_{\bm{r}s},
\end{equation}
from which we can obtain the kinetic energy with respect to the ground state by $E_{\text{el}}^{(T)} = \langle \Psi | H_{\text{el}}^{(T)} | \Psi\rangle$:
\begin{align}   \label{eq:kinetic_energy_electronInSC}
    E_{\text{el}}^{(T)}
\nonumber
&=
    \dfrac{m}{2D^2e^2}
    \int_{\text{SC}}    \dd{V}
    \hat{\bm{J}}^2
    \ev{c_{\bm{r}s}^\dagger c_{\bm{r}s}}{\Psi},
\\\nonumber
&=
    \dfrac{m}{De^2} \dfrac{H_0^2}{2\lambda_L^2 I_1^2(r_w/\lambda_L)}
    \cdot
    l   \int_0^{2\pi}   \dd{\theta}
    \int_0^{r_w}    r\dd{r} I_0^2(r/\lambda_L),
\\\nonumber
&=
    \dfrac{i_s^2}{2}
    \dfrac{m}{De^2} 
    \dfrac{l}{4\pi\lambda_L^2}
        \dfrac{
            I_0^2(r_w/\lambda_L) - I_1^2(r_w/\lambda_L)
        }
        {I_1^2(r_w/\lambda_L)},
\\
&\approx
    \dfrac{i_s^2}{2}
    \dfrac{m}{De^2} 
    \dfrac{l}{\sigma_{\text{eff}}}
    \dfrac{1}{2},
\quad
    \text{for }\lambda_L \ll r_w
\end{align}
in which we use $\sigma_{\text{eff}} \approx 2 \pi r_w \lambda_L$ and adopt $H_0 = \frac{i_s}{2\pi r_w}$ from \cref{eq:supercurrent}.

In addition to the electron kinetic energy, the other source for kinetic inductance is the energy of field inside the superconductor, $E_{\bm{B}^{\rm int}}$:
\begin{align}   \label{eq:Bint_energy}
    E_{\bm{B}^{\rm int}}
\nonumber
&=
    \dfrac{1}{2\mu_0}    \int_\text{SC}  \dd{V}
    \abs{E_{\bm{B}^{\rm int}}}^2,
\\\nonumber
&=
    \dfrac{\mu_0 i_s^2 l}{8\pi r_w}
    \dfrac{1}{I_1^2(r_w/\lambda_L)}
    \cdot
    l \int_0^{2\pi} \dd{\theta}
    \int_0^{r_w}    r\dd{r} I_1^2(r/\lambda_L),
\\\nonumber
&=
    \dfrac{i_s^2}{2}
    \dfrac{\mu_0l}{4\pi r_w}
    \left[
        r_w
        -
        r_w \dfrac{I_0^2(r_w/\lambda_L)}{I_1^2(r_w/\lambda_L)}
        +
        2\lambda_L  \dfrac{I_0(r_w/\lambda_L)}{I_1(r_w/\lambda_L)}
    \right],
\\\nonumber
&\approx
    \dfrac{i_s^2}{2}
    \dfrac{\mu_0l\lambda_L}{4\pi r_w},
\quad
    \text{for }\lambda_L \ll r_w,
\\
&=
    \dfrac{i_s^2}{2}
    \dfrac{m}{De^2} \dfrac{l}{\sigma_\text{eff}}
    \dfrac{1}{2},
\quad
    \because 
    \lambda_L = \sqrt{\dfrac{m}{\mu_0De^2}}.
\end{align}
The equally distributed energy in $E_\text{el}^{(T)}$ and $E_{\bm{B}^\text{int}}$ is the result from the equalpartition theorem in thermal equilibrium, and the sum of them gives\mbox{
\(
    E_\text{el}^{(T)} + E_{\bm{B}^\text{int}}
=
    \frac{1}{2} L_K i_s^2
\)}.
Therefore the kinetic inductance is given by
\begin{equation}    \label{eq:inductance_kinetic}
    L_K
=
    \dfrac{m}{De^2} \dfrac{l}{\sigma_{\text{eff}}}.
\end{equation}

\paragraph*{1D model}

On the other hand, the 1D wire carries a constant supercurrent density \mbox{$\bm{J} = H_0 / \lambda_L = i_s / (2\pi r_w\lambda_L)$}, and the wire has the volume of $\sigma_\text{eff} l = 2\pi r_w \lambda_L l$.
Following the same procedures above, the kinetic energy of the electrons is given by
\begin{align}   \label{eq:KenergyEl_SCcylindrical}
    E_{\text{el}}^{(T)}
&=
    \dfrac{i_s^2}{2}
    \underbrace{
        \dfrac{m}{De^2}
        \dfrac{l}{\sigma_{\text{eff}}},
    }_{=L_K}
\end{align}
which is twice of that of the 2D wire.
Since the vector potential inside the quasi-1D wire is a constant, the magnetic field, which is the curl of the vector potential, is zero.
Namely, the field energy inside the 1D superconducting wire is zero as well.
Hence the only contribution to the kinetic inductance comes from the electrons.

\subsubsection{Total Energy}
The total energy of the system is the summation of the field energy \cref{eq:EMenergy_Bind} and the electron kinetic energy \cref{eq:KenergyEl_SCcylindrical}:
\begin{equation}
    E_{\text{tot}}
=
    E_{\bm{B}^{\text{ext}}}
    +
    E_{\text{el}}^{(T)}
    +
    E_{\bm{B}^{\text{int}}}
=
    \dfrac{i_s^2}{2}
    \underbrace{
        \left(  L_G + L_K \right)
    }_{\equiv L},
\end{equation}
from which we show that the total inductance of the system as $L = L_G + L_K$.
The relation provides an evidence that the total inductance of the circuit has to cover both geometric design and the material properties of the superconducting device.

\end{document}